\documentclass{aa}

\usepackage{txfonts}
\usepackage{subfigure}
\usepackage{graphicx,amsmath}
\usepackage{amssymb}
\usepackage{xcolor}
\usepackage{natbib}
\usepackage{url}
\usepackage{multirow}
\usepackage[para,online,flushleft]{threeparttable}
\usepackage{soul,mathabx}
\bibpunct{(}{)}{;}{a}{}{,}
\usepackage{lscape}
\usepackage{adjustbox,textcomp}
\usepackage{dblfloatfix}
\usepackage{dsfont}
\usepackage{tabularx}
\usepackage{array,multirow,booktabs}
\usepackage{hyperref}
\hypersetup{
    colorlinks=true,
    linkcolor=blue,
    filecolor=blue,
    urlcolor=blue,
    citecolor=blue,
}
\usepackage{braket}
\usepackage{upgreek}
\usepackage{xspace}

\newcommand{\paleos}{\textsc{paleos}\xspace}
\newcommand{\Mearth}{M_\oplus}
\newcommand{\Rearth}{R_\oplus}
\newcommand{\Tsurf}{T_\mathrm{surf}}
\newcommand{\Tpot}{T_\mathrm{pot}}
\newcommand{\Psurf}{P_\mathrm{surf}}
\newcommand{\naad}{\nabla_\mathrm{ad}}
\newcommand{\dd}{\mathrm{d}}

\providecommand{\nolinenumbers}{}

\begin{document}

\title{PALEOS: 
       Multiphase equations of state and mass--radius relations for exoplanet interiors}

\author{Mara~Attia
   \and Tim~Lichtenberg
   \and Ema~Jungov\'{a}
   \and Mariana~Sastre
}

\institute{Kapteyn Astronomical Institute, University of Groningen,
           9747 AD Groningen, The Netherlands\\
           \email{m.attia@rug.nl}
}

\authorrunning{Attia et al.}
\titlerunning{PALEOS: Equations of state for exoplanet interiors}

\date{Received ... / Accepted ...}

\abstract
{
Modeling the interior of a rocky or water-rich exoplanet is fundamentally a thermodynamic closure problem. Every layer's density, temperature gradient, and phase state must follow from an equation of state (EoS) that stays self-consistent across the many orders of magnitude in pressure and temperature spanned between the surface and the core, but existing EoSs are scattered across disciplines, employ different formalisms, and rarely provide the full set of thermodynamic quantities needed for such a model. This presents fundamental hurdles for evolutionary models that rely on a self-consistent thermodynamic treatment of interior phase transitions. We present \paleos{}\thanks{Available at \url{https://github.com/maraattia/PALEOS}} (Planetary Assemblage Layers: Equations of State), an open-source and extensible Python toolkit that consolidates published EoSs for iron and magnesium silicate (MgSiO$_3$) with the composite \textsc{aqua} EoS for water (H$_2$O), which we further correct at high pressure and temperature, into a unified phase-aware and thermally responsive framework covering 17 thermodynamic phases. Vapor and supercritical silicate phases lie outside the current scope of the toolkit. For each material, \paleos analytically derives the density, specific internal energy, entropy, heat capacities, thermal expansion, and adiabatic gradient from the underlying pressure--volume--temperature relations via Maxwell relations. The resulting EoSs are validated for thermodynamic consistency and compiled into lookup tables on regular pressure--temperature grids, which are publicly released. We validate the framework against the preliminary reference Earth model, recovering Earth's radius to 0.3\% and lower-mantle densities to 3\%. Using \paleos, we computed a grid of 17,900 mass--radius relations spanning 0.1--100\,$\Mearth$ for rocky (Fe + MgSiO$_3$) and water-rich (Earth-like core plus H$_2$O envelope) compositions at surface temperatures from 300 to 4000\,K. Because these EoSs treat solid and molten states continuously, thermal expansion remains active from fully solid interiors into the magma-ocean regime. For low-mass silicate planets, thermal expansion inflates the radius by more than 1\% above 1500\,K and by up to 16\% at 4000\,K, a change as large as both the radius spread from bulk composition (the iron--rock--water degeneracy) and the typical transit radius measurement uncertainty. We demonstrate the resulting degeneracy on two ultrashort-period super-Earths, \object{WASP-47\,e} and \object{TOI-1807\,b}, showing that each admits two purely rocky interior solutions indistinguishable in mass and radius but occupying radically different geophysical states: one fully solid and the other hosting a deep magma ocean above a liquid iron core. Phase-aware thermally resolved EoSs are thus essential for translating astronomical observations into exoplanetary geophysics.
}


\maketitle
\nolinenumbers

\section{Introduction}
\label{sec:intro}

Mass and radius are necessary but not sufficient to determine a planet's interior state. The same point in the mass--radius diagram can be occupied by a frozen, fully solid body and by a planet that hosts a deep magma ocean above a liquid metallic core, with consequences that propagate from the core up through the atmosphere \citep{Guimond2024, Lichtenberg2025}. The first JWST observations of ultrashort-period (USP) rocky exoplanets \citep{Hu2024, August2025, Monaghan2025, Teske2025, ParkCoy2026} have made this frozen-versus-molten distinction observationally relevant. Atmospheric retention and composition are sensitive tracers of the underlying thermal state but only insofar as the interior models that interpret them remain thermodynamically self-consistent from surface temperatures of a few hundred Kelvin to multi-megabar core conditions \citep{Lichtenberg2025b, Baumeister2025}. For the majority of known planets, mass and radius are the only bulk properties accessible to observation. Together, they yield the mean density, which provides a first-order constraint on interior composition, that is, whether a planet is predominantly rocky, volatile-rich, or enveloped in a hydrogen--helium atmosphere. However, the mapping from observed mass--radius to internal structure is fundamentally degenerate. Different combinations of the iron core mass fraction (CMF), silicate mantle thickness, and volatile envelope mass can produce the same bulk density \citep{Valencia2006, Seager2007, Rogers2010, Dorn2015}. This compositional degeneracy is compounded by a thermal one. Hotter interiors are less dense, so a large iron-poor planet and a smaller iron-rich but thermally expanded planet can occupy the same point in the mass--radius diagram.

This degeneracy has far-reaching geophysical consequences. Whether the mantle is molten or solid and whether the core is liquid or crystallized governs mantle convection, volcanic outgassing, and ultimately the potential for surface habitability \citep{Driscoll2014, Foley2016, Boukare2022, Boukare2025}. Whether a planet's interior remains molten or solidifies over its evolution strongly alters its capacity for atmospheric retention \citep{Kite2021, Dorn2021, KrissansenTotton2024, Boer2025}, and the coupling runs both ways: A retained atmosphere traps heat, slows cooling, and prolongs the magma ocean stage so that interior thermal state and atmospheric inventory coevolve as a single thermodynamic system \citep{Nicholls2026, Steinmeyer2026}. For short-period exoplanets, this coupling is becoming directly observable, with molten and partially molten surfaces suggested to drive variability in thermal emission and phase curves \citep{Meier2023, Loftus2025, Meier2026, Farhat2026}. Capturing the coupling requires interior models that go beyond density to track the full thermodynamic state of each layer, including entropy, heat capacity, thermal expansivity, and the adiabatic gradient that governs the radial temperature profile, and that resolve the liquid-to-solid transition self-consistently across a vast temperature and pressure range, a feat unresolved by any publicly available equation of state (EoS).

The thermodynamic data needed for such models have accumulated over decades of work in mineral physics, geophysics, shock-compression experiments, and ab initio simulations \citep{Duffy2019}. Self-gravitation and internal heating drive interior pressures and temperatures to extreme values, reaching the terapascal level in the cores of giant planets and massive super-Earths \citep{Helled2011, Militzer2013, Militzer2023, Attia2025}, where planetary materials undergo a cascade of solid--solid phase transitions and melting. The boundaries of these transitions often remain contested between experimental groups \citep[as for iron,][]{Boehler1993, Anzellini2013}. Water adds a further layer of complexity, requiring composite EoS descriptions that stitch together ice polymorphs, liquid, vapor, and supercritical phases across many orders of magnitude in pressure \citep{Senft2008, Haldemann2020}. Efforts to consolidate this record range from comprehensive phase-equilibrium databases \citep{Dong2025} to self-consistent thermodynamic modeling frameworks \citep{Stixrude2005, Stixrude2022}, but each source uses its own formalism and provides a different subset of thermodynamic quantities. Assembling self-consistent multi-material models therefore requires homogenizing these disparate sources, which were never designed to be combined, into a coherent description of planetary interiors.

Several existing tools address parts of this landscape. \textsc{BurnMan} \citep{Cottaar2014} combines user-supplied EoSs but leaves phase-diagram construction and stable-phase selection to the user. \textsc{ExoPlex} \citep{Unterborn2023} solves the interior structure problem but returns the density rather than the broader thermodynamic state, and \textsc{magrathea} \citep{Huang2022} computes mass--radius relations with a modular EoS backend but without phase-resolved output. None of the methods consolidates a multi-material EoS that delivers the density, entropy, heat capacities, thermal expansion, and adiabatic gradient from a single query, with automatic phase selection across the full pressure--temperature domain of planetary interiors. In response to this situation, we present \paleos (Planetary Assemblage Layers: Equations of State), an open-source Python package that fills this gap. It implements EoSs for three materials representative of the core, the mantle, and volatiles -- iron (Fe), magnesium silicate (MgSiO$_3$), and water (H$_2$O) -- spanning 17 thermodynamic phases from 1\,bar to 100\,TPa and 100 to $10^5$\,K, and for each material, it constructs a complete phase diagram with automatic phase selection and derives core thermodynamic quantities -- density ($\rho$), specific internal energy ($u$), specific entropy ($s$), isobaric and isochoric heat capacities ($C_P$ and $C_V$), thermal expansion coefficient ($\alpha$), and adiabatic gradient ($\naad$) -- analytically from the underlying pressure--volume--temperature relations via Maxwell relations. The resulting EoSs are compiled into lookup tables on regular $(P, T)$ grids for efficient interpolation and released publicly on Zenodo.\footnote{\url{https://doi.org/10.5281/zenodo.19000315}}

This paper makes three contributions. First, we present the thermodynamic framework and its application to each material, validating the resulting EoS against the preliminary reference earth model \citep[PREM;][]{Dziewonski1981}, which recovers Earth's radius and lower-mantle densities. Second, we compute a grid of 17,900 mass--radius relations for rocky and water-rich planets at finite surface temperatures from 300 to 4000\,K, released as a second Zenodo dataset.\footnote{\url{https://doi.org/10.5281/zenodo.19221214}} Third, we demonstrate the composition--temperature degeneracy on two USP super-Earths, \object{WASP-47\,e} \citep{Becker2015} and \object{TOI-1807\,b} \citep{Hedges2021}, showing that mass--radius data alone cannot distinguish a cold, fully solid interior from a hot partially molten one. This article is organized as follows. Section~\ref{sec:thermo} introduces the thermodynamic framework. Section~\ref{sec:eos} describes the EoS for each material and Sect.~\ref{sec:prem} validates them against PREM. Section~\ref{sec:mr} presents the mass--radius grid and compares it to existing mass--radius tables. Section~\ref{sec:superearths} applies the framework to \object{WASP-47\,e} and \object{TOI-1807\,b}. We discuss implications and limitations in Sect.~\ref{sec:discussion}.

\section{Thermodynamics}
\label{sec:thermo}

\subsection{General framework}
\label{sec:thermo:framework}

An EoS is an empirical relationship between the pressure ($P$), volume ($V$), and temperature ($T$) of a material. No absolute thermodynamic basis exists for specifying the correct functional form of an EoS for solids \citep{Angel2000}; all formalisms in common use rest on assumptions whose validity can only be assessed against experimental compression and elasticity data. Because no single formalism spans the full pressure--temperature domain of a planetary interior, \paleos adopts the conventional modular decomposition of the total pressure:
\begin{equation}
\label{eq:pvt_decomposition}
P(V, T) = P_\mathrm{cold}(V) + P_\mathrm{th}(V, T) + P_\mathrm{el}(V, T) + P_\mathrm{mag}(V, T),
\end{equation}
where $P_\mathrm{cold}$ is the static-lattice (zero-temperature) pressure, $P_\mathrm{th}$ is the thermal-phonon contribution, $P_\mathrm{el}$ is the electronic excitation pressure, and $P_\mathrm{mag}$ is the magnetic contribution. Not every material requires all four terms. The electronic and magnetic contributions are relevant only for metals at high temperatures \citep{Hillert1978, Dinsdale1991, Iota2007}. Liquid phases replace the phonon model with thermal parametrizations calibrated to their partially randomized structures \citep[e.g.,][]{Wolf2018, Ichikawa2020} since their individual atomic bonds constantly break and reform in response to bulk stresses and localized atomic diffusion \citep{Stebbins1988}. Each term of Eq.~\eqref{eq:pvt_decomposition} can be chosen independently for a given material, so the same framework accommodates both simple two-term models (cold + phonon) and the full four-term expression needed for iron. In the following Sects.~\ref{sec:thermo:cold} and \ref{sec:thermo:thermal}, we present the various formalisms employed in \paleos corresponding to those terms.

From this decomposition, seven output quantities fully specify the thermodynamic state needed for planetary interior structure integration: the density ($\rho$), internal energy ($U$), entropy ($S$), isochoric and isobaric heat capacities ($C_V$ and $C_P$), thermal expansivity ($\alpha$), and adiabatic temperature gradient ($\naad$). For any pressure--temperature point $(P_\mathrm{target}, T_\mathrm{target})$, the volume $V(P_\mathrm{target}, T_\mathrm{target})$ is obtained by numerical root-finding ($P(V, T_\mathrm{target}) = P_\mathrm{target}$), from which $\rho$ follows directly. The remaining six quantities are derived from the first and second partial derivatives of $P(V, T)$:
\begin{align}
\left(\frac{\partial U}{\partial V}\right)_T &= T\left(\frac{\partial P}{\partial T}\right)_V - P, \label{eq:du_dv} \\
\left(\frac{\partial S}{\partial V}\right)_T &= \left(\frac{\partial P}{\partial T}\right)_V, \label{eq:ds_dv} \\
C_V &= \left(\frac{\partial U}{\partial T}\right)_V, \label{eq:cv} \\
C_P &= C_V - T \left(\frac{\partial P}{\partial T}\right)_V^2 \left(\frac{\partial P}{\partial V}\right)_T^{-1}, \label{eq:cp} \\
\alpha &= \frac{1}{V}\left(\frac{\partial V}{\partial T}\right)_P = -\frac{1}{V}\frac{(\partial P/\partial T)_V}{(\partial P/\partial V)_T}, \label{eq:alpha} \\
\naad &= \frac{P}{T}\left(\frac{\partial T}{\partial P}\right)_S = \frac{P\,\alpha}{\rho\,C_P}. \label{eq:nabla_ad}
\end{align}
The heat capacities, thermal expansivity, and adiabatic gradient follow from local derivatives. Internal energy and entropy require isothermal integration of Eqs.~\eqref{eq:du_dv}--\eqref{eq:ds_dv} from a reference volume. In practice, each pressure contribution in Eq.~\eqref{eq:pvt_decomposition} yields a corresponding internal energy $U_i$, so the total one decomposes as $U = \Sigma_i U_i + U_0$, where $U_0$ is a phase-dependent reference offset (Sect.~\ref{sec:eos:refstate}). Entropy follows the same decomposition logic $S = \Sigma_i S_i + S_0$ ($S_0$ also being a phase-dependent offset, Sect.~\ref{sec:eos:refstate}), with the difference that it has no static-lattice contribution $S_\mathrm{cold}$ since it is intrinsically thermal. The bulk modulus $K_T = -V(\partial P/\partial V)_T$, measuring resistance to bulk compression, likewise splits into cold and thermal components, and represents an important auxiliary quantity used for all calculations necessitating volume derivatives of the pressure, such as $\alpha$. In fact, another related auxiliary parameter is $\alpha K_T = (\partial P/\partial T)_V$ and controls the temperature derivative of the pressure, closing the system of structural equations (Eqs.~\eqref{eq:du_dv}--\eqref{eq:nabla_ad}). 

Assembling a composite EoS from diverse cold and thermal formalisms across several orders of magnitude in pressure and temperature demands a quantitative consistency check. Following \citet{Timmes1999} and \citet{Becker2014}, we evaluated the dimensionless consistency parameter
\begin{equation}
\label{eq:delta}
\Delta \equiv 1 - \rho^2 \frac{(\partial S / \partial P)_T}{(\partial \rho / \partial T)_P},
\end{equation}
which vanishes identically for a perfectly consistent EoS via the Maxwell relation $(\partial S/\partial P)_T = -(\partial V/\partial T)_P$. \citet{Haldemann2020} demonstrate that their composite \textsc{aqua} EoS satisfies $\vert\Delta\vert < 0.01$ across most of the pressure--temperature domain; we adopt the same threshold as our consistency criterion.

\subsection{Cold pressure formalisms}
\label{sec:thermo:cold}

The static-lattice pressure $P_\mathrm{cold}(V)$ dominates the total pressure at planetary interior conditions, and its parametrization therefore controls the accuracy of the model. The theoretical development of isothermal EoS proceeds from finite-strain theory \citep{Murnaghan1951, Birch1978, Jeanloz1988}, through empirically motivated interatomic-potential forms \citep{Rose1984}, to formulations incorporating the Thomas--Fermi high-pressure limit \citep{Holzapfel1998}. \paleos employs five formalisms depending on the material and pressure range. All are three-parameter forms, specified by the zero-pressure volume $V_0$, the bulk modulus at zero pressure $K_0 = K_T(P{=}0)$, and its pressure derivative $K_0' = (\partial K_T/\partial P)_{T,\,P=0}$, except the Keane equation which adds a fourth parameter $K_\infty'$. Each formalism implies a distinct value of the second derivative $K_0''$, which controls extrapolation behavior at high compression. The parameters $K_0$ and $K_0'$ are strongly anticorrelated in least-squares fits: an increase in $K_0$ can be compensated by a decrease in $K_0'$ while maintaining comparable residuals \citep{Angel2000}. Consequently, for moderate compressions ($V/V_0 \gtrsim 0.85$), different formalisms yield equivalent fits \citep{Holzapfel1996}. The choice becomes significant only at the high compressions ($V/V_0 \lesssim 0.6$) characteristic of super-Earth interiors.

The third-order Birch--Murnaghan (BM3) equation, historically the first widely adopted finite-strain EoS \citep{Murnaghan1944, Birch1947}, expands the Helmholtz free energy in powers of the Eulerian strain $f_\mathrm{E} = \frac{1}{2}[(V_0/V)^{2/3} - 1]$, which ensures better convergence properties at high compression than the Lagrangian alternative \citep{Angel2000}. The resulting pressure is
\begin{equation}
\label{eq:bm3}
P_\mathrm{cold}(V) = 3 K_0 f_\mathrm{E} (1 + 2f_\mathrm{E})^{5/2} \left[1 + \frac{3}{2}(K_0' - 4)f_\mathrm{E}\right].
\end{equation}
Setting $K_0' = 4$ reduces BM3 to the second order (BM2); this value has no physical basis and merely reflects truncation of the strain expansion. The implied $K_0''$ of BM3 differs from that of other forms, so parameters fitted to one formalism should not be transferred to another at high compression.

The Vinet equation \citep{Stacey1981, Vinet1987} takes a different approach, deriving the cold pressure from a universal scaling of interatomic binding-energy curves:
\begin{equation}
\label{eq:vinet}
P_\mathrm{cold}(V) = 3 K_0 \frac{1 - X}{X^2} \exp\!\left[\eta_\mathrm{V}(1 - X)\right], \quad \eta_\mathrm{V} = \frac{3}{2}(K_0' - 1),
\end{equation}
where $X = (V/V_0)^{1/3}$ is the linear compression ratio. The Vinet form provides a reliable description across all compression ratios \citep{Jeanloz1981} and is particularly well suited at high compressions ($V/V_0 \lesssim 0.6$), where BM3 can develop unphysical oscillations in $K_T$ \citep{Angel2000}.

The remaining three formalisms address the known failure of BM3 and Vinet at extreme compression. The Holzapfel equation \citep[also called adapted polynomial expansion of second order or AP2]{Holzapfel1996, Holzapfel1998} enforces the Thomas--Fermi limit $P \propto V^{-5/3}$ as $V \to 0$:
\begin{equation}
\label{eq:holzapfel}
P_\mathrm{cold}(V) = 3 K_0 \frac{1 - X}{X^5}\left[1 + c_2\,X(1-X)\right]\exp\!\left[c_0(1 - X)\right],
\end{equation}
where $c_0 = -\ln(3K_0/p_\mathrm{FG0}) + c_2$, with $p_\mathrm{FG0}$ the Fermi gas pressure at $V_0$ and $c_2$ a material-dependent correction related to $K_0'$. The Keane equation \citep{Keane1954, Stacey2004} takes a different approach, introducing a finite asymptotic pressure derivative $K_\infty'$ that prevents the unphysical divergence of $K_T'$ at infinite compression:
\begin{equation}
\label{eq:keane}
P_\mathrm{cold}(V) = K_0\left[\frac{K_0'}{K_\infty'^2}\!\left(Y^{K_\infty'} - 1\right) - \!\left(\frac{K_0'}{K_\infty'} - 1\right)\ln Y\right],
\end{equation}
where $Y = V_0/V$. This is the only four-parameter form in \paleos. Finally, the Kunc equation \citep[$k = 5$,][]{Kunc2003} generalizes the Holzapfel form with a tunable exponent $k$ controlling the high-pressure asymptote:
\begin{equation}
\label{eq:kunc}
P_\mathrm{cold}(V) = 3 K_0 \frac{1 - X}{X^k}\exp\!\left[\eta_\mathrm{K}(1-X)\right], \quad \eta_\mathrm{K} = \frac{3}{2}K_0' - k + \frac{1}{2}.
\end{equation}
Setting $k = 2$ recovers the Vinet equation and the default $k = 5$ matches the Holzapfel asymptote. Closed-form expressions for $K_T^\mathrm{cold}$ and $U_\mathrm{cold}$ for each formalism are collected in Appendix~\ref{app:cold}.

\subsection{Thermal models}
\label{sec:thermo:thermal}

For crystalline solids, lattice symmetries constrain interatomic bonding, and the quasiharmonic approximation provides an effective description of thermal properties. The Gr\"{u}neisen parameter $\gamma = \alpha K_T V / C_V$, which couples thermal expansion to elastic properties, depends only on volume. In the Mie--Gr\"{u}neisen framework \citep{Mie1903, Gruneisen1912}, $\gamma$ relates the thermal pressure to the thermal energy:
\begin{equation}
\label{eq:pth}
P_\mathrm{th}(V, T) = \frac{\gamma(V)}{V}\left[U_\mathrm{th}(V, T) - U_\mathrm{th}(V, T_0)\right],
\end{equation}
where $T_0$ is the reference temperature. Two phonon models compute $U_\mathrm{th}$ for Eq.~\eqref{eq:pth} in \paleos. The Mie--Gr\"{u}neisen--Debye (MGD) model, the most widely used thermal EoS for solid minerals, sums over a Debye phonon spectrum \citep{Debye1912}:
\begin{equation}
\label{eq:debye_energy}
U_\mathrm{th}(V, T) = 9 n R T \left(\frac{T}{\Theta_\mathrm{D}}\right)^{\!-3} \int_0^{\Theta_\mathrm{D}/T} \frac{x^3}{\mathrm{e}^x - 1}\,\dd x \equiv 3nRT\,D_3\!\left(\frac{\Theta_\mathrm{D}}{T}\right),
\end{equation}
with $n$ as the number of atoms per formula unit, $R$ as the gas constant, $D_3$ as the third Debye function, and $\Theta_\mathrm{D}(V)$ as the volume-dependent Debye temperature. The isochoric heat capacity $C_V = (\partial U_\mathrm{th}/\partial T)_V$ and entropy $S$, at least their thermal contributions, both follow analytically from $D_3$. At high temperatures ($T \gg \Theta_\mathrm{D}$), $C_V$ tends to the Dulong--Petit limit $3nR$, and the product $\alpha K_T$ becomes nearly temperature-independent \citep{Chopelas1992, Anderson1995}. In this regime, $P_\mathrm{th}$ is approximately linear in $T$, which simplifies extrapolation beyond measured conditions. The Einstein model \citep{Einstein1907} is the alternative to the Debye spectrum, with a single characteristic frequency $\Theta_\mathrm{E}(V)$, yielding 
\begin{equation}
\label{eq:einstein_energy}
U_\mathrm{th}(V, T) = 3nR\left[\frac{\Theta_\mathrm{E}}{2} + \frac{\Theta_\mathrm{E}}{\exp(\Theta_\mathrm{E}/T) - 1}\right]. 
\end{equation}
It is simpler than the MGD model but adequate when the cold pressure dominates ($P_\mathrm{cold} \gg P_\mathrm{th}$). The Debye and Einstein variants of the Mie--Gr\"{u}neisen framework practically coincide at high temperatures characteristic of super-Earth cores \citep{Poirier1991}, both recovering the abovementioned linear regime. Central to both models is the Gr\"{u}neisen parameter $\gamma(V)$, which not only couples the thermal pressure to the thermal energy through Eq.~\eqref{eq:pth} but also determines the characteristic temperature $\Theta(V)$ (Debye or Einstein) via the relation $\dd\ln\Theta/\dd\ln V = -\gamma$. The parametrizations of $\gamma(V)$ and the resulting closed-form expressions for $\Theta(V)$ are given in Appendix~\ref{app:gruneisen}.

The quasiharmonic assumption that $\gamma$ depends only on volume breaks down for liquids. Unlike solids, atoms in liquids retain only short-range order: this structural freedom causes the compressive and thermal properties of liquids to be more sensitive to pressure and temperature than those of their solid counterparts, and the Gr\"{u}neisen parameter itself acquires temperature dependence \citep{Knopoff1969, Knopoff1970}. The RTpress model \citep{Wolf2018} addresses this by decomposing the Helmholtz free energy into three independent components: (1) an isothermal reference compression curve $P(V, T_0)$ described by a standard cold-pressure, (2) a reference adiabatic temperature profile $T_{0S}(V)$ determined by the Gr\"{u}neisen parameter along the adiabat, and (3) a generalized Rosenfeld--Tarazona \citep[RT,][]{Rosenfeld1998} thermal perturbation that captures deviations from the reference adiabat. The thermal perturbation uses the RT power law ($U \propto b\,T^{3/5}$) with a polynomial volume-dependent coefficient $b(V) = \sum_n b_n\,(V/V_0 - 1)^n$, whose coefficients are fitted to molecular dynamics (MD) simulations. All thermodynamic quantities follow analytically from derivatives of the free energy surface, ensuring self-consistency. 

For iron, two additional contributions can be significant at high temperatures. The electronic excitation pressure $P_\mathrm{el}(V, T) \propto e(V) T^2 / V$ arises from thermal excitation of conduction electrons, with a volume-dependent electronic Gr\"{u}neisen parameter $e$ \citep[which can be seen as the simplest form of anharmonicity]{Zharkov1971}. The magnetic contribution, relevant only for select iron polymorphs, is modeled via a piecewise parametric formalism \citep{Dinsdale1991, Jacobs2010} affecting directly the magnetic Helmholtz free energy. It depends only on temperature and therefore does not contribute to the pressure but affects $C_V$ and $S$. Full expressions for all thermal models and their derivatives are given in Appendix~\ref{app:thermal}.

\section{Equations of state}
\label{sec:eos}

\subsection{Iron}
\label{sec:eos:iron}

\begin{figure}
\centering
\includegraphics[width=\columnwidth]{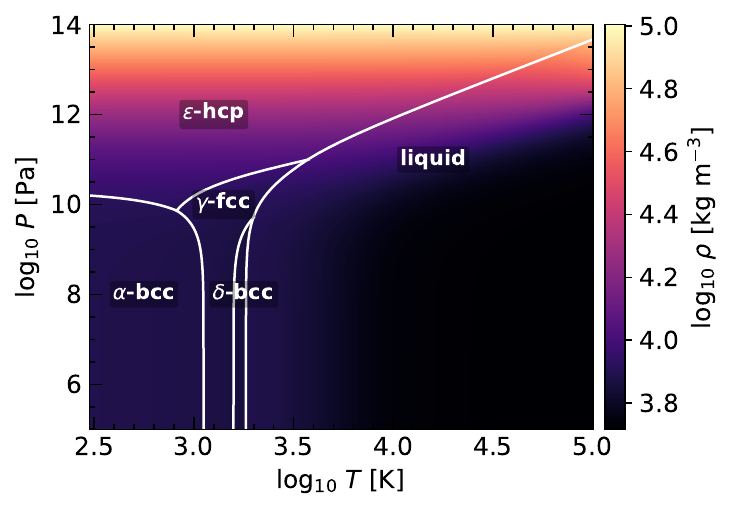}
\caption{Iron EoS implemented in \paleos. Density is shown in $(\log_{10}T,\, \log_{10}P)$ space across the table domain ($300\,\mathrm{K} \le T \le 10^5\,\mathrm{K}$, $1\,\mathrm{bar} \le P \le 100\,\mathrm{TPa}$). White curves are the phase boundaries used in \paleos{}: the four solid--solid transitions $\alpha \leftrightarrow \gamma$, $\delta \leftrightarrow \gamma$, $\alpha \leftrightarrow \varepsilon$, and $\gamma \leftrightarrow \varepsilon$ together with the melting curve $T_\mathrm{melt}^\mathrm{Fe}(P)$. Labels mark the fields of $\alpha$-bcc, $\delta$-bcc, $\gamma$-fcc, $\varepsilon$-hcp, and liquid iron. The color map is continuous across every phase boundary, illustrating the smoothness of the composite EoS over the full planetary domain.}
\label{fig:iron_phase}
\end{figure}

\begin{table*}
\caption{Equation of state parameters for iron phases in \paleos ($n = 1$ for all phases).}
\label{tab:iron_params}
\centering
\begin{threeparttable}
\footnotesize
\begin{tabular}{llrrrrrrrrl}
\toprule
Phase & Formalism & $V_0$ & $K_0$ & $K_0'$ & $T_0$ & $\Theta_0$ & $\gamma_0$ & $\gamma_\infty$ & $q$ or $\beta$ & Reference \\
      &  & \multicolumn{1}{r}{[cm$^3$/mol]} & \multicolumn{1}{r}{[GPa]} & & \multicolumn{1}{r}{[K]} & \multicolumn{1}{r}{[K]} & & & & \\
\midrule
$\alpha/\delta$-bcc & Vinet + Ein.    & 7.092 & 164.0  & 5.50  & 298.15 & 303   & 1.736 & ---   & 1.125 & \citet{Dorogokupets2017} \\
$\gamma$-fcc        & Vinet + Ein.    & 6.929 & 146.2  & 4.67  & 298.15 & 222.5 & 2.203 & ---   & 0.01  & \citet{Dorogokupets2017} \\
$\varepsilon$-hcp ($< 310$\,GPa) & BM3 + MGD           & 6.870 & 129.0  & 6.24  & 300    & 420   & 1.11  & ---   & 0.3   & \citet{Miozzi2020} \\
$\varepsilon$-hcp ($> 310$\,GPa) & Holz.\,$^\dag$ + Ein. & 4.286 & 1145.7$^\dag$ & 3.19$^\dag$ & 300 & 44.6 & 1.408 & 0.827 & 0.826 & \citet{Hakim2018} \\
liquid              & BM3 + linear$^\ddag$ & 9.823 & 49.2   & 4.976 & 8000   & ---   & ---   & ---   & ---   & \citet{Luo2024} \\
\bottomrule
\end{tabular}
\begin{tablenotes}[flushleft]\footnotesize
\item[$\dag$] Holzapfel cold EoS: $K_0$ and $K_0'$ columns list $K_{T0}$ and $c_0$; $V_0$ is the volume at $P_0 = 234.4$\,GPa, $c_2 = -2.40$.
\item[$\ddag$] See Appendix~\ref{app:thermal}.
\end{tablenotes}
\end{threeparttable}
\end{table*}

Iron is the most tightly bound nucleus and the most cosmochemically abundant refractory metal \citep{Burbidge1957, Asplund2021}, and the only abundant element that condenses as a pure elemental metal in the protoplanetary disk \citep{Lodders2003}. That metallic preference, combined with its high density relative to silicates, drives gravitational differentiation: in a sufficiently large embryo, metallic iron sinks to form a dense core \citep{Nimmo2015, Rubie2016}, so the iron phase diagram maps directly onto the layered structure of planetary cores. At ambient conditions iron adopts the body-centered cubic (bcc) structure; with increasing pressure it transforms through face-centered cubic (fcc) to hexagonal close-packed (hcp), the stable solid phase at Earth's inner-core pressures \citep{Tateno2010}, while liquid iron fills the outer core. The same sequence extends to the terapascal pressures of super-Earth cores, where the $\varepsilon$-hcp phase persists to at least 10\,TPa in density functional theory (DFT) calculations \citep{Hakim2018}. No single EoS calibration spans this range: laboratory experiments constrain the low-pressure phases to ${\sim}\,350$\,GPa \citep{Dorogokupets2017}, while ab initio simulations reach multi-TPa conditions but carry their own systematic uncertainties \citep{Luo2024}. Assembling a composite description from these disparate sources, with mutual thermodynamic consistency across phase boundaries, is the central challenge.

The \paleos iron EoS covers five phases: $\alpha$-bcc and $\delta$-bcc, $\gamma$-fcc from \citet{Dorogokupets2017}, $\varepsilon$-hcp from \citet{Miozzi2020} blended with \citet{Hakim2018}, and liquid from \citet{Luo2024}. Solid--solid phase boundaries follow \citet{Dorogokupets2017} and the melting curve follows \citet{Anzellini2013}; the analytical forms are collected in Appendix~\ref{app:boundaries:iron}. The phase diagram is shown in Fig.~\ref{fig:iron_phase}, and the EoS parameters are listed in Table~\ref{tab:iron_params}. The key contribution of \citet{Dorogokupets2017} is the simultaneous optimization of all four phases (three solid polymorphs plus liquid) from a single Helmholtz free energy framework, using a Vinet cold pressure (Eq.~\eqref{eq:vinet}), Einstein thermal model (Eq.~\eqref{eq:einstein_energy}), and electronic Gr\"{u}neisen terms (Eq.~\eqref{eq:grunel}). The EoS is mutually constrained by a large combined experimental dataset spanning static compression in diamond anvil cells, shock-wave Hugoniot data, and ambient calorimetric measurements. This unified approach ensures that the phase boundaries and triple points emerge self-consistently from the free energy surfaces rather than being imposed ad hoc. Three triple points anchor the topology of the phase diagram: bcc--fcc--liquid at $(5.2\;\mathrm{GPa},\, 1991\;\mathrm{K})$, bcc--fcc--hcp at $(7.3\;\mathrm{GPa},\, 820\;\mathrm{K})$, and the geophysically critical fcc--hcp--liquid triple point near $(100\;\mathrm{GPa},\, 3700\;\mathrm{K})$, whose precise location remains debated because of inconsistent pressure scales across experimental groups \citep{Boehler1993, Komabayashi2009, Anzellini2013, Zhang2016, Dorogokupets2017}. 

The $\alpha/\delta$-bcc phase is ferromagnetic below the Curie temperature $T_\mathrm{c} = 1043$\,K. The magnetic free energy (Appendix~\ref{app:magnetic}), modeled following \citet{Dinsdale1991} with saturation moment $B_0 = 2.22\,\upmu_\mathrm{B}$ and structure factor $p = 0.4$, depends only on temperature and therefore affects the heat capacity and entropy but not the pressure. This is physically consequential: the magnetic transition produces a lambda anomaly in $C_P$ near $T_\mathrm{c}$ that influences thermal profiles at shallow depths. Under compression, however, the magnetic moment is suppressed and effectively vanishes in the stability field of the high-pressure phases. The iron of everyday experience is ferromagnetic, but the iron of planetary cores is not. The $\gamma$-fcc phase occupies a narrow stability field between bcc and hcp (Fig.~\ref{fig:iron_phase}), carries no magnetic moment, and exhibits a nearly volume-independent Gr\"{u}neisen parameter ($q = 0.01$, Eq.~\eqref{eq:gamma_power}, Table~\ref{tab:iron_params}), meaning its thermal pressure changes minimally with compression.

The $\varepsilon$-hcp phase dominates planetary cores above ${\sim}\,15$\,GPa and spans the largest pressure range of any iron phase in interior models: from the shallow core of Mars \citep[${\sim}\,40$\,GPa, e.g.,][]{Man2025} through Earth's inner core \citep[${\sim}\,330$\,GPa,][]{Dziewonski1981} to the centers of massive sub-Neptunes \citep[${\sim}\,10$\,TPa,][]{Attia2021}. Covering this range requires two complementary EoS calibrations. \citet{Miozzi2020} provide a BM3 + MGD description calibrated against X-ray diffraction measurements in laser-heated diamond anvil cells to 310\,GPa. This calibration anchors the EoS at Earth-relevant pressures, where experimental data are most abundant and the Debye model provides an adequate phonon description. \paleos reproduces their reported densities at 68, 100, and 135\,GPa to within the experimental scatter. Above 310\,GPa, experimental constraints become sparse and extrapolation of the BM3 cold curve becomes unreliable. \citet{Hakim2018} fill this gap with a DFT-calibrated EoS using a Holzapfel cold pressure (Eq.~\eqref{eq:holzapfel}), which enforces the correct Thomas--Fermi high-compression limit, combined with an Einstein thermal model and an electronic Gr\"{u}neisen contribution (Table~\ref{tab:el_mag_params}). Their calibration extends from 234\,GPa to 10\,TPa, covering the full pressure range of super-Earth cores. We blend the two descriptions via a smoothstep (Hermite cubic) transition in the overlap zone $310 \pm 100$\,GPa, applied directly to each thermodynamic quantity. This ensures continuity in all output quantities \citep[which cannot otherwise be enforced with sole interpolation of free energy, even using a special scheme,][]{Swesty1996, Baturin2019}, at the cost of moderately breaking the underlying Maxwell relations within the narrow transition zone. 

The liquid iron EoS of \citet{Luo2024} combines a BM3 cold pressure with a custom thermal contribution fitted to their ab initio MD simulations at $8000$--$14{,}000$\,K and 50--1300\,GPa. Rather than a Mie--Gr\"{u}neisen relation, the thermal pressure is an ad hoc polynomial in $(V,\,T)$ that is linear in temperature and quadratic in the compression ratio $Y$. This parametrization reproduces the simulated pressures with high fidelity: \paleos recovers the published MD volumes at 250, 500, and 1000\,GPa to within a few percent. The polynomial form, however, introduces a thermodynamic inconsistency: because the thermal energy and entropy derived from the free energy are independent of temperature, the isochoric heat capacity vanishes identically. We restore a physical value by supplementing the free energy with Dulong--Petit correction terms that impose $C_V = 3nR$ (Appendix~\ref{app:luo24}); this correction affects the caloric properties (entropy, heat capacity) but leaves the pressure and density unchanged. We also evaluated the liquid iron EoS of \citet{Ichikawa2020}, which couples a Vinet cold curve to a constant-$\gamma$ Mie--Gr\"{u}neisen thermal model. This formulation is thermodynamically self-consistent by construction, deriving all quantities from a single free energy surface. It fails, however, for temperatures above the reference temperature $T_0 = 8000$\,K at pressures below ${\sim}\,100$\,GPa: the positive thermal pressure raises the minimum achievable total pressure above zero, so the root-finding for $V(P,T)$ has no solution in the physical volume range. This limitation excludes hot-start and giant-impact conditions relevant for young planets, even though it may in fact correspond to a vapor or supercritical phase (as discussed for MgSiO$_3$, Sect.~\ref{sec:eos:mgsio3}). We therefore adopted \citet{Luo2024} despite the need for post hoc thermodynamic corrections. 

The thermodynamic consistency parameter $\vert\Delta\vert$ (Eq.~\eqref{eq:delta}) satisfies $\vert\Delta\vert < 0.01$ for 98.2\% of the $P$--$T$ domain, with a median of $1.4 \times 10^{-7}$. The \citet{Dorogokupets2017} phases ($\alpha$-bcc, $\delta$-bcc, $\gamma$-fcc) achieve the highest consistency, with $\max\vert\Delta\vert < 10^{-4}$ across their entire stability fields, a direct consequence of deriving all quantities from a single Helmholtz free energy. The \citet{Luo2024} liquid yields $\max\vert\Delta\vert \simeq 5 \times 10^{-3}$, acceptable given the abovementioned corrections. The residual inconsistency is localized to the $\varepsilon$-hcp smoothstep transition zone, where blending individual thermodynamic quantities preserves continuity but not the underlying Maxwell relations. This is an acceptable tradeoff given the narrow width of the transition, and the alternative (a sharp discontinuity at $310$\,GPa) would be more problematic for interior structure calculations.

\subsection{Magnesium silicate}
\label{sec:eos:mgsio3}

\begin{figure}
\centering
\includegraphics[width=\columnwidth]{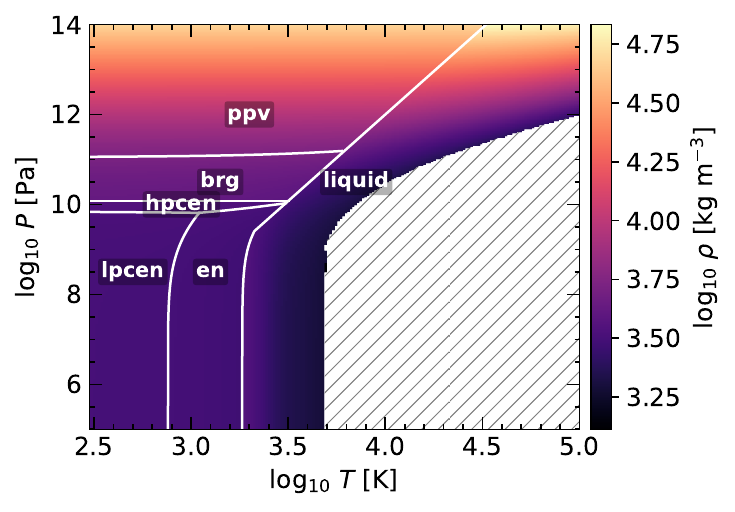}
\caption{Magnesium silicate EoS implemented in \paleos. Density is shown in $(\log_{10}T,\, \log_{10}P)$ space across the same domain as Fig.~\ref{fig:iron_phase}. White curves are the phase boundaries used in \paleos: the lpcen--en, lpcen--hpcen, en--hpcen, and hpcen--brg pyroxene--orthosilicate transitions; the brg--ppv transition; and the melting curve $T_\mathrm{melt}^{\mathrm{MgSiO_3}}(P)$. Labels mark the fields of lpcen, en, hpcen, brg, ppv, and liquid MgSiO$_3$. The hatched gray region in the lower right is the corner of the $(P, T)$ box where the liquid EoS does not converge (vapor and supercritical regime), corresponding to the masked nodes of the public table.}
\label{fig:mgsio3_phase}
\end{figure}

\begin{table}
\caption{Electronic contribution parameters$^\dag$.}
\label{tab:el_mag_params}
\centering
\begin{threeparttable}
\footnotesize
\begin{tabular}{llrr}
\toprule
Material & Phase & $e_0$ & $g$ \\
         &       & \multicolumn{1}{c}{($10^{-6}$\,K$^{-1}$)} & \\
\midrule
Fe         & $\alpha/\delta$-bcc  & 198   & 1.0 \\
Fe         & $\gamma$-fcc         & 198   & 0.5 \\
Fe         & $\varepsilon$-hcp ($> 310$\,GPa)  & 212.1 & 1.891 \\
MgSiO$_3$  & en                   & 13    & 1.0 \\
\bottomrule
\end{tabular}
\begin{tablenotes}[flushleft]\footnotesize
\item[$\dag$] The ``anharmonic-electronic'' $(a_0,\,m)$ of \citet{Hakim2018} and ``anharmonic'' $(a_0,\,m_\mathrm{anh})$ of \citet{Sokolova2022} are formally identical to the electronic G\"uneisen $(e_0,\,g)$ of Eq.~\eqref{eq:grunel}.
\end{tablenotes}
\end{threeparttable}
\end{table}

\begin{table*}
\caption{Equation of state parameters for MgSiO$_3$ phases in \paleos ($n = 5$ for all phases).}
\label{tab:mgsio3_params}
\centering
\begin{threeparttable}
\footnotesize
\begin{tabular}{llrrrrrrrrrl}
\toprule
Phase & Formalism & $V_0$ & $K_0$ & $K_0'$ & $K_\infty'$ & $T_0$ & $\Theta_0$$^\dag$ & $\gamma_0$ & $\gamma_\infty$ & $q$ or $\beta$ & Reference \\
      &  & \multicolumn{1}{r}{[cm$^3$/mol]} & \multicolumn{1}{r}{[GPa]} & & & \multicolumn{1}{r}{[K]} & \multicolumn{1}{r}{[K]} & & & & \\
\midrule
lpcen  & Kunc + Ein.  & 31.280 & 112.8 & 6.27  & ---   & 298.15 & 327\,/\,973 & 0.98  & --- & 0.6  & \citet{Sokolova2022} \\
en     & Kunc + Ein.  & 31.350 & 106.2 & 7.80  & ---   & 298.15 & 327\,/\,973 & 1.0   & --- & 1.0  & \citet{Sokolova2022} \\
hpcen  & Kunc + Ein.  & 30.310 & 128.4 & 4.52  & ---   & 298.15 & 327\,/\,973 & 0.98  & --- & 0.6  & \citet{Sokolova2022} \\
brg    & Vinet + MGD      & 24.408 & 262.3 & 4.044 & ---   & 300    & 1000        & 1.675 & --- & 1.39 & \citet{Wolf2015} \\
ppv    & Keane + MGD      & 24.724 & 205.4 & 5.069 & 2.627 & 300    & 995         & 1.495 & 0.818 & 1.97 & \citet{Sakai2016} \\
liquid & Vinet + RT$^\ddag$ & 43.215$^\S$ & 13.53 & 6.767 & --- & 3000 & ---         & 0.158 & --- & ---  & \citet{Wolf2018} \\
\bottomrule
\end{tabular}
\begin{tablenotes}[flushleft]\footnotesize
\item[$\dag$] Dual Einstein temperatures $\Theta_{01}/\Theta_{02}$ with mode weights $(6.785,\,8.215)$.
\item[$\ddag$] \citet{Luo2025} parameter set: $\gamma_P' = -1.710$, $m = 0.6$, $b_0 = 1.763$, $b_1 = 0.982$, $b_2 = 2.11$, $b_3 = 0.37$, $b_4 = 1.9$.
\item[$\S$] Equivalent to $14.352$\,\AA$^3$\,atom$^{-1}$ in the RTpress atomic basis.
\end{tablenotes}
\end{threeparttable}
\end{table*}

After oxygen, magnesium and silicon are the two most abundant rock-forming elements in the solar composition \citep{Asplund2021, Lodders2003}, and the enstatite composition MgSiO$_3$ is the natural mantle endmember: it matches the near-unity Mg/Si molar ratio inferred for bulk Earth \citep{McDonough1995} and for rocky exoplanets whose host stars share solar-like refractory ratios. At lower-mantle pressures MgSiO$_3$ adopts the perovskite structure known as bridgmanite (brg), the most abundant mineral in Earth by volume, which fills the region from the 660\,km seismic discontinuity to the D$''$ layer near the core--mantle boundary \citep[CMB,][]{Ringwood1991}. Under higher compression brg transforms to postperovskite (ppv), associated with the seismic anisotropy and sharp velocity gradients of the lowermost mantle \citep{Murakami2004, Stixrude2005, Murakami2024}, and beyond the melting curve liquid MgSiO$_3$ represents the silicate magma oceans of young, recently impacted planets and short-period exoplanets \citep{Schaefer2018, Boukare2025b}. The \paleos implementation covers six phases from ambient conditions to extreme pressures (Fig.~\ref{fig:mgsio3_phase}, Table~\ref{tab:mgsio3_params}); as for iron, assembling them into a single thermodynamic surface requires choices about source calibrations and compositional simplifications.

Earth's upper mantle is not pure MgSiO$_3$ but a multicomponent assemblage of olivine, pyroxene, garnet, and their high-pressure dissociation products \citep{Irifune1993}. The transitions between these mineral systems produce the seismic discontinuities at 410\,km (olivine to wadsleyite) and 660\,km (ringwoodite to brg plus ferropericlase), and the partitioning of iron between coexisting phases further modulates the density profile at each depth. Multicomponent Gibbs free energy minimization codes model this full mineralogy \citep[e.g.,][]{Connolly2009}, tracking equilibrium phase assemblages along self-consistent adiabats \citep[see][for a recent compilation]{Dong2025}. A single-component MgSiO$_3$ description sacrifices this fine structure: the 410 and 660\,km discontinuities disappear, iron partitioning is absent, and effects such as aluminum incorporation into brg and the iron spin transition in ferropericlase are excluded. For mass--radius calculations, however, the planetary radius depends on the volume-averaged density profile, in which the lower mantle (dominated by brg regardless of compositional model) contributes the largest share. The bulk compressibility of pure MgSiO$_3$ and that of a full pyrolite assemblage yield nearly identical mass--radius curves at the percent level \citep{Dorn2017, Haldemann2024}, and the single-component approach avoids the additional free parameters and interphase calibration uncertainties inherent in multicomponent descriptions.

The upper mantle in \paleos is represented by three pyroxene polymorphs from \citet{Sokolova2022}: low-pressure clinoenstatite (lpcen, monoclinic $P2_1/c$), orthoenstatite (en, orthorhombic $Pbca$), and high-pressure clinoenstatite (hpcen, monoclinic $C2/c$). Each combines a Kunc cold pressure (Eq.~\eqref{eq:kunc}) with a two-mode Einstein thermal model whose characteristic temperatures and mode weights are listed in Table~\ref{tab:mgsio3_params}. The three phases were simultaneously calibrated within a unified Helmholtz free energy framework, analogous to the \citet{Dorogokupets2017} approach for iron, ensuring that the phase boundaries emerge self-consistently from the free energy surfaces. One of the three pyroxenes carries an electronic Gr\"{u}neisen contribution (Table~\ref{tab:el_mag_params}, en) while lpcen/hpcen show no measurable electronic contribution in their stability field. The three polymorphs meet at a triple point $(6.5\;\mathrm{GPa},\, 1100\;\mathrm{K})$ from which three phase boundaries emanate; all boundary expressions are collected in Appendix~\ref{app:boundaries:mgsio3}. Numerical fidelity is confirmed by round-trip pressure tests: solving for $V(P,T)$ and recomputing $P$ yields relative residuals below $10^{-8}$ across the full 0.1--20\,GPa calibration range. The transition from hpcen to brg is set at 12\,GPa, the empirical upper validity limit of the \citet{Sokolova2022} parametrization. Above this pressure, brg becomes the stable solid phase.

Brg and ppv together span the pressure range from the base of the upper mantle to the centers of super-Earth mantles. The brg EoS follows \citet{Wolf2015}, who performed a Bayesian reanalysis of static compression data \citep{Tange2012} in laser-heated diamond anvil cells with neon pressure medium, covering 30--130\,GPa and temperatures up to 2500\,K. The formulation combines a Vinet cold pressure (Eq.~\eqref{eq:vinet}) with a MGD thermal model (Eq.~\eqref{eq:debye_energy}). The 0\% Fe (pure MgSiO$_3$) parameter set is adopted, and the reference volume $V_0$ corresponds to an orthorhombic $Pbnm$ unit cell with $Z = 4$ (number of formula units per unit cell with volume $V_\mathrm{cell} = 162.12$\,\AA$^3$). The ppv EoS follows \citet{Sakai2016}, whose ab initio calibration employs a Keane cold pressure (Eq.~\eqref{eq:keane}), ensuring physically correct high-compression behavior at terapascal pressures, combined with a MGD thermal model and an Al'tshuler Gr\"{u}neisen parametrization (Eq.~\eqref{eq:gamma_altshuler}) that provides a meaningful asymptotic value $\gamma_\infty$. The reference volume corresponds to a $Cmcm$ cell with $Z = 4$ ($V_\mathrm{cell} = 164.22$\,\AA$^3$). The brg--ppv boundary follows \citet{Ono2005}. At pressures above approximately 800--1000\,GPa, ab initio calculations predict that ppv dissociates into its constituent oxides MgO and SiO$_2$ \citep{Tsuchiya2011, Niu2015, Umemoto2017}. \paleos does not implement this dissociation because reliable thermodynamic data for the high-pressure oxide phases remain sparse. The omission becomes relevant for planets exceeding roughly 5--10\,$M_\oplus$, where central mantle pressures enter the predicted dissociation regime.

The liquid phase employs the RTpress analytic free energy model of \citet{Wolf2018}, which decomposes the Helmholtz energy into a Vinet cold curve, a finite-strain Gr\"{u}neisen reference adiabat, and thermal excursions parameterized by five heat capacity coefficients $b_0$--$b_4$. \citet{Luo2025} reparametrized this formalism from ab initio MD simulations of 32 MgSiO$_3$ formula units spanning 0--1200\,GPa and 2200--14\,000\,K; the resulting parameters are listed in Table~\ref{tab:mgsio3_params}.\footnote{During implementation, we identified that the $b_0$--$b_4$ coefficients are dimensionless despite their tabulated units of eV/atom. The Boltzmann constant in the RTpress implementation of \citet{Wolf2018} already absorbs the eV$\to$GPa\,\AA$^3$ conversion factor, so dividing the \citet{Luo2025} coefficients by this factor (as suggested by the tabulated units) suppresses the thermal pressure by a factor of ${\sim}\,160$.} Liquid silicate is the dominant mantle phase in the immediate aftermath of giant impacts, where a substantial fraction of the mantle (or its entirety) may be molten \citep{Greenwood2005, ElkisTanton2012, Schaefer2018, Boukare2025b}. The melting curve separating liquid from solid combines that of \citet{Belonoshko2005} below ${\sim}\,2.55$\,GPa with the lower bound of \citet{Fei2021} above; all boundary expressions are collected in Appendix~\ref{app:boundaries:mgsio3}.

The hatched region in Fig.~\ref{fig:mgsio3_phase} at low pressures and temperatures above approximately 5000\,K marks the domain where the RTpress model does not converge. This regime corresponds to vapor and supercritical silicate, phases that \paleos does not currently implement. The critical point of MgSiO$_3$ lies near 6000--7000\,K at approximately 1\,kbar \citep{Caracas2023, Caracas2024}, well below the temperatures reached during giant impacts: the minimum impact velocity for onset of silicate vaporization is approximately 6\,km\,s$^{-1}$ \citep{Saurety2025}, making partial to complete mantle vaporization a common outcome of late-stage accretion rather than an exceptional event. Modeling these conditions requires coupling to a dedicated vapor EoS and constitutes an important future extension of \paleos for studies of impact-generated thermal states and early planetary evolution. The analogous gap in the iron EoS is less pressing: planetary cores are mechanically shielded from direct impact energy deposition by the surrounding silicate mantle, and the critical point of iron is substantially higher \citep[${\sim}\,9000$\,K at 4--7\,kbar,][]{Li2020}. Partial core vaporization can nevertheless occur in mantle-stripping collisions, where core material is exposed to low-pressure release paths and decompresses into the supercritical regime \citep{Nakajima2015}.

The thermodynamic consistency parameter $\vert\Delta\vert$ (Eq.~\eqref{eq:delta}) satisfies $\vert\Delta\vert < 0.01$ for 100\% of the valid $P$--$T$ domain, with a median of $3.5 \times 10^{-8}$. The \citet{Sokolova2022} pyroxenes achieve the highest consistency, with hpcen reaching $\max\vert\Delta\vert \simeq 3 \times 10^{-9}$, a direct consequence of deriving all quantities from a unified Helmholtz free energy. The liquid shows the largest residuals among the six phases, though still well within the $10^{-2}$ threshold. The overall consistency surpasses that of the iron EoS (100\% versus 98.2\% within the threshold) because the MgSiO$_3$ description contains no analog of the $\varepsilon$-hcp smoothstep blending zone that concentrates the residual iron inconsistency.

\subsection{Water}
\label{sec:eos:water}

Water is the most abundant condensable volatile in planet-forming disks \citep{Lodders2003} and the dominant ice beyond the snow line. Its role in planetary interiors spans a wide range, from the deep oceans or high-pressure ice mantles of surmised water-rich sub-Neptunes such as K2-18\,b \citep{Benneke2017, Benneke2019} to the few percent of mass that nominally rocky super-Earths can sequester as water dissolved in silicate melts, altering their bulk density and thermal budget \citep{Dorn2021, Luo2024}. Unlike iron and MgSiO$_3$, whose phase diagrams each span two or three crystal structures plus a melt, the water phase diagram extends from ice-Ih through ice-VII and X to superionic water, crossing liquid, vapor, supercritical, and plasma regimes from ${\sim}\,1$\,Pa to hundreds of TPa, so modeling it within a single framework demands a composite EoS that stitches together fundamentally different data sources.

\textsc{aqua} \citep{Haldemann2020} provides such a framework. It assembles seven underlying EoS into a single pressure--temperature table covering $0.1$\,Pa to $400$\,TPa and 150 to $10^5$\,K: \citet{Wagner2002} for liquid and steam, \citet{Feistel2006} for ice-Ih, \citet{Journaux2020} for ice-II through VI, \citet{French2015} for ice-VII and X, \citet{Brown2018}  for low-density vapor, and \citet{Mazevet2019} for the supercritical and superionic regime. \paleos\ interpolates bilinearly the \textsc{aqua} table in log space and regrids it to match the resolution used for iron and MgSiO$_3$ (Sect.~\ref{sec:eos:tables}). Derived quantities not directly tabulated are obtained from standard thermodynamic relations: $\alpha = -(1/\rho)\,(\partial\rho/\partial T)_P$ via centered finite differences on the interpolated density, $C_P = \alpha\,P / (\rho\,\naad)$ from the tabulated adiabatic gradient, and $C_V$ from $C_P$ and the speed of sound $w$ via $C_V = C_P^2 / (C_P + T \alpha^2 w^2)$. Because \textsc{aqua} already handles phase stitching, reference-state calibration, and interpolation-region assignment, the water EoS enters \paleos\ at far lower implementation cost than the phase-by-phase Helmholtz constructions required for iron and MgSiO$_3$, at the expense of reduced control over individual phase descriptions.

We applied two corrections to the Helmholtz free energy parametrization of \citet{Mazevet2019} used in \textsc{aqua}. The first addresses a sign error in the published Eq.~(13), where the first two terms of $F_T$ carry the wrong sign \citep[as already flagged in, e.g.,][]{Aguichine2025}. The second accounts for a revision of the reference entropy $S_0$ from $4.9\,n_\mathrm{at}\,k_\mathrm{B}$ to $9.8\,n_\mathrm{at}\,k_\mathrm{B}$, where $k_\mathrm{B}$ is the Boltzmann constant and $n_\mathrm{at}$ the number of atoms per kilogram. Since the free energy contains a $-S_0\,T$ term and the \textsc{aqua} table was built with the old value, the corrective contribution is $-4.9\,n_\mathrm{at}\,k_\mathrm{B}\,T$. Both terms depend only on temperature, so the pressure and density are identically unaffected. The combined correction to the free energy is
\begin{equation}
\label{eq:fshift_water}
F_\mathrm{shift}(T) = 2\,n_\mathrm{at}\left[b_1\,\tau\ln(1+\tau^{-2}) + b_2\,\tau\arctan\tau\right] - 4.9\,n_\mathrm{at}\,k_\mathrm{B}\,T,
\end{equation}
where $\tau$ is a dimensionless temperature and $b_1$, $b_2$ are fitted coefficients. The correction propagates analytically to entropy via $S_\mathrm{shift} = -\partial F_\mathrm{shift}/\partial T$ and to internal energy via $U_\mathrm{shift} = F_\mathrm{shift} + T\,S_\mathrm{shift}$. The $S_0$ term contributes a constant entropy shift of $+4.9\,n_\mathrm{at}\,k_\mathrm{B} \approx +6784$\,J\,kg$^{-1}$\,K$^{-1}$ but cancels exactly in $U_\mathrm{shift}$, while the sign-error term adds a temperature-dependent correction to both. Analytical derivatives of $F_\mathrm{shift}$ match finite differences to ${\sim}\,10^{-10}$ relative error, and the thermodynamic identity $U = -T^2\,\partial(F/T)/\partial T$ is satisfied to the same precision.

These corrections must only be applied where \textsc{aqua} actually populated its table from \citet{Mazevet2019} to avoid corrupting grid points backed by a different EoS. Because the \textsc{aqua} phase column collapses the supercritical, superionic, and plasma codes into a single label, the \citet{Mazevet2019} patch \citep[Region 7 in the nomenclature of][]{Haldemann2020} cannot be selected from the tabulated phase alone, and is reconstructed analytically from the three blend curves of \textsc{aqua} that stitch it to its neighbors: ice-X for $T \leq 2250$\,K, the \citet{Brown2018} supercritical fluid for $1800 \leq T \leq 5500$\,K, and the CEA ideal gas for $T > 5500$\,K. Each transition defines a ramp $w_i(P, T) \in [0, 1]$ that is linear in $\log_{10}P$ across its blend band, equal to zero on the neighbor side and to unity on the \citet{Mazevet2019} side. The three ramps combine via $w_7 = \max(w_3, w_5, w_6)$, and the additive shifts in entropy and internal energy are multiplied by $w_7$. Two refinements complete the prescription. First, the $\max$ rule creates a narrow overlap strip at $T \in [1800, 2250]$\,K where the $5 \leftrightarrow 7$ and $3 \leftrightarrow 7$ blend bands coexist, which would otherwise contaminate ice-VII and ice-X grid points with the Eq.~\eqref{eq:fshift_water} shift. We therefore forced $w_5$ and $w_6$ to zero wherever the tabulated phase code is not supercritical while leaving $w_3$ unconditionally active because ice-VII and ice-X are the legitimate low-pressure neighbors of the $3 \leftrightarrow 7$ blend. Second, \citet{Mazevet2019} is itself defined only down to $300$\,K, but \textsc{aqua} extends it to lower temperatures by evaluating it at the $300$\,K isotherm \citep[Sect.~2.3.7]{Haldemann2020}. To mirror that prescription, the shift functions are evaluated at $\max(T, 300\,\mathrm{K})$ in the public methods, while $w_3$ remains active down to the bottom of the \textsc{aqua} grid because its blend factor is pressure-only. 

\begin{figure}
\centering
\includegraphics[width=\columnwidth]{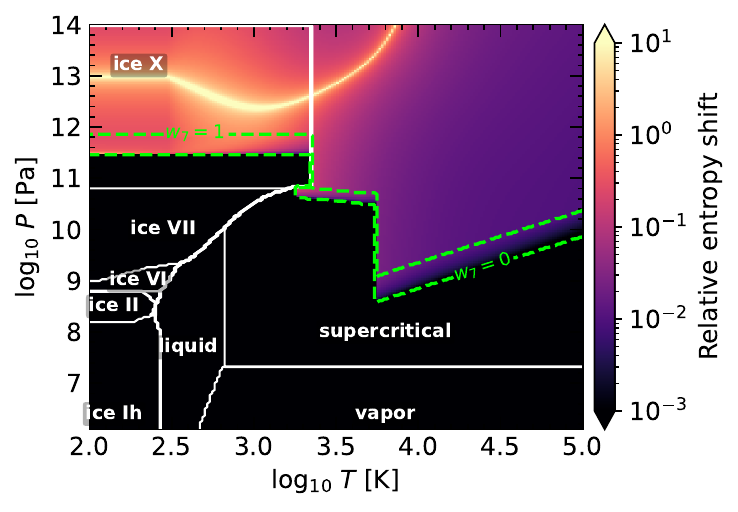}
\caption{Relative entropy shift applied by the \citet{Mazevet2019} correction (Eq.~\eqref{eq:fshift_water}) across the water phase diagram in $(\log_{10}T,\, \log_{10}P)$ space. The color map is logarithmic between $10^{-3}$ and $10$, with the upper bound saturating along a narrow ridge where the \textsc{aqua} baseline entropy nearly vanishes. Solid white lines mark the phase boundaries of the underlying \textsc{aqua} table; labels identify the fields of ices, liquid, vapor, and supercritical water. Dashed lime contours with inline labels mark the analytic gate $w_7$ that selects where the patch is applied: $w_7 = 0$ outside the patch (no shift) and $w_7 = 1$ inside its core (full correction). The figure shows that the correction is concentrated in a localized strip of the supercritical regime between ice X and the \citet{Brown2018} fluid, leaving the rest of the \textsc{aqua} table untouched.}
\label{fig:water_entropy_shift}
\end{figure}

Figure~\ref{fig:water_entropy_shift} maps the relative entropy shift and the analytic gate. The median unsigned shift is ${\sim}\,4.7\%$ in entropy and ${\sim}\,0.7\%$ in internal energy for pure Region 7 ($w_7 = 1$), and ${\sim}\,1.1\%$ and ${\sim}\,2.0\%$ in the transition bands. The entropy tails are heavy, reaching an order of magnitude in the ice-X zone and along an isentropic ridge into the supercritical phase, but only because the raw \textsc{aqua} entropy there is small compared with the $T$-clamped shift $S_\mathrm{shift}(300\,\mathrm{K}) \approx 2570$\,J\,kg$^{-1}$\,K$^{-1}$, not because the absolute correction is large. The internal-energy shift stays below ${\sim}\,8\%$ at the 99th percentile because the $S_0$ part cancels exactly in $U_\mathrm{shift}$, and the adiabatic gradient is negligibly affected because $\partial^2 F_\mathrm{shift}/\partial T^2$ is small, so the correction acts as a nearly constant entropy offset that barely modifies $(\partial S/\partial T)_P = C_P/T$ entering $\naad = \alpha\,P / (\rho\,C_P)$. As \textsc{aqua} is a composite table rather than a single free energy surface, the Maxwell consistency parameter $\Delta$ (Eq.~\eqref{eq:delta}) need not vanish: its median across the grid is $3 \times 10^{-3}$, with $71\%$ of points below $10^{-2}$ and the largest deviations at the stitched region boundaries \citep{Haldemann2020}. The offset also preserves entropy monotonicity: specific entropy still rises with temperature across every isobaric phase-boundary crossing in both the raw and corrected tables, so the second-law jump on heating stays positive at all transitions. The only departures are within a single phase, confined to the ice-X and supercritical bands that are already present in the raw \textsc{aqua} entropies and are slightly reduced, not introduced, by the correction.

\citet{CanoAmoros2026} independently analyzed the same sign error and built a more comprehensive revision of the \textsc{aqua} table: they incorporate DFT--MD data for the superionic phase \citep{French2016}, restitch the interpolation between EoS sources, and resolve a first-order liquid--superionic transition with a latent heat of ${\sim}\,1000$\,J\,kg$^{-1}$\,K$^{-1}$. Along a Uranus-like adiabat their entropies depart from the original \textsc{aqua} by up to $\pm 10\%$, giving deep interiors ${\sim}\,650$\,K colder and a latent-heat release that could help power Neptune's luminosity. Cross-checking the \paleos-corrected fields against their table on a $200 \times 200$ log-uniform sample stratified by the Region-7 weight $w_7$, we find machine-precision agreement outside the patch ($w_7 = 0$), where the two approaches share the same \textsc{aqua} baseline, and a median entropy difference of ${\sim}\,8\%$ inside pure Region 7 ($w_7 = 1$), with the closest match along isotherms above ${\sim}\,3000$\,K. The residual differences are a matter of scope, as our correction (Eq.~\eqref{eq:fshift_water}) targets the two \citet{Mazevet2019} errors analytically whereas \citet{CanoAmoros2026} replace and restitch the underlying data. For the mass--radius calculations of Sect.~\ref{sec:mr} the distinction is immaterial, since the correction acts only on entropy and internal energy and leaves the density and $\naad$ that drive the structure integration identical to the \textsc{aqua} baseline; the revised entropies matter instead for ice-giant thermal evolution, where \citet{CanoAmoros2026} now represent the state of the art.

\subsection{Reference state calibration}
\label{sec:eos:refstate}

Internal energy and entropy are defined only up to additive constants, and each of the EoS calibrations imported in Sects.~\ref{sec:eos:iron} and \ref{sec:eos:mgsio3} carries its own, arbitrary integration convention. In a single-phase model these constants drop out of every observable, and their values are immaterial. In a phase-aware, thermally coupled model they are not. The Clausius--Clapeyron relation ties the slope of each phase boundary to the ratio $\Delta S/\Delta V$ across it, and the second law demands that an isobar crossing a boundary on heating carry a positive entropy jump. This is not a numerical nicety: the magnitude of $\Delta S$ across a melting curve is the latent heat of fusion per unit temperature, and it governs the energy budget of the mush phase between solidus and liquidus that thermal-evolution models based on mixing length theory rely on to describe partial melting of rocky mantles and cores \citep{Bower2019, Lichtenberg2021, Lichtenberg2025c}. Raw integration constants carry no knowledge of this requirement. The calibration layer of \paleos therefore replaces each phase-specific Gibbs energy $G_\mathrm{raw}(P, T)$ by $G_\mathrm{raw}(P, T) + (U_0 - T_0 S_0)$ and tunes the offsets against that target. A secondary purpose of the same layer is to align the three phases of a chosen reference triple point, so that their Gibbs surfaces meet at a common $(P, T)$ and the piecewise-arbitrary shifts inherited from the individual sources do not propagate into the composite thermodynamic surface of Sect.~\ref{sec:eos:tables}.

Three constraints fix the offsets, in order of how tightly the calibration is tied to them. The primary constraint is the sign of $\Delta S$ across every solid--solid and solid--liquid boundary of the \paleos phase diagram. Along each boundary we evaluate the entropy jump between the two adjacent phases and require $\Delta S > 0$ on heating with a numerical safety margin of roughly $1\;\mathrm{J\,mol^{-1}\,K^{-1}}$ against EoS imprecision. Where the raw integration constants violate this sign, an $S_0$ offset is added (or depressed) to restore it, and Clausius--Clapeyron acts as a consistency check that the calibrated $\Delta S / \Delta V$ reproduces the slope $\dd P/\dd T$ of the boundary given by the independent expressions of Appendix~\ref{app:boundaries:iron} and \ref{app:boundaries:mgsio3}. The second constraint pins the absolute scale of $\Delta S$ along each melting branch, which the sign alone leaves free. Laboratory calorimetry supplies the entropy of fusion anchors at ambient pressure: $\Delta S_\mathrm{fus} = 1.16\,k_\mathrm{B}$ per atom for iron \citep{Zhang2015}, and $\Delta S_\mathrm{fus} = 41.99\;\mathrm{J\,mol^{-1}\,K^{-1}} \approx 1.01\,k_\mathrm{B}$ per atom for MgSiO$_3$ \citep{Stebbins1984}. For MgSiO$_3$ the pyroxene--liquid anchor is met at the calorimetric precision of the experiment; for iron the $P \to 0$ anchor drifts upward to roughly $1.70\,k_\mathrm{B}$ per atom, so that the second-law margin is preserved along the entire hcp--liquid branch at interior pressures. The third constraint is the alignment of three coexisting phases at a single chosen triple point: setting $G$ equal on all three there removes three piecewise-arbitrary constants at once and selects the deepest accessible triple point as the nucleus of the calibration. This is a local statement at one $(P, T)$. It does not enforce continuity of $G$ at every phase boundary and is not meant to. Continuity elsewhere is inherited through the same $S_0$ and $U_0$ offsets but remains a calibration residual rather than a mathematical identity.

\begin{table}
\caption{Reference-state calibration constants for the Fe and MgSiO$_3$ phases of \paleos.}
\label{tab:ref_state}
\centering
\footnotesize
\begin{tabular}{lrr}
\toprule
Phase & $U_0$ [kJ/mol] & $S_0$ [J\,mol$^{-1}$\,K$^{-1}$] \\
\midrule
\multicolumn{3}{l}{Iron} \\
\cmidrule(lr){1-3}
$\alpha/\delta$-bcc             & $0.000$      & $0.000$    \\
$\gamma$-fcc                    & $4.470$      & $0.000$    \\
$\varepsilon$-hcp ($<310$\,GPa)    & $13.771$     & $8.000$    \\
$\varepsilon$-hcp ($>310$\,GPa)    & $1183.406$   & $-44.401$  \\
liquid                          & $337.536$    & $153.099$  \\
\midrule
\multicolumn{3}{l}{MgSiO$_3$} \\
\cmidrule(lr){1-3}
lpcen  & $-1565.100$  & $0.060$    \\
en     & $-1564.600$  & $0.000$    \\
hpcen  & $-1559.100$  & $4.389$    \\
brg    & $0.000$      & $0.000$    \\
ppv    & $11.927$     & $-0.698$   \\
liquid & $5987.008$   & $820.303$  \\
\bottomrule
\end{tabular}
\end{table}

For iron, we aligned at the $\gamma$-fcc--$\varepsilon$-hcp--liquid triple point at $(98.5\;\mathrm{GPa},\, 3712\;\mathrm{K})$ because $\gamma$-fcc, $\varepsilon$-hcp, and liquid iron are the three phases that populate planetary cores. The shallower bcc--fcc--liquid triple point at $(5.2\;\mathrm{GPa},\, 1991\;\mathrm{K})$ lies outside every (planetary-sized) core and is deliberately not used as the alignment point. The $\alpha/\delta$-bcc and $\gamma$-fcc phases inherit the raw \citet{Dorogokupets2017} convention, $U_0^\mathrm{bcc} = S_0^\mathrm{bcc} = 0$, $U_0^\mathrm{fcc} = 4.470\;\mathrm{kJ\,mol^{-1}}$, $S_0^\mathrm{fcc} = 0$, which the unified Helmholtz framework already renders mutually consistent across the bcc--fcc topology at low pressure. The \citet{Miozzi2020} description of $\varepsilon$-hcp below $310\;\mathrm{GPa}$ requires an entropy lift: its raw absolute entropy underpredicts $S$ at low temperature relative to the \citet{Dorogokupets2017} bcc surface by up to ${\sim}\,7\;\mathrm{J\,mol^{-1}\,K^{-1}}$, producing a second-law-violating jump across the $\alpha \leftrightarrow \varepsilon$ boundary; adding $S_0^\mathrm{hcp} = +8.0\;\mathrm{J\,mol^{-1}\,K^{-1}}$ restores $\Delta S > 0$ with the nominal margin, and $U_0^\mathrm{hcp}$ is then fixed by the triple-point alignment. The high-pressure extension of \citet{Hakim2018} inherits the same lift in lockstep, so that $S$ stays continuous across the $310 \pm 100\;\mathrm{GPa}$ smoothstep blend, while $U_0$ shifts by $T_0\,\Delta S_0$ to preserve the internal energy through the same blend. The \citet{Luo2024} liquid closes the iron calibration: $S_0^\mathrm{liq}$ is set so that $\Delta S_\mathrm{fus}$ at $P \to 0$ matches the \citet{Zhang2015} anchor of $1.16\,k_\mathrm{B}$ per atom, then lifted by a further ${\sim}\,0.5\,k_\mathrm{B}$ per atom so that $\Delta S_\mathrm{fus}$ remains positive and above the numerical margin all the way along the \citet{Anzellini2013} melting curve at interior pressures, and $U_0^\mathrm{liq}$ is fixed by the triple-point alignment. The calibration delivers $\Delta S > 0$ across every $\alpha \leftrightarrow \gamma$, $\alpha \leftrightarrow \varepsilon$, $\gamma \leftrightarrow \varepsilon$, and $\varepsilon \leftrightarrow \mathrm{liq}$ crossing permitted by the phase diagram, with the $\varepsilon$--liquid branch tightest (${\sim}\,0.14\,k_\mathrm{B}$ per atom at $1.5\;\mathrm{TPa}$), while Gibbs residuals at the anchor triple point fall below $10^{-3}\;\mathrm{J\,mol^{-1}}$. The calibrated offsets are collected in Table~\ref{tab:ref_state}.

For MgSiO$_3$ we aligned at the brg--ppv--liquid triple point at $(155.68\;\mathrm{GPa},\, 6168\;\mathrm{K})$, computed numerically as the intersection of the \citet{Ono2005} brg--ppv boundary with the \citet{Fei2021} melting curve. The shallow pyroxene triple point at $(6.5\;\mathrm{GPa},\, 1100\;\mathrm{K})$ is not used as the alignment point, as enforcing the alignment there forces an anomalous pyroxene--liquid fusion entropy that overshoots the \citet{Stebbins1984} calorimetric datum by a factor of approximately five. The \citet{Wolf2015} brg phase is the local reference, $U_0^\mathrm{brg} = S_0^\mathrm{brg} = 0$, playing the role in the silicate calibration that $\gamma$-fcc plays for iron. The raw entropy gap across the $\mathrm{ppv} \leftrightarrow \mathrm{brg}$ boundary along \citet{Ono2005} is already positive by ${\sim}\,0.8\;\mathrm{J\,mol^{-1}\,K^{-1}}$; a small depression $S_0^\mathrm{ppv} = -0.70\;\mathrm{J\,mol^{-1}\,K^{-1}}$ opens the $\geq 1\;\mathrm{J\,mol^{-1}\,K^{-1}}$ safety margin without disturbing the brg side of the boundary, and $U_0^\mathrm{ppv} = +11.93\;\mathrm{kJ\,mol^{-1}}$ is then fixed by the triple-point alignment. The \citet{Sokolova2022} pyroxenes are calibrated locally, with en playing the role of local anchor ($S_0^\mathrm{en} = 0$) and the clinoenstatites carrying lifts $S_0^\mathrm{lpcen} = +0.06\;\mathrm{J\,mol^{-1}\,K^{-1}}$ and $S_0^\mathrm{hpcen} = +4.39\;\mathrm{J\,mol^{-1}\,K^{-1}}$ tuned to keep $\Delta S > 0$ across the pyroxene--pyroxene boundaries of Appendix~\ref{app:boundaries:mgsio3}, while the $U_0$ values are adopted directly from Table~2 of \citet{Sokolova2022}. No triple-point alignment is imposed on the pyroxene group itself, mirroring the iron case where no alignment is imposed at the bcc--fcc--hcp triple point. The \citet{Wolf2018} liquid closes the silicate calibration: $S_0^\mathrm{liq}$ is set so that $\Delta S_\mathrm{fus}$ matches the \citet{Stebbins1984} pyroxene--liquid datum of $41.99\;\mathrm{J\,mol^{-1}\,K^{-1}}$ at $1\;\mathrm{bar},\,1831\;\mathrm{K}$ to within the percent-level precision of the experiment, which simultaneously delivers $\Delta S_\mathrm{fus} > 450\;\mathrm{J\,mol^{-1}\,K^{-1}}$ along the brg--liquid and ppv--liquid branches at interior pressures, well above the second-law threshold, and $U_0^\mathrm{liq} = +5.987\;\mathrm{MJ\,mol^{-1}}$ is fixed by the triple-point alignment. The calibration achieves $\Delta S > 0$ across every solid--solid and solid--liquid boundary of the MgSiO$_3$ phase diagram, with Gibbs residuals at the anchor triple point below $2 \times 10^{-9}\;\mathrm{J\,mol^{-1}}$. The full set of offsets, listed in Table~\ref{tab:ref_state}, enters the tabulation of Sect.~\ref{sec:eos:tables} as part of the composite thermodynamic surface.

\subsection{Tabulation}
\label{sec:eos:tables}

Evaluating the composite thermodynamic surface of Sects.~\ref{sec:eos:iron}--\ref{sec:eos:water} at a single $(P,\,T)$ is not cheap: at every query, the cold pressure must be inverted for the volume by a bracketed root-finder, and the thermal contributions rely on numerical integrals. Interior-structure calculations compound that cost without mercy: they integrate coupled differential equations (e.g., Eq.~\eqref{eq:structure_odes}) over hundreds of substeps per planet, wrap those integrations inside bisection schemes on the outer radius, or sweep tens of thousands of planets per mass--radius grid, so that direct evaluation in the inner loop becomes impractical even with an optimized backend. Lookup tables resolve this tension. Precomputing each thermodynamic field on a regular grid and interpolating online reduces the per-query cost by two orders of magnitude, while the interpolation accuracy is bounded by the grid refinement and can be tightened below the intrinsic uncertainties of the underlying calibrations. The tables\footnote{\url{https://doi.org/10.5281/zenodo.19000315}} are therefore one of the main deliverable artifacts of the present work. They encapsulate the entire composite construction of Sects.~\ref{sec:eos:iron}--\ref{sec:eos:water}, including the phase-by-phase parameter choices, the smoothstep blending of $\varepsilon$-hcp iron, the reference-state alignment of Sect.~\ref{sec:eos:refstate}, the \citet{Mazevet2019} entropy and internal-energy corrections, and the Maxwell-consistency diagnostics, into a single flat product that any interior code can consume with a few lines of bilinear interpolation in $(\log_{10} P,\,\log_{10} T)$. Downstream users therefore inherit the full thermodynamic surface of \paleos without reimplementing the formalisms of Sect.~\ref{sec:eos}, and without carrying forward the typos and sign errors that tend to propagate silently through bespoke in-house EoS modules.

Each table is built on a grid that is uniform in $(\log_{10} P,\,\log_{10} T)$, the natural coordinate system for thermodynamic fields whose magnitudes span many orders across the planetary domain. The grid resolution was fixed empirically: for each candidate spacing we evaluated density on the full grid, drew 5000 random control points from the interior of the domain, and measured the relative interpolation error of a bilinear query against a direct EoS evaluation at the same point. A resolution of 150 points per decade keeps the 99th-percentile density error below $10^{-4}$ and the corresponding thermal expansion error below $10^{-2}$, while yielding table files in the tens of megabytes. Finer grids offer diminishing returns against the intrinsic uncertainties of the source calibrations, whereas coarser grids begin to imprint visible bumps on the derivative-sensitive fields ($\alpha$, $C_P$, $\naad$) that govern thermal profiles. We adopt 150 points per decade as the operating resolution across all three materials. We additionally provide high-resolution tables at 600 points per decade for calculations requiring ultraprecise quantities.

The iron and MgSiO$_3$ tables share the same box, $1\;\mathrm{bar}$ to $100\;\mathrm{TPa}$ in pressure and $300$ to $10^5\;\mathrm{K}$ in temperature, producing a $1351 \times 380$ grid. The lower bounds are deliberately conservative: solid Fe and MgSiO$_3$ are essentially incompressible below $1\;\mathrm{bar}$, so that pressure acts as an effective zero-pressure datum, while the $300\;\mathrm{K}$ floor matches the coldest rocky-planet surfaces and lies at or above the calibration floor of every source in Sects.~\ref{sec:eos:iron}--\ref{sec:eos:mgsio3}. The upper bounds are deliberately generous: $100\;\mathrm{TPa}$ sits more than a decade above the cores of the most massive super-Earths and super-Mercuries, and $10^5\;\mathrm{K}$ offers matching headroom on the thermal axis. These upper corners lie beyond the calibration windows of the individual phases, and the ppv field in particular enters the dissociation regime above roughly $1\;\mathrm{TPa}$ (Sect.~\ref{sec:eos:mgsio3}), which \paleos does not model. We nevertheless extend the tables to the full box rather than truncate at the most restrictive calibration, relying on cold pressure formalisms that stay well-behaved under extreme compression: the Holzapfel form for $\varepsilon$-hcp iron \citep{Hakim2018} enforces the Thomas--Fermi limit and the Keane form for ppv \citep{Sakai2016} imposes a finite $K_\infty'$ (Eqs.~\eqref{eq:holzapfel} and \eqref{eq:keane}), so neither extrapolation diverges as a BM3 or Vinet extension would. We verified that the extrapolated fields remain smooth and thermodynamically consistent (Eq.~\eqref{eq:delta}) there, but the intent in these upper corners is numerical robustness and seamless interior coverage, not physical fidelity: at $100\;\mathrm{TPa}$ and $10^5\;\mathrm{K}$, matter enters the warm-dense and dissociation regimes that the formalisms of Sect.~\ref{sec:eos} do not capture, and predictive modeling there would require dedicated first-principles EoS outside the scope of \paleos{}.

Each table stores ten columns: pressure, temperature, density, internal energy, entropy, both heat capacities, thermal expansivity, adiabatic gradient, and the label of the stable phase at that $(P,\,T)$. The iron table is strictly rectangular. The MgSiO$_3$ table is punctured at the nodes where the RTpress liquid fails to converge, corresponding to the hatched vapor and supercritical regime of Fig.~\ref{fig:mgsio3_phase}: of the grid nodes in the $1351 \times 380$ box, about 65\% carry a valid entry, leaving a nonrectangular footprint that downstream interpolators must handle with masking or nearest-neighbor fill. The water table is regridded from \textsc{aqua} at the same $150$ points per decade but inherits the broader \textsc{aqua} domain, $0.1\;\mathrm{Pa}$ to $100\;\mathrm{TPa}$ and $100$ to $10^5\;\mathrm{K}$, yielding a $2251 \times 451$ grid. All three tables are publicly released on Zenodo alongside this paper, together with the notebooks used to generate them.\footnote{The \paleos package, including the EoS implementations, notebooks, and Zenodo links can be found at \url{https://github.com/maraattia/PALEOS}.}

\section{Validation against PREM}
\label{sec:prem}

\begin{figure*}
\centering
\includegraphics[width=\textwidth]{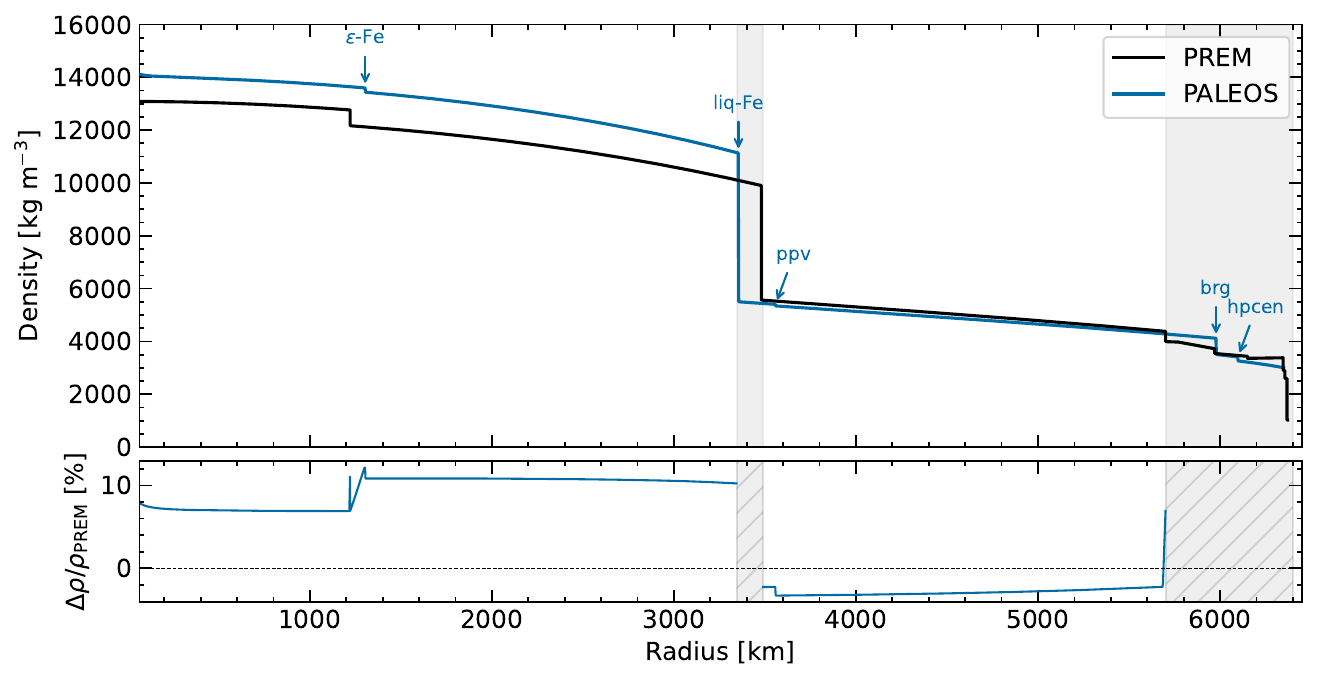}
\caption{Radial density profile of Earth for \paleos (blue) against PREM (black) for the two-layer Earth-like configuration of Sect.~\ref{sec:prem} ($M = 1\,\Mearth$, CMF $= 0.325$, $\Psurf = 1$\,bar, $\Tpot = 1600$\,K, $T_\mathrm{CMB} = 4370$\,K). Arrows mark phase transitions identified by \paleos: en $\to$ hpcen $\to$ brg $\to$ ppv in the mantle and liquid Fe $\to$ solid $\varepsilon$-hcp Fe at the ICB. The gray-shaded regions (upper mantle above $r = 5700$\,km, and a $\pm 10$\,km buffer around each model's CMB) are excluded from the quantitative residuals quoted in the text because they reflect the deliberate olivine-as-pyroxene proxy and the CMB-radius mismatch rather than EoS errors. Bulk sound speed residuals, derived from the same \paleos thermodynamic output, are not plotted but discussed numerically in the text. The composite \paleos profile recovers the dual-phase iron core automatically, with residuals consistent with the missing chemistry of the deliberate two-component model (light elements in the core, FeO and minor oxides in the mantle).}
\label{fig:prem}
\end{figure*}

The composite thermodynamic surface assembled in Sects.~\ref{sec:eos:iron}--\ref{sec:eos:water} is ultimately a falsifiable prediction. When applied to Earth's observed mass and bulk composition, it must reproduce Earth's radial structure within the biases introduced by our deliberate chemical simplifications, namely, a pure iron core, a pure MgSiO$_3$ mantle, and no volatile envelope. We benchmarked against PREM \citep{Dziewonski1981}, the standard seismologically constrained profile of Earth's interior. The two-layer reduction isolates the iron and silicate subsystems so that any systematic offset traces back to a specific physics decision in Sects.~\ref{sec:eos:iron}--\ref{sec:eos:mgsio3}, and it sets the accuracy baseline carried forward into the rest of the paper.

\citet{Dziewonski1981} built PREM from a joint inversion of body-wave travel times, surface-wave dispersion, roughly one thousand free-oscillation eigenfrequencies, and Earth's mass and moment of inertia, yielding a one-dimensional spherically symmetric profile of density $\rho$, compressional and shear wave speeds $V_P$ and $V_S$, and attenuation $Q$, with fixed discontinuities at the major mantle transitions, the CMB (2891\,km), and the inner core boundary (ICB; 5149.5\,km). One subtlety matters for comparison with a thermodynamic code: PREM constrains density and elastic wave speeds directly, but pressure, temperature, and gravity are not tabulated and must be reconstructed from $\rho(r)$ by hydrostatic integration. We therefore restricted our quantitative tests to the quantities PREM actually constrains, the density $\rho(r)$ and the bulk sound speed $V_\phi = \sqrt{K_S/\rho}$ (with $K_S$ the isentropic bulk modulus), obtained from $V_P$ and $V_S$ via $V_\phi^2 = V_P^2 - \frac{4}{3} V_S^2$.

We configure a two-layer Earth with total mass $M = 1\,\Mearth$, CMF $= 0.325$, no water envelope, and surface pressure $\Psurf = 1$\,bar. The temperature boundary condition introduces $\Tpot = 1600$\,K, a concept that will govern every subsequent calculation. $\Tpot$ is the mantle potential temperature: the adiabatic anchor of the convecting interior at the reference pressure $\Psurf$, the temperature the convecting mantle would have if adiabatically decompressed along a hypothetical zero-conduction path to $\Psurf$. It is not the physical surface temperature. Earth's lithosphere is conductive rather than adiabatic \citep[e.g.,][]{Goes2020}, and anchoring the interior adiabat at the 300\,K ground temperature would drive the mantle to unphysically cold temperatures at depth. Earth's actual lower-mantle geotherm is reproduced by an adiabat anchored at approximately 1600\,K at 1\,bar \citep{McKenzie1988}, and that is the value we adopt. More broadly, $\Tpot$ parameterizes the thermal state of the convecting interior decoupled both from the uppermost thermal boundary layer and from the stellar irradiation that sets the equilibrium temperature $T_\mathrm{eq}$. Section~\ref{sec:superearths} exploits this freedom to expose an exoplanet composition--temperature degeneracy that a classical isothermal analysis at $T_\mathrm{eq}$ cannot recover. To reproduce Earth's dual-phase core, we imposed a thermal discontinuity at the CMB, $T_\mathrm{CMB} = 4370$\,K, which models the D$''$ boundary layer as a free parameter rather than solving for it self-consistently, as discussed below. The three coupled hydrostatic structure equations
\begin{equation}
\frac{\dd m}{\dd r} = 4\pi r^2 \rho, \quad
\frac{\dd P}{\dd r} = -\frac{Gm\rho}{r^2}, \quad
\frac{\dd T}{\dd r} = \naad \frac{T}{P} \frac{\dd P}{\dd r},
\label{eq:structure_odes}
\end{equation}
where $r$ is the radial coordinate and $m$ the mass enclosed within $r$, were integrated inward from the surface, and the outer radius $R$ was determined by bisection on the residual $m(r = 0)$. The aim was to verify that the composite EoS captures the first-order behavior of an Earth-like interior, not to reproduce PREM exactly. The deliberate Fe + MgSiO$_3$ chemistry leaves percent-level density residuals from the missing core light elements and minor mantle oxides, well below the 5--20\% mass and radius uncertainties typical of small-exoplanet observations \citep{Parc2024}.

The integration still recovers Earth's total radius to within 0.33\%: $R = 6350$\,km against PREM's 6371\,km. The CMB lands at 3356\,km, 124\,km shallower than PREM's 3480\,km, with a CMB pressure of 142.8\,GPa (a 5.0\% overshoot relative to 136\,GPa) and a central pressure of 451\,GPa (a 24\% overshoot relative to the PREM-reconstructed 364\,GPa). The central temperature reaches 7055\,K. The phase sequence from surface to center matches the expected Earth mineralogy without any tuning beyond the CMB thermal jump: en in the uppermost mantle, hpcen across the pyroxene transition at approximately 12\,GPa, brg through the bulk of the lower mantle, and ppv immediately above the CMB. The presence of ppv at the base of the mantle is itself a first-order ingredient of lower-mantle dynamics, where its lowered viscosity and elevated thermal conductivity reshape the planform of CMB-driven convection \citep{Tosi2010}. Liquid iron fills the outer core, and solid $\varepsilon$-hcp iron forms the inner core. The dual-phase core is the most telling qualitative success of the composite EoS, because it emerges automatically once the iron melting curve of Sect.~\ref{sec:eos:iron} intersects the interior adiabat at the right pressure, with no further adjustment. The $T_\mathrm{CMB}$ value that yields this structure is tightly constrained: outside a narrow 4350--4400\,K window the core degenerates to either entirely solid or liquid. This range rather corresponds to the higher ends of the $T_\mathrm{CMB}$ estimates \citep[which remain poorly constrained,][]{Frost2022} and is a consequence of the absence of light elements in the core \citep{Hirose2021, Shahar2026}: their presence would have otherwise shifted the Fe melting curve to lower temperatures \citep{Stixrude2014}, mechanically lowering the $T_\mathrm{CMB}$ window. Importantly, the global radius is nearly blind to this thermal tuning: scanning $T_\mathrm{CMB}$ from 3500 to 5000\,K moves $R$ by less than 0.3\%. Total radius is therefore a poor lever for inferring the thermal state of the core, and the dual-phase structure test is the only informative one.

Figure~\ref{fig:prem} compares the \paleos density profile with PREM, with two regions excluded from the quantitative residuals (gray bands). The first is the upper mantle ($r > 5701$\,km), where the pyroxene polymorph series adopted in Sect.~\ref{sec:eos:mgsio3} as a proxy for olivine does not reproduce the 410 and 660\,km seismic discontinuities; the disagreement in this depth range reflects a mineralogical choice rather than an EoS error. Still, density differences remain acceptable there ($\sim$ 20\%) with a reasonable qualitative agreement. The second excluded region is a $\pm 10$\,km buffer around each model's CMB, where the mismatch in CMB radius makes point-by-point interpolation physically meaningless. In the lower mantle the \paleos profile is systematically low by a median of 3.0\% with a maximum deviation of 3.3\%, consistent with the absence of iron, calcium, and aluminum in the silicate: natural brg incorporates roughly 10\,mol\% Fe together with minor CaO and Al$_2$O$_3$ components \citep{McDonough1995}, all of which raise the density by a few percent. The core is systematically dense, by a median of 10.6\% and up to 12.1\%, with a central density of $13\,712\;\mathrm{kg\,m^{-3}}$ against PREM's $13\,089\;\mathrm{kg\,m^{-3}}$. At the ICB conditions of 330\,GPa and approximately 6000\,K, the composite $\varepsilon$-hcp surface lies $\sim$ 4\% above PREM's $12\,763\;\mathrm{kg\,m^{-3}}$. This overshoot is the observational signature of the light elements \citep[S, Si, O, C, H at a combined 8--10\,wt\%,][]{McDonough1995} dissolved in Earth's iron core, that is, the classical ICB density deficit. Independent estimates from pure-iron EoS frameworks bracket our value: \citet{Dorogokupets2017} report a 4.4\% deficit and \citet{Fei2016} report 3.6\%. \paleos lies squarely within this range, which both validates the pure-Fe implementation and quantifies the amount of chemistry that a terrestrial-fidelity model would need to add.

The second independent check against PREM is the bulk sound speed $V_\phi$, which we assemble from the \paleos thermodynamic output via $K_T = \rho(C_P - C_V)/(T\alpha^2)$ and $K_S = K_T C_P/C_V$. Because $V_\phi$ depends on three second-order thermodynamic quantities ($C_P$, $C_V$, and $\alpha$), the test is strictly tighter than density: errors in EoS derivatives that cancel in $\rho$ can accumulate in $V_\phi$. The agreement is nonetheless encouraging. The lower mantle shows a median offset of $+3.6$\% with a maximum of 5.4\%; the core shows $+3.9$\% with a maximum of 5.1\%. Both residuals lie in the range predicted by the same missing-chemistry argument that explains the density mismatch, and neither points to a failure of the thermal models of Sect.~\ref{sec:thermo:thermal}. We do not separate $V_P$ and $V_S$, because \paleos does not implement a shear modulus: rigidity is a property of the lattice rather than of the bulk thermodynamic surface, and it lies outside the scope of a compositional EoS. Taken together, the \paleos composite surface reproduces Earth's density to within the biases predicted by its deliberate two-component chemistry and its bulk sound speed to the 4--5\% level. These biases, a 3\% deficit in the lower mantle and a 10\% excess in the core, are carried forward unchanged into the mass--radius grid of Sect.~\ref{sec:mr} and into the composition inferences of Sect.~\ref{sec:superearths}, where they must be borne in mind when translating measured bulk properties into a compositional statement.

\section{Mass--radius relations}
\label{sec:mr}

\subsection{Background and scope}
\label{sec:mr:background}

The mass--radius relation is the workhorse of exoplanet interior characterization: a transiting, mass-confirmed planet delivers only $M$ and $R$, and any inference about composition or thermal state must flow through a structural model that closes the underdetermined problem. For more than a decade the standard has been the isothermal rocky--icy three-layer grid \citep{Valencia2006, Seager2007, Swift2012, Zeng2013, Zeng2016}, later extended to water and gas envelopes \citep{Zeng2019}. These adopt finite-strain equations of state fitted at ambient temperature and freeze the interior at 300\,K. The PREM-calibrated formula of \citet{Zeng2016}, $R/\Rearth = (1.07 - 0.21\,\mathrm{CMF})(M/\Mearth)^{1/3.7}$, is the standard rocky benchmark because it absorbs Earth's thermal profile implicitly through the PREM density fit, and its extension to other planets assumes that same profile; the authors note that the temperature effect on the mantle density jump is ``secondary and thus ignored.'' For cool, dry super-Earths at Earth-like insolation this is defensible, since the adiabatic rise across a silicate mantle barely reaches 2000\,K and the associated thermal expansion is diluted by self-compression. It breaks for the population that fills the transit catalogs: most known rocky exoplanets orbit inside 0.1\,AU, many have equilibrium temperatures above 1500\,K, and a subset sit on USP orbits with $T_\mathrm{eq} \gtrsim 2500$\,K, while tidal heating, impact energy, and primordial contraction can drive the silicate potential temperature $\Tpot$ (Sect.~\ref{sec:prem}) well into the magma-ocean regime \citep{Lichtenberg2025}. The classical isothermal grids cannot represent this population.

The modern mass--radius literature has pushed in three parallel directions, all focused on the envelope rather than the interior. For irradiated rocky planets carrying condensed water, \citet{Turbet2020} coupled a radiative--convective climate model to a Zeng-style core and showed that above the runaway-greenhouse threshold a steam atmosphere inflates the radius by factors of a few, cutting previously inferred water mass fractions (WMF) of TRAPPIST-1 b, c, and d \citep{Gillon2016, Dorn2018} by more than an order of magnitude. For sub-Neptunes, \citet{Aguichine2025} embedded a layered interior in a thick H$_2$O envelope and tracked the coupled radius evolution, emphasizing that the WMF inferred from $(M, R)$ is degenerate with thermal history. For H/He-enveloped gas dwarfs, \citet{Tang2025} resolved interior thermal pressure through the Gr\"{u}neisen parameter and a Lindemann-parameterized silicate solidus. All three treat the interior as an inner boundary condition for the envelope, and none revisits the dry-rocky grid beneath it, which remains the cold isothermal scaling of \citet{Zeng2016}.

The \paleos grid targets this underserved regime. Building on the EoS of Sect.~\ref{sec:eos} and the PREM-validated adiabatic integrator of Sect.~\ref{sec:prem}, we compute self-consistent mass--radius relations for two families: a rocky family (Fe core plus MgSiO$_3$ mantle) spanning pure silicate to pure iron, and a water-rich family that adds an H$_2$O envelope to an Earth-like rocky core. The distinguishing feature is not thermal expansion in isolation, which \citet{Tang2025} already resolves for sub-Neptunes, but its coupling to the explicit multiphase treatment of silicate and iron melting (Sects.~\ref{sec:eos:iron} and \ref{sec:eos:mgsio3}), which engages precisely where the isothermal grids fail. H/He and steam atmospheres are out of scope and better served by \citet{Turbet2020}, \citet{Aguichine2025}, and \citet{Tang2025}, to which the \paleos grid is complementary. The complete grid comprises 17\,900 models (9\,900 rocky and 8\,000 water-rich), released on Zenodo\footnote{\url{https://doi.org/10.5281/zenodo.19221214}}; 17\,898 converged at the $10^{-5}$ radius tolerance, the two failures occurring at $M = 0.1\,\Mearth$ with high WMF, where the bracket search finds no sign change in the shooting residual before the envelope becomes unphysical. The remainder of this section presents the two families in turn, with the rocky grid as the principal showcase.

\subsection{Rocky planets}
\label{sec:mr:rocky}

\begin{figure}
\centering
\includegraphics[width=\columnwidth]{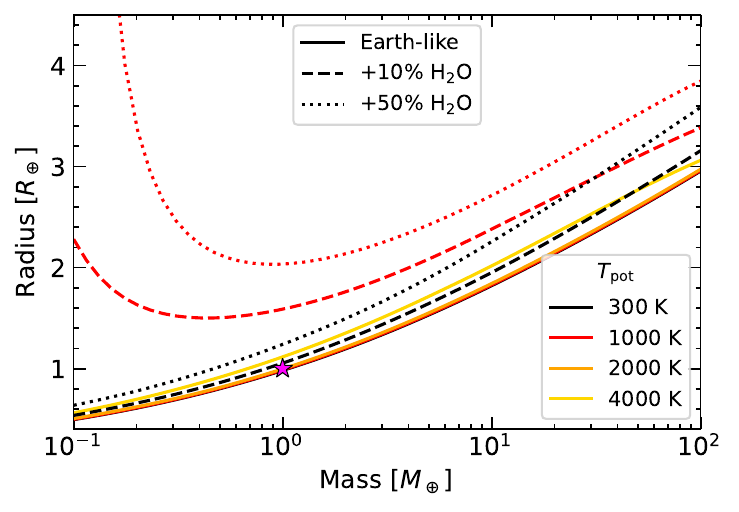}
\caption{Temperature dependence of \paleos mass--radius curves for three composition families across the planet-mass range $0.1 \le M / \Mearth \le 100$: Earth-like rocky (CMF $= 0.325$, solid lines), Earth-like rocky core plus a 10\,wt\% water envelope (dashed), and Earth-like rocky core plus a 50\,wt\% water envelope (dotted). The line color encodes the mantle potential temperature $\Tpot$, ranging from 300\,K (black) to 4000\,K (yellow). The two water-rich families are restricted to 300 and 1000\,K to keep the figure readable. The magenta star marks Earth. The figure shows that thermal expansion produces percent-level radius shifts at the rocky end and steepens markedly with water content, especially near the liquid--vapor transition of the envelope.}
\label{fig:mr_thermal}
\end{figure}

The rocky grid resolves 22 CMF values ($= 0.00, 0.05, \ldots, 1.00$ supplemented by the Earth-like value 0.325), 50 planet masses logarithmically spaced between 0.1 and 100\,$\Mearth$, and nine potential temperatures (300, 500, 1000, 1500, 2000, 2500, 3000, 3500, 4000\,K). Each model integrates the structure equations from $\Psurf = 1$\,bar inward along an adiabat anchored at the surface value of $\Tpot$, so that $\Tpot$ plays the same role here as in the PREM validation of Sect.~\ref{sec:prem}: it is the mantle temperature extrapolated adiabatically to the reference surface pressure, and Earth's $1600\,\mathrm{K}$ value anchors the cool end of our range while the $4000\,\mathrm{K}$ endpoint brackets the magma-ocean state expected during late-stage accretion and for the hottest USP planets. Grid construction and convergence metrics are documented in Sect.~\ref{sec:mr:background}; all subsequent analysis draws on the resulting $(\Tpot, \mathrm{CMF}, M) \to R$ tabulation.

The cold reference layer at $\Tpot = 300$\,K provides a direct sanity check against the canonical Zeng grids. Residuals against \citet{Zeng2016, Zeng2019} at fixed 300\,K are bounded by $\pm 0.5$\% for silicate-dominated compositions (CMF $\leq 0.3$) with a root mean square of 1\%, well within the propagated uncertainty of the underlying EoS calibrations. For iron-dominated compositions the agreement degrades: a pure iron planet in \paleos is systematically smaller than the Zeng curve by up to 6\%, a discrepancy that tracks the same 10\% core density excess identified against PREM in Sect.~\ref{sec:prem} and reflects the stiffer \citet{Dorogokupets2017}+\citet{Hakim2018} blend adopted here versus the PREM-fitted BM2 core of \citet{Zeng2016}. The iron EoS is the largest single source of bias carried forward into inferred compositions, and it should be borne in mind whenever CMF values above roughly 0.5 are being interpreted. Across the composition axis, the radius span from pure MgSiO$_3$ to pure Fe grows monotonically with mass, from 28\% of the Earth-like radius at $0.5\,\Mearth$ to 37\% at $20\,\Mearth$, reflecting the increasing leverage of bulk density on radius at higher self-compression. These 300\,K curves reproduce the classical picture and consolidate confidence in the grid before the temperature axis is engaged.

Thermal expansion is negligible at 300\,K but becomes a first-order effect on the mass--radius relation as $\Tpot$ rises (Fig.~\ref{fig:mr_thermal}). Along an adiabat, thermal pressure raises the equilibrium specific volume at every depth, and the cumulative radius response integrates these local expansions from surface to core. The magnitude depends on the stiffness profile of each layer: lower-mantle brg and ppv are stiff ($K_T \gtrsim 200\,\mathrm{GPa}$, $q \sim 1.4$--2.0) and absorb thermal pressure with only modest volume change, while upper-mantle pyroxenes are softer and, beyond the melting curve, give way to an RTpress liquid whose thermal expansivity is several times larger. The same depth-dependent thermal expansivity exerts a first-order control on mantle dynamics in two-dimensional convection models that adopt $P$--$T$-dependent parametrizations of $\alpha$ and the thermal conductivity across the major upper- and lower-mantle phases \citep{Tosi2013}, lending an independent geodynamics motivation to retaining the full $P$--$T$ dependence of $\alpha$ in the mass--radius budget. The grid reflects this ordering cleanly. For silicate-rich compositions the radius increase relative to 300\,K exceeds 1\% above $\Tpot \approx 1500\,\mathrm{K}$ and reaches 16.5\% at 4000\,K for a pure MgSiO$_3$ planet of $0.27\,\Mearth$. Earth-like compositions (CMF $= 0.325$) expand by up to 14.4\% at 4000\,K for a $0.31\,\Mearth$ planet, while pure iron expands by at most 5.9\% at $10.5\,\Mearth$. The effect is strongest at low masses because weaker self-compression leaves more dynamic range for thermal expansion to register. At $\Mearth$ a few-thousand-K rise in $\Tpot$ shifts the radius by several percent, comparable to the measurement uncertainty of the best-characterized super-Earths. By $20\,\Mearth$ the thermal signal is partially smothered by self-compression, though it remains at the percent level for silicate-rich compositions.

Underlying the thermal expansion signal is the more abrupt contribution of phase changes, and here \paleos departs from the prior literature most sharply. Once the adiabat crosses the melting curve (Fig.~\ref{fig:mgsio3_phase}, Appendix~\ref{app:boundaries:mgsio3}), an RTpress liquid silicate layer replaces crystalline brg or pyroxene in the upper mantle, and, for $\Tpot \gtrsim 3500\,\mathrm{K}$ at low and intermediate masses, the molten layer propagates into the lower mantle. The liquid silicate EoS of \citet{Wolf2018} with the \citet{Luo2025} parameter set yields substantially lower densities than the crystalline phases at the same $(P, T)$, so the mass--radius curve lifts off the cold benchmark more steeply than smooth thermal expansion would predict once melting engages. The analogous transition in the core, where the \citet{Luo2024} liquid replaces $\varepsilon$-hcp at the iron melting curve, is less dramatic in the radius budget but is the physically correct description for young or highly irradiated planets whose inner cores have not yet crystallized. Neither \citet{Zeng2016, Zeng2019} nor the more recent \citet{Tang2025} sub-Neptune models capture this regime self-consistently: the former grids assume solid silicate throughout, and the latter parameterize silicate melting via a Lindemann relation inside an envelope-dominated radius budget where the effect is marginal. For dry super-Earths the same phase changes dominate the thermal signal, and ignoring them biases inferred CMFs by several percent for hot targets, a bias that the case studies of Sect.~\ref{sec:superearths} exploit directly.

The practical implication for exoplanet population work is that the commonly tabulated 300\,K rocky mass--radius curves are not neutral priors for hot planets. USP super-Earths with $T_\mathrm{eq} \gtrsim 1500\,\mathrm{K}$ occupy the regime in which thermal expansion and partial melting shift the predicted radius by several percent at fixed composition, a displacement of the same order as the current radius uncertainties of the best-characterized targets and larger than the mass precision of most \citep{Parc2024}. Treating these planets with cold scalings therefore injects a systematic bias into any inference of bulk Fe/Mg/Si, and the bias sits on top of, not instead of, the conventional composition--mass degeneracy. The \paleos grid addresses this by making $\Tpot$ an explicit axis alongside mass and composition: an observed $(M, R)$ is no longer mapped to a point in CMF but to a curve in the two-dimensional (CMF, $\Tpot$) plane, the shape of which encodes the interplay between thermal expansion, silicate melting, and iron-core stiffness. Section~\ref{sec:superearths} puts this apparatus to work on two USP super-Earths of current observational interest. The tabulated grid, released on Zenodo at per-mille radius resolution across the full $(M, \mathrm{CMF}, \Tpot)$ parameter space, is designed to support both forward predictions for individual targets and population-level studies that marginalize over the thermal axis.

\subsection{Water-rich planets}
\label{sec:mr:water}

The water-rich grid mirrors the structure of the rocky family but replaces the composition axis with a WMF ($= 0.05, 0.10, \ldots, 1.00$, 20 values), layering an H$_2$O envelope on top of an Earth-like rocky core (CMF$_\mathrm{rocky} = 0.325$). Surface temperatures span $\Tsurf \in \{300, 400, \ldots, 1000\}\,\mathrm{K}$ at 100\,K resolution, and the 50-mass axis is shared with the rocky grid, yielding 8\,000 models bounded by $\Psurf = 1$\,bar. This family is a deliberate dry-interior counterpart to the rocky grid rather than an attempt at a steam-atmosphere model. At $\Tsurf = 300\,\mathrm{K}$ the curves agree with \citet{Zeng2019} at the 1\% level for both 100\% and 50\% H$_2$O compositions, validating the \textsc{aqua}-based water EoS adopted in Sect.~\ref{sec:eos:water} against the canonical cold benchmark and confirming that within the classical isothermal paradigm \paleos and the Zeng grids are interchangeable.

The agreement dissolves as soon as the water layer is warmed. \paleos truncates the envelope at $\Psurf = 1\,\mathrm{bar}$, while \citet{Zeng2019} adopt 1\,mbar, and at $\Tsurf \gtrsim 500\,\mathrm{K}$ this three-order-of-magnitude pressure offset straddles the saturation curve of water: at 1\,mbar the outer envelope is fully gaseous and contributes a geometrically thick, low-density layer, whereas at 1\,bar it remains liquid or supercritical and contributes a compact, high-density one. The resulting radii diverge by factors of several, exceeding 700\% for a pure-water $0.1\,\Mearth$ planet at 1000\,K. The divergence is not a physics disagreement but a definition of what constitutes the planetary surface, and the two prescriptions converge above roughly $10\,\Mearth$ where self-gravity compresses the envelope and erases the memory of the truncation pressure. The divergence is also the entry point for the modern envelope literature: above the runaway-greenhouse threshold identified by \citet{Turbet2020}, neither a 1\,bar nor a 1\,mbar convention is physically meaningful, because the steam atmosphere becomes optically thick and its structure is set by radiative--convective balance rather than by a bolted-on isotherm or adiabat. For sub-Neptunes that retain such atmospheres over gigayears, the coupled interior--envelope evolution modeled by \citet{Aguichine2025} becomes the appropriate framework. The \paleos water-rich grid therefore inhabits a deliberately narrow physical niche: condensed or supercritical water envelopes under modest irradiation, where the 1\,bar truncation is a calibration choice and the interior rather than the atmosphere sets the radius.

Within that niche, the thermal sensitivity of water-rich models is dramatically larger than for rocky ones, reflecting the softness of the water EoS near the liquid--vapor transition. At $M \sim 0.1\,\Mearth$ a pure-water planet at surface temperature $\Tsurf = 1000\,\mathrm{K}$ is some 17 times larger than its 300\,K counterpart, an inflation that diminishes to a few percent by $5\,\Mearth$ as self-gravity dominates the envelope budget. The same physics that produces this dramatic response, however, is precisely what makes the 1\,bar truncation least defensible. The runaway regime lies just beyond the domain probed by the grid, and the inflation curve steepens asymptotically as it is approached. This is the physically honest outer boundary of the dry-interior paradigm, and beyond it the problem passes from interior physics to atmospheric radiative transfer (\citealt{Turbet2020}) and to coupled thermal evolution (\citealt{Aguichine2025}). The \paleos water-rich tables are released with this upper bound on $\Tsurf$ in view and should be consumed with the explicit understanding that they stop where the steam atmosphere begins.

\section{Internal structure of super-Earths}
\label{sec:superearths}

\begin{figure*}
\centering
\includegraphics[width=\textwidth]{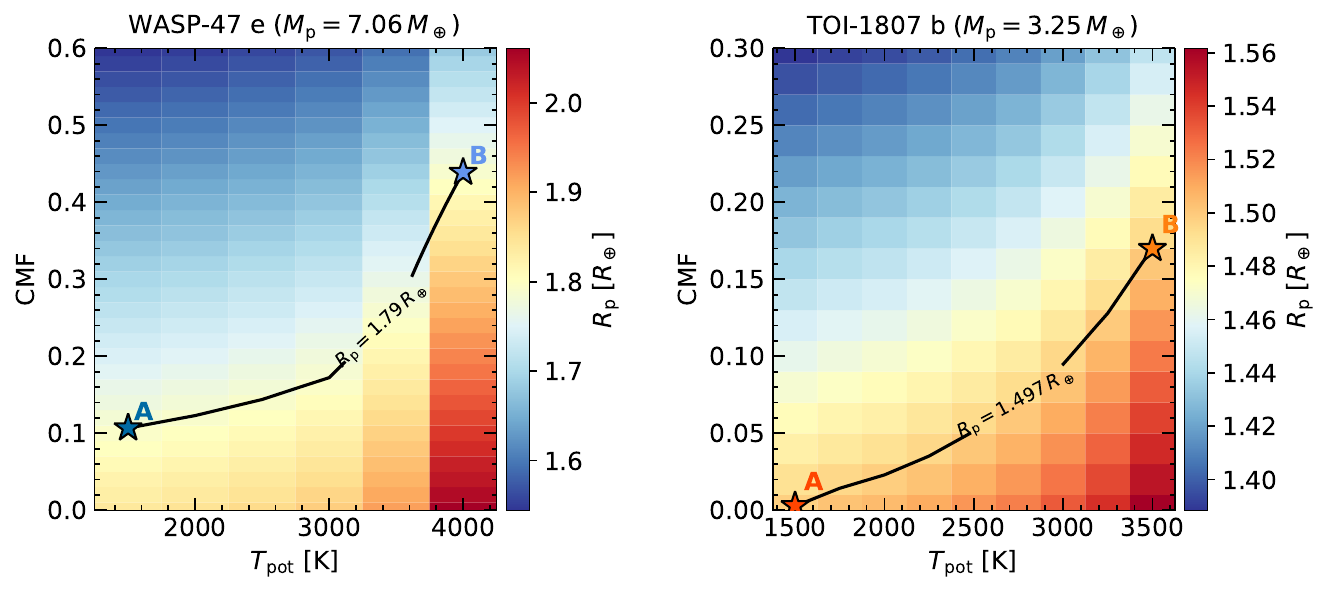}
\caption{Composition--temperature degeneracy for \object{WASP-47\,e} ($M = 7.06\,\Mearth$, left) and \object{TOI-1807\,b} ($M = 3.25\,\Mearth$, right). Each panel shows the \paleos-computed planet radius (color coded) on a (CMF, $\Tpot$) grid at a fixed mass and surface pressure $\Psurf = 1$\,bar, with the CMF spanning silicate-dominated to iron-dominated compositions and $\Tpot$ spanning the rocky range adopted in Sect.~\ref{sec:mr}. The black contour traces the isoradius locus at the observed value of each target, labeled inline. Stars mark the two representative solutions discussed in the text: cool iron-poor solutions A (lower-left) and hot iron-rich solutions B (upper-right). Every point along the black contour is a distinct interior model that reproduces the observed mass and radius exactly, demonstrating that $(M, R)$ alone cannot pin down composition once thermal expansion is allowed.}
\label{fig:degeneracy}
\end{figure*}

\subsection{The composition--temperature degeneracy}
\label{sec:superearths:degeneracy}

As demonstrated in Sect.~\ref{sec:mr}, the classical reading of a rocky mass--radius relation, in which a measured $(M, R)$ is projected onto an isothermal grid to recover a single CMF, is a convenience the underlying physics does not support. The curves of \citet{Seager2007}, \citet{Zeng2013}, \citet{Zeng2016}, and \citet{Zeng2019} fix the silicate and iron endmembers to their ambient solid densities at 300\,K, leaving thermal expansion either absent or frozen into a correction that never changes phase, because the EoS compilations they rely on carry solid backbones only. \citet{Unterborn2019} anticipated the consequence, showing that plausible shifts in the mantle potential temperature already blur the isothermal classification at current observational precision, but their analysis stops below the silicate solidus. \paleos crosses that boundary: the continuous solid-plus-melt treatment of iron \citep{Luo2024} and silicates \citep{Wolf2018} keeps thermal expansion a legitimate thermodynamic degree of freedom all the way into the magma-ocean regime, where it becomes a first-order control on any compositional inference drawn from a hot super-Earth.

Our contribution sits alongside, not against, the atmosphere-centric literature on this degeneracy. \citet{Dorn2017} formalized the Bayesian retrieval of interior structure from mass, radius, refractory abundances, and envelope properties, but retains an isothermal rocky interior by construction, with the mantle potential temperature fixed. \citet{Wilkinson2024} coupled a self-consistent radiative atmosphere to an interior EoS with a variable intrinsic temperature through the \textsc{hades} model, but in the sub-Neptune regime where the envelope carries the radius response and the rocky core underneath stays effectively frozen. For bare super-Earths, neither approach opens the thermal axis of the rocky interior itself. That axis, visible only through a mass--radius grid that tracks silicate and iron melt (Sect.~\ref{sec:mr}), is the one \paleos makes available, orthogonal to the atmosphere-side strategy rather than a competitor to it.

We recall from Sect.~\ref{sec:prem} that $\Tpot$ is the mantle potential temperature, not the photospheric or equilibrium temperature, and that it need not equal, and can far exceed, $T_\mathrm{eq}$: Earth itself sets the scale, with a present-day $\Tpot \approx 1600$\,K against $T_\mathrm{eq} \approx 255$\,K. For close-in and young rocky worlds the budget is richer. \citet{Lichtenberg2021} found that volatile-dependent thermal blanketing sustains global magma oceans for $10^5$ to more than $10^8$\,yr, with $\Tpot$ above 2000\,K for much of that interval, and \citet{Bower2019, Bower2022} obtained $\Tpot \approx 2000$--2700\,K immediately after accretion with a secular cooling timeline of 1--100\,Myr, extended to Gyr for more volatile-rich compositions \citep{Kite2020, Calder2025, Nicholls2026}. Past tidal dissipation during circularization of USP planets \citep{Farhat2025, Hallatt2026}, radiative heating by a thin outgassed mineral atmosphere \citep{Seidler2025, vanBuchem2025}, and a possible conductive surface boundary layer \citep{Nicholls2024} extend the window further. None of these scenarios is exotic, and each supplies a mechanism by which $\Tpot$ can be held well above $T_\mathrm{eq}$ on planets young enough or close enough that the interior has not yet relaxed. Surface pressure, by contrast, plays no role: varying $\Psurf$ from 1 to 100\,bar changes the rocky radius by less than $0.01$\%, because interior pressures of hundreds of GPa dominate the structure integration.

\begin{figure}
\centering
\includegraphics[width=\columnwidth]{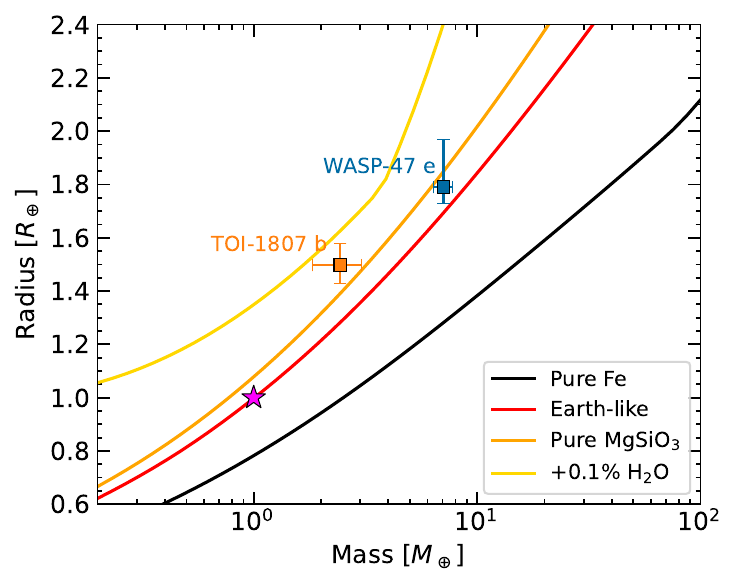}
\caption{Mass--radius diagram showing the two USP super-Earths analyzed in Sect.~\ref{sec:superearths}: \object{WASP-47\,e} (blue square, $M_\mathrm{p} = 7.06^{+0.71}_{-0.68}\,\Mearth$, $R_\mathrm{p} = 1.79^{+0.18}_{-0.06}\,\Rearth$) and \object{TOI-1807\,b} (orange square, $M_\mathrm{p} = 2.44 \pm 0.60\,\Mearth$, $R_\mathrm{p} = 1.497^{+0.081}_{-0.068}\,\Rearth$). Error bars are the $1\sigma$ observational uncertainties on each measurement. \paleos composition lines computed at $\Tpot = 2000$\,K are overlaid for reference: pure Fe (black), Earth-like (red, CMF $= 0.325$), pure MgSiO$_3$ (orange), and an Earth-like rocky core plus a 0.1\,wt\% water envelope (gold). The magenta star marks Earth. Both planets sit in the regime where the mass--radius pair alone cannot distinguish a hot iron-poor interior from a cold iron-rich one (Fig.~\ref{fig:degeneracy}).}
\label{fig:mr_targets}
\end{figure}

We sweep the forward model across a (CMF, $\Tpot$) grid at fixed $(M, \Psurf)$ and locate the isoradius contour at the observed value. Figure~\ref{fig:degeneracy} shows this for the two USP super-Earths analyzed below. Each panel is the computed radius surface, and the black contour marks the points that reproduce the measured radius. This contour is a one-dimensional family of (CMF, $\Tpot$) pairs all equally consistent with $(M, R)$, not a point estimate broadened by error bars. Moving along it from low to high $\Tpot$ trades silicate thermal expansion against core iron content, and each point corresponds to a distinct interior phase state (Sects.~\ref{sec:superearths:wasp47e} and \ref{sec:superearths:toi1807b}). We selected \object{WASP-47\,e} and \object{TOI-1807\,b} because they cover complementary masses (${\sim}\,7$ and ${\sim}\,3\,\Mearth$), have tight $(M, R)$ measurements, and share the USP architecture that makes elevated $\Tpot$ physically plausible. Their positions relative to the $\Tpot = 2000$\,K composition lines are shown in Fig.~\ref{fig:mr_targets}.

\subsection{WASP-47\,e}
\label{sec:superearths:wasp47e}

\object{WASP-47\,e} \citep{Becker2015} is the innermost member of a tightly packed four-planet system around a G-dwarf, discovered alongside the original \object{WASP-47\,b} hot Jupiter \citep{Hellier2012} and subsequently refined by high-precision transit photometry. We adopt the joint solution of \citet{Howard2025}: $M_\mathrm{p} = 7.06^{+0.71}_{-0.68}\,\Mearth$, $R_\mathrm{p} = 1.79^{+0.18}_{-0.06}\,\Rearth$, orbital period $P = 0.790$\,d, and equilibrium temperature $T_\mathrm{eq} = 1992$\,K. On a classical mass--radius diagram (Fig.~\ref{fig:mr_targets}) the planet sits between the Earth-like (CMF\,$= 0.325$) and pure MgSiO$_3$ lines at $\Tpot = 2000$\,K. An isothermal reading returns CMF\,$= 0.0$--0.3 and labels \object{WASP-47\,e} a silicate-dominated super-Earth with subterrestrial iron, closing the question. This is the null hypothesis the \paleos analysis addresses.

Opening the thermal axis changes the story. We computed 279 converged forward integrations spanning CMF\,$= 0.00$--$0.60$ and $\Tpot = 300$--$4000$\,K at the nominal $M = 7.06\,\Mearth$; the left panel of Fig.~\ref{fig:degeneracy} renders the resulting radius surface. The isoradius contour runs from CMF\,$= 0.074$ at 300\,K to CMF\,$= 0.439$ at 4000\,K, a factor-of-six swing in CMF along a single degenerate family. Two features of the trajectory deserve emphasis. First, the hot end of the contour exceeds Earth's CMF of 0.325, which places the hot \paleos solution outside any possible isothermal classification: no grid that fixes $\Tpot$ can generate a \object{WASP-47\,e}-class planet with a superterrestrial iron fraction. Second, the contour steepens sharply above $\Tpot \approx 3000$\,K, reflecting the accelerating thermal expansion of the silicate mantle as it approaches and crosses the melting curve. This is the same mechanism that broadens the rocky mass--radius curves in Fig.~\ref{fig:mr_thermal}, viewed here in the conjugate (CMF, $\Tpot$) projection at fixed $(M, R)$. Because the fractional mass uncertainty is ${\sim}\,10$\% and the measured radius sits comfortably within the rocky envelope at 7\,$\Mearth$, the full degenerate family is accessible without invoking a mass shift.

The cold end of the family, which we label Solution~A, corresponds to CMF\,$= 0.107$ and $\Tpot = 1500$\,K. The interior is fully crystalline: the mantle resolves into the four-phase en\,$\to$\,hpcen\,$\to$\,brg\,$\to$\,ppv sequence built up in Sect.~\ref{sec:eos:mgsio3}, and a solid $\varepsilon$-hcp iron core occupies 33\% of the planetary radius. Central conditions roughly reach 2\,TPa and 5400\,K, comfortably below the iron melting curve at that pressure (Fig.~\ref{fig:iron_phase}), so no part of the core is molten. The CMB sits deep and cold, with no magma ocean and a fully solid core. This endmember is nearest in spirit to the classical cold inference, but it still carries a CMF three times smaller than the isothermal Earth-like estimate would predict at the same radius, because a 1500\,K silicate is already detectably less dense than its 300\,K counterpart.

The hot end, Solution~B, at CMF\,$= 0.439$ and $\Tpot = 4000$\,K, is qualitatively different. The entire silicate envelope is molten: no solid mantle phase survives the adiabat, and the radius is carried by a deep magma ocean stacked on top of a compact iron core that occupies 53\% of the radius. Within the core, the structure splits. The outer core is liquid, but the innermost iron reenters the solid field and crystallizes as $\varepsilon$-hcp even at $T_\mathrm{center} \approx 20\,000$\,K, because the melting curve of iron rises more steeply with pressure above 2\,TPa than the adiabat does (Sect.~\ref{sec:eos:iron}). The result is an inner-core/outer-core architecture directly analogous to Earth's but at pressures an order of magnitude larger, a purely mass-driven outcome: the core-pressure range extends far enough to recross the iron melting curve from below. The fully molten mantle sustains outgassing of a mineral atmosphere, which feeds back into the thermal blanketing that keeps $\Tpot$ high, and within the mass--radius constraint alone Solution~B is indistinguishable from the cold Solution~A.

\subsection{TOI-1807\,b}
\label{sec:superearths:toi1807b}

\object{TOI-1807\,b} \citep{Hedges2021} pushes the same logic into a regime where the classical reading fails more conspicuously. The planet is a ${\sim}\,300$\,Myr USP rocky world ($P = 0.549$\,d) orbiting a K-dwarf, with $M_\mathrm{p} = 2.44 \pm 0.60\,\Mearth$, $R_\mathrm{p} = 1.497^{+0.081}_{-0.068}\,\Rearth$, and $T_\mathrm{eq} = 1694$\,K \citep{Polanski2024}. On the mass--radius diagram of Fig.~\ref{fig:mr_targets}, it sits just above the pure MgSiO$_3$ line at $\Tpot = 2000$\,K, in a location that classical isothermal curves can only explain by introducing a small volatile layer of order 0.1\% H$_2$O. That inference, delivered by every isothermal grid currently in use, is an artifact of the assumed thermal state rather than a genuine requirement from the data. A hot rocky interior at the same $(M, R)$ is thermodynamically permissible and, as we will show, physically well motivated for a planet of this age and irradiation level.

Two adjustments are needed to read the observation cleanly. The mass uncertainty is substantial (${\sim}\,25$\% fractional, dominated by the K-dwarf activity of the host), and the central radius sits above the pure MgSiO$_3$ line at every temperature we considered once $M$ is fixed at the central value. We therefore worked within the uncertainty envelope at $M = 3.25\,\Mearth$ ($+1.35\sigma$), a physical-plausibility shift that brings the rocky regime into reach without conflicting with the measurement. At this mass we computed 144 converged integrations spanning CMF\,$= 0.00$--$0.30$ and $\Tpot = 1500$--$3500$\,K; the right panel of Fig.~\ref{fig:degeneracy} shows the resulting radius surface. The isoradius contour of $R = 1.497\,\Rearth$ climbs monotonically from CMF\,$= 0.003$ at 1500\,K to CMF\,$= 0.170$ at 3500\,K. Along this contour the volatile envelope disappears entirely. No ice, no water, and no atmospheric layer is required: a pure silicate-plus-iron interior, warmed to a potential temperature consistent with the planet's youth, reproduces the observed radius exactly.

Solution~A, at the cold end (CMF\,$= 0.003$, $\Tpot = 1500$\,K), is a nearly coreless silicate planet. The same four-phase mantle sequence (en\,$\to$\,hpcen\,$\to$\,brg\,$\to$\,ppv) descends onto a vestigial $\varepsilon$-hcp iron core that carries only 10\% of the radius ($R_\mathrm{CMB} = 0.147\,\Rearth$). Central conditions reach just 520\,GPa and 3400\,K, an order of magnitude softer than Solution~A of \object{WASP-47\,e} in both pressure and temperature. The interior is fully solid, with no magma ocean and little prospect of surface volcanism once the young residual heat has radiated away. It is also, crucially, a composition so far from solar refractory ratios (CMF\,$\approx 0.003$ versus Earth's 0.325 or the chondritic estimate near 0.32) that formation under standard accretion pathways \citep[e.g.,][]{Raymond2014} would be difficult to arrange. Solution~A is admissible on radius alone, but not particularly plausible on composition.

Solution~B, at the hot end (CMF\,$= 0.170$, $\Tpot = 3500$\,K), reverses the picture. A surface magma ocean extends inward to ${\sim}\,37$\,GPa (${\sim}\,900$\,km depth), beneath which solid brg and ppv take over. The iron core (450--950\,GPa, 8060--11\,500\,K) is fully liquid across its entire extent, 39\% of the planetary radius, because the lower core pressures never cross the iron melting curve from below, in sharp contrast to \object{WASP-47\,e} Solution~B. The planet hosts an active magma ocean above a fully liquid core. The physical case for a high $\Tpot$ on a 300\,Myr USP planet is direct: postaccretion heat enhanced by the ultrashort orbit, which supplies an additional source through past tidal dissipation during circularization. Earth, at the same age, exceeded 2000\,K in mantle potential temperature \citep{Abbott1994} despite an order of magnitude less irradiation. Solution~B is therefore not a contrived endpoint but the configuration most consistent with the planet's stellar age and orbital architecture, and Solution~A is the one that needs extra justification.

\subsection{Synthesis: Two mass--radius twins, four geophysical states}
\label{sec:superearths:synthesis}

\begin{figure*}
\centering
\includegraphics[width=\textwidth]{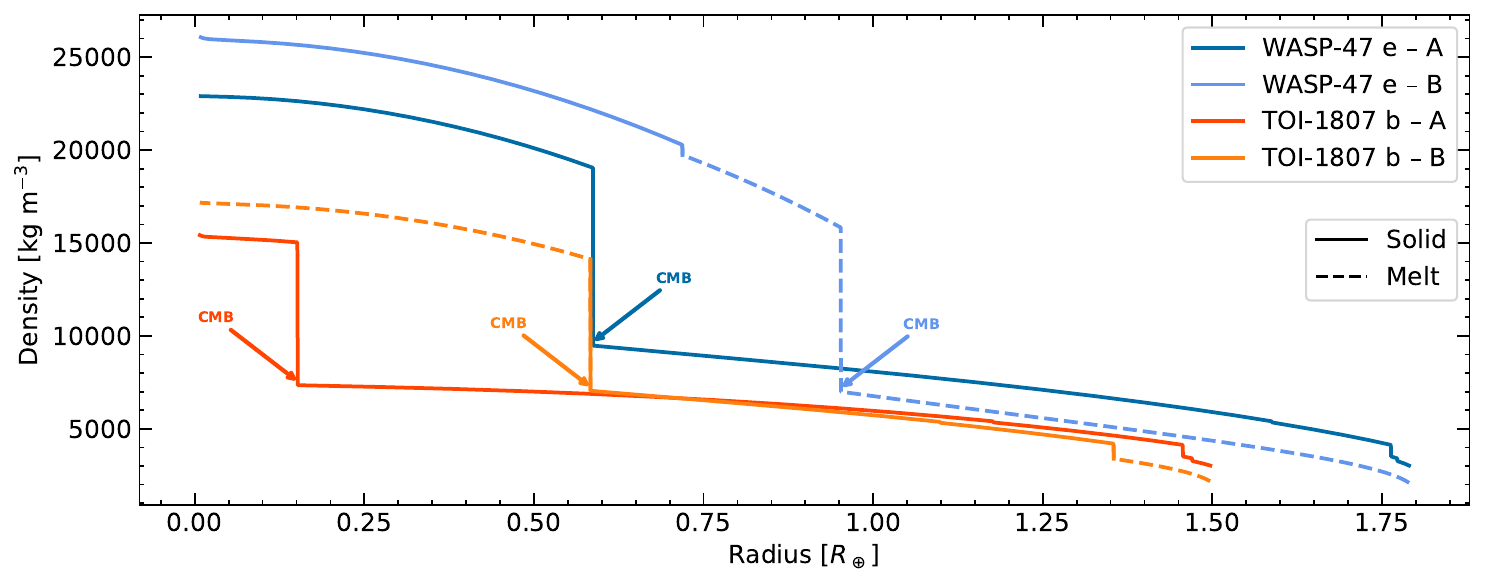}
\caption{Radial density profiles for the four degenerate interior solutions identified in Fig.~\ref{fig:degeneracy} and Sects.~\ref{sec:superearths:wasp47e}--\ref{sec:superearths:toi1807b}. Blue shades show \object{WASP-47\,e} at $M_\mathrm{p} = 7.06\,\Mearth$: solution A (dark blue, CMF $= 0.107$, $\Tpot = 1500$\,K) and solution B (light blue, CMF $= 0.439$, $\Tpot = 4000$\,K). Orange shades show \object{TOI-1807\,b} at $M_\mathrm{p} = 3.25\,\Mearth$: solution A (dark orange, CMF $= 0.003$, $\Tpot = 1500$\,K) and solution B (light orange, CMF $= 0.170$, $\Tpot = 3500$\,K). Solid line segments mark solid mineral phases; dashed segments mark melt phases, so the extent of the dashed segments can be directly read as the magma-ocean depths of the hot solutions. Arrows mark the CMB radius for each profile, taken from the integrated structure rather than from a phase-label reading. Within each pair, the A and B solutions share the same $(M, R)$ but assign radically different fractions of the planetary radius to the iron core and to the silicate mantle.}
\label{fig:profiles}
\end{figure*}

Figure~\ref{fig:profiles} collects the radial density profiles of all four solutions and makes the geophysical content of the degeneracy visible at a glance. Within each pair the two solutions share a mass and radius to within measurement precision but differ in central pressure and temperature by large factors (Table~\ref{tab:comparison}) and assign the CMB to radically different fractions of the planetary radius. Solid lines mark solid phases and dashed lines mark melt, so the dashed extents read off the magma-ocean depths of the two hot solutions directly. The figure makes parametrically the same point as the surface and contour plots of Fig.~\ref{fig:degeneracy}. Once $\Tpot$ is allowed to vary over the range its physical drivers support, a mass and a radius determine very little about the interior, and the classical one-to-one map from $(M, R)$ to the composition is a feature of the cold isothermal assumption rather than of the data.

Therefore, breaking the degeneracy is not a mass--radius problem. It is a problem for observables that see the thermal state directly: day-side thermal emission with broadband or spectroscopic sensitivity to mineral atmospheres at short wavelengths \citep[e.g.,][]{Schaefer2012, Zilinskas2022}, transit measurements of the tidal Love number whose monotonic scaling with the normalized mass moment of the interior makes it a direct discriminant of the core radius fraction at the 10\%-level differences seen here \citep{Padovan2018}, and, on a longer horizon, magnetic-field detection either through near-ultraviolet transit asymmetries caused by standoff bow shocks around planetary magnetospheres \citep{Vidotto2011} or through direct star--planet-interaction radio emission from the cool host stars \citep{Callingham2021}. The interior thermal axis \paleos opens supplies the prior that such follow-up will need, converting an $(M, R)$ pair into a physically grounded (CMF, $\Tpot$) prior for downstream retrievals. The complementary self-consistent atmosphere coupling that Sect.~\ref{sec:discussion} returns to, in the spirit of \citet{Lichtenberg2021, Lichtenberg2025c}, is the mechanism that closes that prior into a posterior.

\begin{table}
\caption{Comparison of the two degenerate interior solutions for \object{WASP-47\,e} and \object{TOI-1807\,b}.}
\label{tab:comparison}
\centering
\begin{tabular}{lcccc}
\toprule
 & \multicolumn{2}{c}{WASP-47\,e} & \multicolumn{2}{c}{TOI-1807\,b} \\
 & A (cool) & B (hot) & A (cool) & B (hot) \\
\midrule
$M_\mathrm{p}$ [$\Mearth$] & 7.06 & 7.06 & 3.25 & 3.25 \\
CMF & 0.107 & 0.439 & 0.003 & 0.170 \\
$\Tpot$ [K] & 1500 & 4000 & 1500 & 3500 \\
$R_\mathrm{CMB}/R_\mathrm{p}$ & 0.33 & 0.53 & 0.10 & 0.39 \\
Mantle & Solid & Fully liquid & Solid & Partial melt \\
Core & Solid & Inner + outer & Solid & Liquid \\
$P_\mathrm{center}$ [GPa] & 2019 & 3242 & 521 & 954 \\
$T_\mathrm{center}$ [K] & 5391 & 20\,010 & 3440 & 11\,500 \\
\bottomrule
\end{tabular}
\end{table}

\section{Discussion and conclusions}
\label{sec:discussion}

We have presented \paleos, an open-source framework that consolidates EoS for iron, magnesium silicate, and water into a unified, phase-aware, and thermally responsive toolkit for exoplanet interiors. Built on a single thermodynamic skeleton, it exposes all its seven thermodynamic quantities analytically, treats solid and molten states on the same footing across the full $(P, T)$ plane of rocky and water-rich planets, and makes the mantle potential temperature $\Tpot$ a controllable input rather than a hidden assumption, delivering three products: the thermodynamic and EoS framework (Sects.~\ref{sec:thermo} and \ref{sec:eos}), the temperature-dependent mass--radius tables (Sect.~\ref{sec:mr}), and the $\Tpot$-driven degeneracy analysis of two USP super-Earths (Sect.~\ref{sec:superearths}). Validation against PREM recovers Earth's radius to 0.3\% and lower-mantle densities to 3\%, sufficient for exoplanet-scale modeling where composition dominates over EoS errors, with the residual ${\sim}\,10$\% core overshoot the known signature of neglected core light elements. The rocky grid is, to our knowledge, the first to treat $\Tpot$ as a free parameter across 300--4000\,K: thermal expansion becomes nonnegligible above $\Tpot \sim 1500$\,K and dominates the radius budget at low masses above 3000\,K, while for water-rich planets the surface-pressure choice (1\,bar versus 1\,mbar) changes the radius by orders of magnitude at low masses and high temperatures, a sensitivity that should be stated explicitly in any compilation. The case studies of WASP-47\,e and TOI-1807\,b concretize the resulting limitation: two planets with the same $(M, R)$ can occupy entirely different phase states, from fully solid to partially molten with a liquid iron core, along a (CMF, $\Tpot$) tradeoff in which a hotter mantle inflates the silicate and demands more iron to match the radius. For USP planets, $\Tpot$ spanning 1500--4000\,K is motivated by primordial heat retention, tidal dissipation, and mineral-atmosphere greenhouse heating, so the bias on an inferred CMF can reach $\Delta \mathrm{CMF} \sim 0.3$, and phase-aware EoS with a continuous solid-to-melt bridge are essential for mapping this degeneracy rather than burying it in an implicit cold-interior assumption.

Beyond these two targets, \paleos is designed as forward-model infrastructure for the next decade of characterization. PLATO \citep{Rauer2025} will deliver percent-level radii for thousands of rocky and transitional planets, lifting interior-model systematics above the observational floor unless they are controlled, which a phase-aware treatment of the silicate and iron melting curves and an explicit surface-pressure choice help to narrow, while Ariel \citep{Tinetti2018} and JWST refine temperature--pressure profiles and compositions for much of the same population and anchor $\Tpot$ through the interior-side boundary of coupled atmosphere--interior models \citep{Tosi2017, Lichtenberg2021, Lichtenberg2025c}. This emphasis on melt and high-$\Tpot$ regimes matches where near-term data will land, since magma-ocean-dominated short-period planets are the accessible laboratory for the coming years \citep{Piette2023, Hu2024, Zhang2024, Teske2025}. The same machinery operates inside statistical interior retrievals: since \citet{Rogers2010} and \citet{Dorn2017}, population-level characterization has been an inverse problem in which the EoS is the forward model for a Bayesian likelihood, and the \paleos tables suit this use case, with queries in tens of microseconds, every state variable thermodynamically consistent by construction, and interpolation well-behaved across solid--solid and solid--melt transitions, so the likelihood surface carries no glitches from piecewise cold EoS. Combined with the $\Tpot$ scan of Sect.~\ref{sec:superearths}, this enables retrievals that treat the mantle thermal state on the same footing as composition rather than fixing an implicit $\Tpot = 300$\,K prior, with the TRAPPIST-1 analysis of \citet{Agol2021} and its PLATO and Ariel successors as the natural testbed.

Several limitations frame the current release. \paleos uses pure endmember compositions, with no light elements in iron cores and no Fe, Ca, or Al in silicate mantles, though mixed compositions can be approximated by ideal lattice mixing \citep[e.g.,][]{Wolf2015} or the additive-volume law; the pyroxene proxy for olivine sacrifices upper-mantle fidelity (the 410 and 660\,km seismic discontinuities) with negligible impact on total radius; no shear modulus is provided, precluding comparison with seismic S-wave velocities; above ${\sim}\,1$\,TPa ppv is predicted to dissociate into oxides \citep{Dong2025}, which we retain for lack of reliable dissociation-product EoS; and the volatile inventory stops at water by design, so H/He sub-Neptunes fall outside scope and are better served by dedicated models such as those of \citet{Tang2025}. Natural extensions include Fe-bearing silicate compositions, volatiles beyond water (CO$_2$, NH$_3$), and coupling to atmosphere models in the spirit of \citet{Lichtenberg2021,Lichtenberg2025c} for self-consistent surface boundary conditions, alongside release of the \paleos tables as a drop-in forward model for retrievals of the forthcoming PLATO and Ariel samples. The aim is not a better or novel EoS, but a continuous and extensible formulation of the underlying thermodynamic properties over large pressure and temperature scales, intended as a reusable basis for interior modeling. The \paleos source code and EoS lookup tables are publicly available, and the mass--radius grid is released as a companion Zenodo dataset, providing a validated foundation for interpreting the growing census of rocky and water-rich exoplanets as the next generation of characterization missions comes online.

\section*{Data availability}

The \paleos source code is publicly available on GitHub (\url{https://github.com/maraattia/PALEOS}). The EoS lookup tables and the mass--radius grid are archived on Zenodo at \url{https://doi.org/10.5281/zenodo.19000315} and \url{https://doi.org/10.5281/zenodo.19221214}, respectively.

\begin{acknowledgements}
We thank the anonymous referee for their positive and helpful review. We offer our warm thanks to Francesca Miozzi for insightful discussions about solid-state thermodynamics. M.A. is supported by the Swiss National Science Foundation through the Postdoc.Mobility fellowship, grant number 230229. This research is supported by the Branco Weiss Foundation, the European Research Council (ERC) under the European Union's Horizon Europe research and innovation program (MagmaWorlds, 101219807), the Alfred P. Sloan Foundation (AEThER, G-2025-25284), NASA’s Nexus for Exoplanet System Science research coordination network (Alien Earths, 80NSSC21K0593), and the NWO NWA--ORC PRELIFE Consortium (NWA.1630.23.013). We thank the Center for Information Technology of the University of Groningen for providing access to the H\'{a}br\'{o}k high performance computing cluster. We made use of the Claude Code Command-Line Interpreter (Anthropic, 2024) for code assistance.
\end{acknowledgements}

\bibliographystyle{aa}
\bibliography{references}

@ARTICLE{Abbott1994,
       author = {{Abbott}, Dallas and {Burgess}, Lee and {Longhi}, John and {Smith}, Walter H.~F.},
        title = "{An empirical thermal history of the Earth's upper mantle}",
      journal = {\jgr},
         year = 1994,
        month = jul,
       volume = {99},
       number = {B7},
        pages = {13,835-13,850},
          doi = {10.1029/94JB00112},
       adsurl = {https://ui.adsabs.harvard.edu/abs/1994JGR....9913835A}
}

@ARTICLE{Agol2021,
       author = {{Agol}, Eric and {Dorn}, Caroline and {Grimm}, Simon L. and {Turbet}, Martin and {Ducrot}, Elsa and {Delrez}, Laetitia and {Gillon}, Micha{\"e}l and {Demory}, Brice-Olivier and {Burdanov}, Artem and {Barkaoui}, Khalid and et al.},
        title = "{Refining the Transit-timing and Photometric Analysis of TRAPPIST-1: Masses, Radii, Densities, Dynamics, and Ephemerides}",
      journal = {\psj},
         year = 2021,
        month = feb,
       volume = {2},
       number = {1},
          eid = {1},
        pages = {1},
          doi = {10.3847/PSJ/abd022},
archivePrefix = {arXiv},
       eprint = {2010.01074},
 primaryClass = {astro-ph.EP},
       adsurl = {https://ui.adsabs.harvard.edu/abs/2021PSJ.....2....1A}
}

@ARTICLE{Aguichine2025,
       author = {{Aguichine}, Artyom and {Batalha}, Natalie and {Fortney}, Jonathan J. and {Nettelmann}, Nadine and {Owen}, James E. and {Kempton}, Eliza M.-R.},
        title = "{Evolution of Steam Worlds: Energetic Aspects}",
      journal = {\apj},
         year = 2025,
        month = aug,
       volume = {988},
       number = {2},
          eid = {186},
        pages = {186},
          doi = {10.3847/1538-4357/add935},
archivePrefix = {arXiv},
       eprint = {2412.17945},
 primaryClass = {astro-ph.EP},
       adsurl = {https://ui.adsabs.harvard.edu/abs/2025ApJ...988..186A}
}

@ARTICLE{Altshuler1987,
       author = {{Al'tshuler}, L.~V. and {Brusnikin}, S.~E. and {Kuz'menkov}, E.~A.},
        title = "{Isotherms and Gr{\"u}neisen functions for 25 metals}",
      journal = {Journal of Applied Mechanics and Technical Physics},
         year = 1987,
        month = jan,
       volume = {28},
       number = {1},
        pages = {129-141},
          doi = {10.1007/BF00918785},
       adsurl = {https://ui.adsabs.harvard.edu/abs/1987JAMTP..28..129A}
}

@ARTICLE{Anderson1995,
       author = {{Anderson}, Orson L. and {Masuda}, Koji and {Isaak}, Donald G.},
        title = "{A new thermodynamic approach for high-pressure physics}",
      journal = {Physics of the Earth and Planetary Interiors},
         year = 1995,
        month = jan,
       volume = {91},
       number = {1},
        pages = {3-16},
          doi = {10.1016/0031-9201(95)03044-W},
       adsurl = {https://ui.adsabs.harvard.edu/abs/1995PEPI...91....3A}
}

@ARTICLE{Angel2000,
       author = {{Angel}, R.~J.},
        title = "{Equations of State}",
      journal = {Reviews in Mineralogy and Geochemistry},
         year = 2000,
        month = jan,
       volume = {41},
       number = {1},
        pages = {35-59},
          doi = {10.2138/rmg.2000.41.2},
       adsurl = {https://ui.adsabs.harvard.edu/abs/2000RvMG...41...35A}
}

@ARTICLE{Anzellini2013,
       author = {{Anzellini}, S. and {Dewaele}, A. and {Mezouar}, M. and {Loubeyre}, P. and {Morard}, G.},
        title = "{Melting of Iron at Earth{\textquoteright}s Inner Core Boundary Based on Fast X-ray Diffraction}",
      journal = {Science},
         year = 2013,
        month = apr,
       volume = {340},
       number = {6131},
        pages = {464-466},
          doi = {10.1126/science.1233514},
       adsurl = {https://ui.adsabs.harvard.edu/abs/2013Sci...340..464A}
}

@ARTICLE{Asplund2021,
       author = {{Asplund}, M. and {Amarsi}, A.~M. and {Grevesse}, N.},
        title = "{The chemical make-up of the Sun: A 2020 vision}",
      journal = {\aap},
         year = 2021,
        month = sep,
       volume = {653},
          eid = {A141},
        pages = {A141},
          doi = {10.1051/0004-6361/202140445},
archivePrefix = {arXiv},
       eprint = {2105.01661},
 primaryClass = {astro-ph.SR},
       adsurl = {https://ui.adsabs.harvard.edu/abs/2021A&A...653A.141A}
}

@ARTICLE{Attia2021,
       author = {{Attia}, M. and {Bourrier}, V. and {Eggenberger}, P. and {Mordasini}, C. and {Beust}, H. and {Ehrenreich}, D.},
        title = "{The JADE code: Coupling secular exoplanetary dynamics and photo-evaporation}",
      journal = {\aap},
         year = 2021,
        month = mar,
       volume = {647},
          eid = {A40},
        pages = {A40},
          doi = {10.1051/0004-6361/202039452},
archivePrefix = {arXiv},
       eprint = {2103.02627},
 primaryClass = {astro-ph.EP},
       adsurl = {https://ui.adsabs.harvard.edu/abs/2021A&A...647A..40A}
}

@ARTICLE{Attia2025,
       author = {{Attia}, M. and {Bourrier}, V. and {Bolmont}, E. and {Mignon}, L. and {Delisle}, J.-B. and {Beust}, H. and {Hara}, N.~C. and {Mordasini}, C.},
        title = "{The JADE code: II. Modeling the coupled orbital and atmospheric evolution of GJ 436 b to constrain its migration and companion}",
      journal = {\aap},
         year = 2025,
        month = oct,
       volume = {702},
          eid = {A132},
        pages = {A132},
          doi = {10.1051/0004-6361/202555239},
archivePrefix = {arXiv},
       eprint = {2509.07938},
 primaryClass = {astro-ph.EP},
       adsurl = {https://ui.adsabs.harvard.edu/abs/2025A&A...702A.132A}
}

@ARTICLE{August2025,
       author = {{August}, P.~C. and {Buchhave}, L.~A. and {Diamond-Lowe}, H. and {Mendon{\c{c}}a}, J.~M. and {Gressier}, A. and {Rathcke}, A.~D. and {Allen}, N.~H. and {Fortune}, M. and {Jones}, K.~D. and {Meier Vald{\'e}s}, E.~A. and {Demory}, B.-O. and {Espinoza}, N. and {Fisher}, C.~E. and {Gibson}, N.~P. and {Heng}, K. and {Hoeijmakers}, J. and {Hooton}, M.~J. and {Kitzmann}, D. and {Prinoth}, B. and {Eastman}, J.~D. and {Barnes}, R.},
        title = "{Hot Rocks Survey I: A possible shallow eclipse for LHS 1478 b}",
      journal = {\aap},
         year = 2025,
        month = mar,
       volume = {695},
          eid = {A171},
        pages = {A171},
          doi = {10.1051/0004-6361/202452611},
archivePrefix = {arXiv},
       eprint = {2410.11048},
 primaryClass = {astro-ph.EP},
       adsurl = {https://ui.adsabs.harvard.edu/abs/2025A&A...695A.171A}
}

@ARTICLE{Baturin2019,
       author = {{Baturin}, V.~A. and {D{\"a}ppen}, W. and {Oreshina}, A.~V. and {Ayukov}, S.~V. and {Gorshkov}, A.~B.},
        title = "{Interpolation of equation-of-state data}",
      journal = {\aap},
         year = 2019,
        month = jun,
       volume = {626},
          eid = {A108},
        pages = {A108},
          doi = {10.1051/0004-6361/201935669},
archivePrefix = {arXiv},
       eprint = {1905.08303},
 primaryClass = {astro-ph.SR},
       adsurl = {https://ui.adsabs.harvard.edu/abs/2019A&A...626A.108B}
}

@ARTICLE{Baumeister2025,
       author = {{Baumeister}, Philipp and {Miozzi}, Francesca and {Guimond}, Claire Marie and {Steinmeyer}, Marie-Luise and {Dorn}, Caroline and {Karato}, Shun-Ichiro and {Bolmont}, {\'E}meline and {Revol}, Alexandre and {Thamm}, Alexander and {Noack}, Lena},
        title = "{Fundamentals of Interior Modelling and Challenges in the Interpretation of Observed Rocky Exoplanets}",
      journal = {\ssr},
         year = 2025,
        month = dec,
       volume = {221},
       number = {8},
          eid = {123},
        pages = {123},
          doi = {10.1007/s11214-025-01248-5},
archivePrefix = {arXiv},
       eprint = {2511.10269},
 primaryClass = {astro-ph.EP},
       adsurl = {https://ui.adsabs.harvard.edu/abs/2025SSRv..221..123B}
}

@ARTICLE{Benneke2017,
       author = {{Benneke}, Bj{\"o}rn and {Werner}, Michael and {Petigura}, Erik and {Knutson}, Heather and {Dressing}, Courtney and {Crossfield}, Ian J.~M. and {Schlieder}, Joshua E. and {Livingston}, John and {Beichman}, Charles and {Christiansen}, Jessie and {Krick}, Jessica and {Gorjian}, Varoujan and {Howard}, Andrew W. and {Sinukoff}, Evan and {Ciardi}, David R. and {Akeson}, Rachel L.},
        title = "{Spitzer Observations Confirm and Rescue the Habitable-zone Super-Earth K2-18b for Future Characterization}",
      journal = {\apj},
         year = 2017,
        month = jan,
       volume = {834},
       number = {2},
          eid = {187},
        pages = {187},
          doi = {10.3847/1538-4357/834/2/187},
archivePrefix = {arXiv},
       eprint = {1610.07249},
 primaryClass = {astro-ph.EP},
       adsurl = {https://ui.adsabs.harvard.edu/abs/2017ApJ...834..187B}
}

@ARTICLE{Benneke2019,
       author = {{Benneke}, Bj{\"o}rn and {Wong}, Ian and {Piaulet}, Caroline and {Knutson}, Heather A. and {Lothringer}, Joshua and {Morley}, Caroline V. and {Crossfield}, Ian J.~M. and {Gao}, Peter and {Greene}, Thomas P. and {Dressing}, Courtney and et al.},
        title = "{Water Vapor and Clouds on the Habitable-zone Sub-Neptune Exoplanet K2-18b}",
      journal = {\apjl},
         year = 2019,
        month = dec,
       volume = {887},
       number = {1},
          eid = {L14},
        pages = {L14},
          doi = {10.3847/2041-8213/ab59dc},
archivePrefix = {arXiv},
       eprint = {1909.04642},
 primaryClass = {astro-ph.EP},
       adsurl = {https://ui.adsabs.harvard.edu/abs/2019ApJ...887L..14B}
}

@ARTICLE{Birch1947,
       author = {{Birch}, Francis},
        title = "{Finite Elastic Strain of Cubic Crystals}",
      journal = {Physical Review},
         year = 1947,
        month = jun,
       volume = {71},
       number = {11},
        pages = {809-824},
          doi = {10.1103/PhysRev.71.809},
       adsurl = {https://ui.adsabs.harvard.edu/abs/1947PhRv...71..809B}
}

@ARTICLE{Birch1978,
       author = {{Birch}, Francis},
        title = "{Finite strain isotherm and velocities for single-crystal and polycrystalline NaCl at high pressures and 300{\textdegree}K}",
      journal = {\jgr},
         year = 1978,
        month = mar,
       volume = {83},
       number = {B3},
        pages = {1257-1268},
          doi = {10.1029/JB083iB03p01257},
       adsurl = {https://ui.adsabs.harvard.edu/abs/1978JGR....83.1257B}
}

@ARTICLE{Becker2014,
       author = {{Becker}, Andreas and {Lorenzen}, Winfried and {Fortney}, Jonathan J. and {Nettelmann}, Nadine and {Sch{\"o}ttler}, Manuel and {Redmer}, Ronald},
        title = "{Ab Initio Equations of State for Hydrogen (H-REOS.3) and Helium (He-REOS.3) and their Implications for the Interior of Brown Dwarfs}",
      journal = {\apjs},
         year = 2014,
        month = dec,
       volume = {215},
       number = {2},
          eid = {21},
        pages = {21},
          doi = {10.1088/0067-0049/215/2/21},
archivePrefix = {arXiv},
       eprint = {1411.4010},
 primaryClass = {astro-ph.EP},
       adsurl = {https://ui.adsabs.harvard.edu/abs/2014ApJS..215...21B}
}

@ARTICLE{Becker2015,
       author = {{Becker}, Juliette C. and {Vanderburg}, Andrew and {Adams}, Fred C. and {Rappaport}, Saul A. and {Schwengeler}, Hans Martin},
        title = "{WASP-47: A Hot Jupiter System with Two Additional Planets Discovered by K2}",
      journal = {\apjl},
         year = 2015,
        month = oct,
       volume = {812},
       number = {2},
          eid = {L18},
        pages = {L18},
          doi = {10.1088/2041-8205/812/2/L18},
archivePrefix = {arXiv},
       eprint = {1508.02411},
 primaryClass = {astro-ph.EP},
       adsurl = {https://ui.adsabs.harvard.edu/abs/2015ApJ...812L..18B}
}

@ARTICLE{Belonoshko2005,
       author = {{Belonoshko}, A.~B. and {Skorodumova}, N.~V. and {Rosengren}, A. and {Ahuja}, R. and {Johansson}, B. and {Burakovsky}, L. and {Preston}, D.~L.},
        title = "{High-Pressure Melting of MgSiO$_{3}$}",
      journal = {\prl},
         year = 2005,
        month = may,
       volume = {94},
       number = {19},
          eid = {195701},
        pages = {195701},
          doi = {10.1103/PhysRevLett.94.195701},
       adsurl = {https://ui.adsabs.harvard.edu/abs/2005PhRvL..94s5701B}
}

@ARTICLE{Boehler1993,
       author = {{Boehler}, R.},
        title = "{Temperatures in the Earth's core from melting-point measurements of iron at high static pressures}",
      journal = {\nat},
         year = 1993,
        month = jun,
       volume = {363},
       number = {6429},
        pages = {534-536},
          doi = {10.1038/363534a0},
       adsurl = {https://ui.adsabs.harvard.edu/abs/1993Natur.363..534B}
}

@ARTICLE{Boer2025,
       author = {{Boer}, Iris D. and {Nicholls}, Harrison and {Lichtenberg}, Tim},
        title = "{Absence of a Runaway Greenhouse Limit on Lava Planets}",
      journal = {\apj},
         year = 2025,
        month = jul,
       volume = {987},
       number = {2},
          eid = {172},
        pages = {172},
          doi = {10.3847/1538-4357/add69f},
archivePrefix = {arXiv},
       eprint = {2505.11149},
 primaryClass = {astro-ph.EP},
       adsurl = {https://ui.adsabs.harvard.edu/abs/2025ApJ...987..172B}
}

@ARTICLE{Boukare2022,
       author = {{Boukar{\'e}}, Charles-{\'E}douard and {Cowan}, Nicolas B. and {Badro}, James},
        title = "{Deep Two-phase, Hemispherical Magma Oceans on Lava Planets}",
      journal = {\apj},
         year = 2022,
        month = sep,
       volume = {936},
       number = {2},
          eid = {148},
        pages = {148},
          doi = {10.3847/1538-4357/ac8792},
archivePrefix = {arXiv},
       eprint = {2205.02864},
 primaryClass = {astro-ph.EP},
       adsurl = {https://ui.adsabs.harvard.edu/abs/2022ApJ...936..148B}
}

@ARTICLE{Boukare2025,
       author = {{Boukar{\'e}}, Charles-{\'E}douard and {Lemasquerier}, Daphn{\'e} and {Cowan}, Nicolas B. and {Samuel}, Henri and {Badro}, James and {Dang}, Lisa and {Falco}, Aur{\'e}lien and {Charnoz}, S{\'e}bastien},
        title = "{The role of interior dynamics and differentiation on the surface and in the atmosphere of lava planets}",
      journal = {Nature Astronomy},
         year = 2025,
        month = jul,
       volume = {9},
        pages = {1511-1522},
          doi = {10.1038/s41550-025-02617-4},
archivePrefix = {arXiv},
       eprint = {2308.13614},
 primaryClass = {astro-ph.EP},
       adsurl = {https://ui.adsabs.harvard.edu/abs/2025NatAs...9.1511B}
}

@ARTICLE{Boukare2025b,
       author = {{Boukar{\'e}}, Charles-{\'E}douard and {Schaefer}, Laura K. and {Rizo}, Hanika},
        title = "{The Earth's magma ocean: Processes and current interpretations from an interdisciplinary perspective}",
      journal = {Physics of the Earth and Planetary Interiors},
         year = 2025,
        month = dec,
       volume = {369},
          eid = {107463},
        pages = {107463},
          doi = {10.1016/j.pepi.2025.107463},
       adsurl = {https://ui.adsabs.harvard.edu/abs/2025PEPI..36907463B}
}

@ARTICLE{Bower2019,
       author = {{Bower}, Dan J. and {Kitzmann}, Daniel and {Wolf}, Aaron S. and {Sanan}, Patrick and {Dorn}, Caroline and {Oza}, Apurva V.},
        title = "{Linking the evolution of terrestrial interiors and an early outgassed atmosphere to astrophysical observations}",
      journal = {\aap},
         year = 2019,
        month = nov,
       volume = {631},
          eid = {A103},
        pages = {A103},
          doi = {10.1051/0004-6361/20193571010.31223/osf.io/ctqe3},
       adsurl = {https://ui.adsabs.harvard.edu/abs/2019A&A...631A.103B}
}

@ARTICLE{Bower2022,
       author = {{Bower}, Dan J. and {Hakim}, Kaustubh and {Sossi}, Paolo A. and {Sanan}, Patrick},
        title = "{Retention of Water in Terrestrial Magma Oceans and Carbon-rich Early Atmospheres}",
      journal = {\psj},
         year = 2022,
        month = apr,
       volume = {3},
       number = {4},
          eid = {93},
        pages = {93},
          doi = {10.3847/PSJ/ac5fb1},
archivePrefix = {arXiv},
       eprint = {2110.08029},
 primaryClass = {astro-ph.EP},
       adsurl = {https://ui.adsabs.harvard.edu/abs/2022PSJ.....3...93B}
}

@ARTICLE{Brown2018,
       author = {{Brown}, J. Michael},
        title = "{Local basis function representations of thermodynamic surfaces: Water at high pressure and temperature as an example}",
      journal = {Fluid Phase Equilibria},
         year = 2018,
        month = may,
       volume = {463},
        pages = {18-31},
          doi = {10.1016/j.fluid.2018.02.001},
       adsurl = {https://ui.adsabs.harvard.edu/abs/2018FlPEq.463...18B}
}

@ARTICLE{vanBuchem2025,
       author = {{van Buchem}, C.~P.~A. and {Zilinskas}, M. and {Miguel}, Y. and {van Westrenen}, W.},
        title = "{LavAtmos 2.0: Incorporating volatile species in vaporisation models}",
      journal = {\aap},
         year = 2025,
        month = mar,
       volume = {695},
          eid = {A154},
        pages = {A154},
          doi = {10.1051/0004-6361/202450992},
archivePrefix = {arXiv},
       eprint = {2408.10863},
 primaryClass = {astro-ph.EP},
       adsurl = {https://ui.adsabs.harvard.edu/abs/2025A&A...695A.154V}
}

@ARTICLE{Burbidge1957,
       author = {{Burbidge}, E. Margaret and {Burbidge}, G.~R. and {Fowler}, William A. and {Hoyle}, F.},
        title = "{Synthesis of the Elements in Stars}",
      journal = {Reviews of Modern Physics},
         year = 1957,
        month = oct,
       volume = {29},
       number = {4},
        pages = {547-650},
          doi = {10.1103/RevModPhys.29.547},
       adsurl = {https://ui.adsabs.harvard.edu/abs/1957RvMP...29..547B}
}

@ARTICLE{Calder2025,
       author = {{Calder}, Robb and {Shorttle}, Oliver and {Nicholls}, Harrison and {Lichtenberg}, Tim and {Guimond}, Claire-Marie},
        title = "{Most rocky sub-Neptunes are molten: mapping the solidification shoreline for gas dwarf exoplanets}",
      journal = {\mnras},
         year = 2026,
        month = jun,
       volume = {549},
       number = {3},
          eid = {stag1007},
        pages = {stag1007},
          doi = {10.1093/mnras/stag1007},
}

@ARTICLE{Callingham2021,
       author = {{Callingham}, J.~R. and {Vedantham}, H.~K. and {Shimwell}, T.~W. and {Pope}, B.~J.~S. and {Davis}, I.~E. and {Best}, P.~N. and {Hardcastle}, M.~J. and {R{\"o}ttgering}, H.~J.~A. and {Sabater}, J. and {Tasse}, C. and et al.},
        title = "{The population of M dwarfs observed at low radio frequencies}",
      journal = {Nature Astronomy},
         year = 2021,
        month = dec,
       volume = {5},
        pages = {1233-1239},
          doi = {10.1038/s41550-021-01483-0},
archivePrefix = {arXiv},
       eprint = {2110.03713},
 primaryClass = {astro-ph.SR},
       adsurl = {https://ui.adsabs.harvard.edu/abs/2021NatAs...5.1233C}
}

@ARTICLE{CanoAmoros2026,
       author = {{Cano Amoros}, M. and {Nettelmann}, N. and {Tosi}, N. and {Mazevet}, S.},
        title = "{Revised entropy of the AQUA equation of state}",
      journal = {\aap},
         year = 2026,
        month = apr,
       volume = {708},
          eid = {A324},
        pages = {A324},
          doi = {10.1051/0004-6361/202557580},
       adsurl = {https://ui.adsabs.harvard.edu/abs/2026A&A...708A.324C}
}

@ARTICLE{Caracas2023,
       author = {{Caracas}, Razvan and {Stewart}, Sarah T.},
        title = "{No magma ocean surface after giant impacts between rocky planets}",
      journal = {Earth and Planetary Science Letters},
         year = 2023,
        month = apr,
       volume = {608},
          eid = {118014},
        pages = {118014},
          doi = {10.1016/j.epsl.2023.118014},
       adsurl = {https://ui.adsabs.harvard.edu/abs/2023E&PSL.60818014C}
}

@ARTICLE{Caracas2024,
       author = {{Caracas}, Razvan},
        title = "{The thermal equation of state of the magma Ocean}",
      journal = {Earth and Planetary Science Letters},
         year = 2024,
        month = jul,
       volume = {637},
          eid = {118724},
        pages = {118724},
          doi = {10.1016/j.epsl.2024.118724},
       adsurl = {https://ui.adsabs.harvard.edu/abs/2024E&PSL.63718724C}
}

@ARTICLE{Chopelas1992,
       author = {{Chopelas}, A. and {Boehler}, R.},
        title = "{Thermal expansivity in the lower mantle}",
      journal = {\grl},
         year = 1992,
        month = oct,
       volume = {19},
       number = {19},
        pages = {1983-1986},
          doi = {10.1029/92GL02144},
       adsurl = {https://ui.adsabs.harvard.edu/abs/1992GeoRL..19.1983C}
}

@ARTICLE{Connolly2009,
       author = {{Connolly}, J.~A.~D.},
        title = "{The geodynamic equation of state: What and how}",
      journal = {Geochemistry, Geophysics, Geosystems},
         year = 2009,
        month = oct,
       volume = {10},
       number = {10},
          eid = {Q10014},
        pages = {Q10014},
          doi = {10.1029/2009GC002540},
       adsurl = {https://ui.adsabs.harvard.edu/abs/2009GGG....1010014C}
}

@ARTICLE{Cottaar2014,
       author = {{Cottaar}, Sanne and {Heister}, Timo and {Rose}, Ian and {Unterborn}, Cayman},
        title = "{BurnMan: A lower mantle mineral physics toolkit}",
      journal = {Geochemistry, Geophysics, Geosystems},
         year = 2014,
        month = apr,
       volume = {15},
       number = {4},
        pages = {1164-1179},
          doi = {10.1002/2013GC005122},
       adsurl = {https://ui.adsabs.harvard.edu/abs/2014GGG....15.1164C}
}

@ARTICLE{Debye1912,
       author = {{Debye}, P.},
        title = "{Zur Theorie der spezifischen W{\"a}rmen}",
      journal = {Annalen der Physik},
         year = 1912,
        month = jan,
       volume = {344},
       number = {14},
        pages = {789-839},
          doi = {10.1002/andp.19123441404},
       adsurl = {https://ui.adsabs.harvard.edu/abs/1912AnP...344..789D}
}

@article{Dinsdale1991,
    title = {{SGTE} data for pure elements},
    volume = {15},
    issn = {0364-5916},
    url = {https://www.sciencedirect.com/science/article/pii/036459169190030N},
    doi = {10.1016/0364-5916(91)90030-N},
    number = {4},
    urldate = {2026-04-01},
    journal = {Calphad},
    author = {Dinsdale, A. T.},
    month = oct,
    year = {1991},
    pages = {317--425},
}

@ARTICLE{Dong2025,
       author = {{Dong}, Junjie and {Mardaru}, Gabriel-Darius and {Asimow}, Paul D. and {Stixrude}, Lars P. and {Fischer}, Rebecca A.},
        title = "{Structure and Melting of Fe, MgO, SiO$_{2}$, and MgSiO$_{3}$ in Planets: Database, Inversion, and Phase Diagram}",
      journal = {\psj},
         year = 2025,
        month = apr,
       volume = {6},
       number = {4},
          eid = {103},
        pages = {103},
          doi = {10.3847/PSJ/adc717},
archivePrefix = {arXiv},
       eprint = {2503.21734},
 primaryClass = {astro-ph.EP},
       adsurl = {https://ui.adsabs.harvard.edu/abs/2025PSJ.....6..103D}
}

@ARTICLE{Dorn2015,
       author = {{Dorn}, Caroline and {Khan}, Amir and {Heng}, Kevin and {Connolly}, James A.~D. and {Alibert}, Yann and {Benz}, Willy and {Tackley}, Paul},
        title = "{Can we constrain the interior structure of rocky exoplanets from mass and radius measurements?}",
      journal = {\aap},
         year = 2015,
        month = may,
       volume = {577},
          eid = {A83},
        pages = {A83},
          doi = {10.1051/0004-6361/201424915},
archivePrefix = {arXiv},
       eprint = {1502.03605},
 primaryClass = {astro-ph.EP},
       adsurl = {https://ui.adsabs.harvard.edu/abs/2015A&A...577A..83D}
}

@ARTICLE{Dorn2017,
       author = {{Dorn}, Caroline and {Venturini}, Julia and {Khan}, Amir and {Heng}, Kevin and {Alibert}, Yann and {Helled}, Ravit and {Rivoldini}, Attilio and {Benz}, Willy},
        title = "{A generalized Bayesian inference method for constraining the interiors of super Earths and sub-Neptunes}",
      journal = {\aap},
         year = 2017,
        month = jan,
       volume = {597},
          eid = {A37},
        pages = {A37},
          doi = {10.1051/0004-6361/201628708},
archivePrefix = {arXiv},
       eprint = {1609.03908},
 primaryClass = {astro-ph.IM},
       adsurl = {https://ui.adsabs.harvard.edu/abs/2017A&A...597A..37D}
}

@ARTICLE{Dorn2018,
       author = {{Dorn}, Caroline and {Mosegaard}, Klaus and {Grimm}, Simon L. and {Alibert}, Yann},
        title = "{Interior Characterization in Multiplanetary Systems: TRAPPIST-1}",
      journal = {\apj},
         year = 2018,
        month = sep,
       volume = {865},
       number = {1},
          eid = {20},
        pages = {20},
          doi = {10.3847/1538-4357/aad95d},
archivePrefix = {arXiv},
       eprint = {1808.01803},
 primaryClass = {astro-ph.EP},
       adsurl = {https://ui.adsabs.harvard.edu/abs/2018ApJ...865...20D}
}

@ARTICLE{Dorn2021,
       author = {{Dorn}, Caroline and {Lichtenberg}, Tim},
        title = "{Hidden Water in Magma Ocean Exoplanets}",
      journal = {\apjl},
         year = 2021,
        month = nov,
       volume = {922},
       number = {1},
          eid = {L4},
        pages = {L4},
          doi = {10.3847/2041-8213/ac33af},
archivePrefix = {arXiv},
       eprint = {2110.15069},
 primaryClass = {astro-ph.EP},
       adsurl = {https://ui.adsabs.harvard.edu/abs/2021ApJ...922L...4D}
}

@ARTICLE{Dorogokupets2017,
       author = {{Dorogokupets}, P.~I. and {Dymshits}, A.~M. and {Litasov}, K.~D. and {Sokolova}, T.~S.},
        title = "{Thermodynamics and Equations of State of Iron to 350{\,}GPa and 6000{\,}K}",
      journal = {Scientific Reports},
         year = 2017,
        month = mar,
       volume = {7},
          eid = {41863},
        pages = {41863},
          doi = {10.1038/srep41863},
       adsurl = {https://ui.adsabs.harvard.edu/abs/2017NatSR...741863D}
}

@ARTICLE{Driscoll2014,
       author = {{Driscoll}, P. and {Bercovici}, D.},
        title = "{On the thermal and magnetic histories of Earth and Venus: Influences of melting, radioactivity, and conductivity}",
      journal = {Physics of the Earth and Planetary Interiors},
         year = 2014,
        month = nov,
       volume = {236},
        pages = {36-51},
          doi = {10.1016/j.pepi.2014.08.004},
       adsurl = {https://ui.adsabs.harvard.edu/abs/2014PEPI..236...36D}
}

@ARTICLE{Duffy2019,
       author = {{Duffy}, Thomas S. and {Smith}, Raymond F.},
        title = "{Ultra-High Pressure Dynamic Compression of Geological Materials}",
      journal = {Frontiers in Earth Science},
         year = 2019,
        month = feb,
       volume = {7},
          eid = {23},
        pages = {23},
          doi = {10.3389/feart.2019.00023},
       adsurl = {https://ui.adsabs.harvard.edu/abs/2019FrEaS...7...23D}
}

@ARTICLE{Dziewonski1981,
       author = {{Dziewonski}, Adam M. and {Anderson}, Don L.},
        title = "{Preliminary reference Earth model}",
      journal = {Physics of the Earth and Planetary Interiors},
         year = 1981,
        month = jun,
       volume = {25},
       number = {4},
        pages = {297-356},
          doi = {10.1016/0031-9201(81)90046-7},
       adsurl = {https://ui.adsabs.harvard.edu/abs/1981PEPI...25..297D}
}

@article{Einstein1907,
author = {Einstein, A.},
title = {Die Plancksche Theorie der Strahlung und die Theorie der spezifischen Wärme},
journal = {Annalen der Physik},
volume = {327},
number = {1},
pages = {180-190},
doi = {https://doi.org/10.1002/andp.19063270110},
url = {https://onlinelibrary.wiley.com/doi/abs/10.1002/andp.19063270110},
eprint = {https://onlinelibrary.wiley.com/doi/pdf/10.1002/andp.19063270110},
year = {1907}
}

@ARTICLE{ElkisTanton2012,
       author = {{Elkins-Tanton}, Linda T.},
        title = "{Magma Oceans in the Inner Solar System}",
      journal = {Annual Review of Earth and Planetary Sciences},
         year = 2012,
        month = may,
       volume = {40},
       number = {1},
        pages = {113-139},
          doi = {10.1146/annurev-earth-042711-105503},
       adsurl = {https://ui.adsabs.harvard.edu/abs/2012AREPS..40..113E}
}

@ARTICLE{Farhat2025,
       author = {{Farhat}, Mohammad and {Auclair-Desrotour}, Pierre and {Bou{\'e}}, Gwena{\"e}l and {Lichtenberg}, Tim and {Laskar}, Jacques},
        title = "{Tides on Lava Worlds: Application to Close-in Exoplanets and the Early Earth─Moon System}",
      journal = {\apj},
         year = 2025,
        month = feb,
       volume = {979},
       number = {2},
          eid = {133},
        pages = {133},
          doi = {10.3847/1538-4357/ad9b93},
archivePrefix = {arXiv},
       eprint = {2412.07285},
 primaryClass = {astro-ph.EP},
       adsurl = {https://ui.adsabs.harvard.edu/abs/2025ApJ...979..133F}
}

@ARTICLE{Farhat2026,
       author = {{Farhat}, Mohammad and {Chiang}, Eugene},
        title = "{Magma Ocean Waves and Thermal Variability on Lava Worlds}",
      journal = {\apj},
         year = 2026,
        month = may,
       volume = {1003},
       number = {2},
          eid = {208},
        pages = {208},
          doi = {10.3847/1538-4357/ae6504},
}

@ARTICLE{Fei2016,
       author = {{Fei}, Yingwei and {Murphy}, Caitlin and {Shibazaki}, Yuki and {Shahar}, Anat and {Huang}, Haijun},
        title = "{Thermal equation of state of hcp-iron: Constraint on the density deficit of Earth's solid inner core}",
      journal = {\grl},
         year = 2016,
        month = jul,
       volume = {43},
       number = {13},
        pages = {6837-6843},
          doi = {10.1002/2016GL069456},
       adsurl = {https://ui.adsabs.harvard.edu/abs/2016GeoRL..43.6837F}
}

@ARTICLE{Fei2021,
       author = {{Fei}, Yingwei and {Seagle}, Christopher T. and {Townsend}, Joshua P. and {McCoy}, Chad A. and {Boujibar}, Asmaa and {Driscoll}, Peter and {Shulenburger}, Luke and {Furnish}, Michael D.},
        title = "{Melting and density of MgSiO$_{3}$ determined by shock compression of bridgmanite to 1254GPa}",
      journal = {Nature Communications},
         year = 2021,
        month = jan,
       volume = {12},
          eid = {876},
        pages = {876},
          doi = {10.1038/s41467-021-21170-y},
       adsurl = {https://ui.adsabs.harvard.edu/abs/2021NatCo..12..876F}
}

@ARTICLE{Feistel2006,
       author = {{Feistel}, Rainer and {Wagner}, Wolfgang},
        title = "{A New Equation of State for H$_{2}$O Ice Ih}",
      journal = {Journal of Physical and Chemical Reference Data},
         year = 2006,
        month = jun,
       volume = {35},
       number = {2},
        pages = {1021-1047},
          doi = {10.1063/1.2183324},
       adsurl = {https://ui.adsabs.harvard.edu/abs/2006JPCRD..35.1021F}
}

@ARTICLE{Foley2016,
       author = {{Foley}, Bradford J. and {Driscoll}, Peter E.},
        title = "{Whole planet coupling between climate, mantle, and core: Implications for rocky planet evolution}",
      journal = {Geochemistry, Geophysics, Geosystems},
         year = 2016,
        month = may,
       volume = {17},
       number = {5},
        pages = {1885-1914},
          doi = {10.1002/2015GC006210},
archivePrefix = {arXiv},
       eprint = {1711.06801},
 primaryClass = {astro-ph.EP},
       adsurl = {https://ui.adsabs.harvard.edu/abs/2016GGG....17.1885F}
}

@ARTICLE{French2015,
       author = {{French}, Martin and {Redmer}, Ronald},
        title = "{Construction of a thermodynamic potential for the water ices VII and X}",
      journal = {\prb},
         year = 2015,
        month = jan,
       volume = {91},
       number = {1},
          eid = {014308},
        pages = {014308},
          doi = {10.1103/PhysRevB.91.014308},
       adsurl = {https://ui.adsabs.harvard.edu/abs/2015PhRvB..91a4308F}
}

@ARTICLE{French2016,
       author = {{French}, Martin and {Desjarlais}, Michael P. and {Redmer}, Ronald},
        title = "{Ab initio calculation of thermodynamic potentials and entropies for superionic water}",
      journal = {\pre},
         year = 2016,
        month = feb,
       volume = {93},
       number = {2},
          eid = {022140},
        pages = {022140},
          doi = {10.1103/PhysRevE.93.022140},
       adsurl = {https://ui.adsabs.harvard.edu/abs/2016PhRvE..93b2140F}
}

@ARTICLE{Frost2022,
       author = {{Frost}, Daniel A. and {Avery}, Margaret S. and {Buffett}, Bruce A. and {Chidester}, Bethany A. and {Deng}, Jie and {Dorfman}, Susannah M. and {Li}, Zhi and {Liu}, Lijun and {Lv}, Mingda and {Martin}, Joshua F.},
        title = "{Multidisciplinary Constraints on the Thermal-Chemical Boundary Between Earth's Core and Mantle}",
      journal = {Geochemistry, Geophysics, Geosystems},
         year = 2022,
        month = mar,
       volume = {23},
       number = {3},
          eid = {e2021GC009764},
        pages = {e2021GC009764},
          doi = {10.1029/2021GC009764},
       adsurl = {https://ui.adsabs.harvard.edu/abs/2022GGG....2309764F}
}

@ARTICLE{Gillon2016,
       author = {{Gillon}, Micha{\"e}l and {Jehin}, Emmanu{\"e}l and {Lederer}, Susan M. and {Delrez}, Laetitia and {de Wit}, Julien and {Burdanov}, Artem and {Van Grootel}, Val{\'e}rie and {Burgasser}, Adam J. and {Triaud}, Amaury H.~M.~J. and {Opitom}, Cyrielle and et al.},
        title = "{Temperate Earth-sized planets transiting a nearby ultracool dwarf star}",
      journal = {\nat},
         year = 2016,
        month = may,
       volume = {533},
       number = {7602},
        pages = {221-224},
          doi = {10.1038/nature17448},
archivePrefix = {arXiv},
       eprint = {1605.07211},
 primaryClass = {astro-ph.EP},
       adsurl = {https://ui.adsabs.harvard.edu/abs/2016Natur.533..221G}
}

@ARTICLE{Goes2020,
       author = {{Goes}, Saskia and {Hasterok}, Derrick and {Schutt}, Derek L. and {Kl{\"o}cking}, Marthe},
        title = "{Continental lithospheric temperatures: A review}",
      journal = {Physics of the Earth and Planetary Interiors},
         year = 2020,
        month = sep,
       volume = {306},
          eid = {106509},
        pages = {106509},
          doi = {10.1016/j.pepi.2020.106509},
       adsurl = {https://ui.adsabs.harvard.edu/abs/2020PEPI..30606509G}
}

@book{Gopal1966,
    series = {International cryogenics monograph series},
    title = {Specific {Heats} at {Low} {Temperatures}},
    isbn = {978-0-608-05758-3},
    url = {https://books.google.nl/books?id=jAhRAAAAMAAJ},
    publisher = {Plenum Press},
    author = {{Gopal}, E.S.R.},
    year = {1966},
    lccn = {65011339},
}

@ARTICLE{Greenwood2005,
       author = {{Greenwood}, Richard C. and {Franchi}, Ian A. and {Jambon}, Albert and {Buchanan}, Paul C.},
        title = "{Widespread magma oceans on asteroidal bodies in the early Solar System}",
      journal = {\nat},
         year = 2005,
        month = jun,
       volume = {435},
       number = {7044},
        pages = {916-918},
          doi = {10.1038/nature03612},
       adsurl = {https://ui.adsabs.harvard.edu/abs/2005Natur.435..916G}
}

@ARTICLE{Gruneisen1912,
       author = {{Gr{\"u}neisen}, E.},
        title = "{Theorie des festen Zustandes einatomiger Elemente}",
      journal = {Annalen der Physik},
         year = 1912,
        month = jan,
       volume = {344},
       number = {12},
        pages = {257-306},
          doi = {10.1002/andp.19123441202},
       adsurl = {https://ui.adsabs.harvard.edu/abs/1912AnP...344..257G}
}

@ARTICLE{Guimond2024,
       author = {{Guimond}, Claire Marie and {Wang}, Haiyang and {Seidler}, Fabian and {Sossi}, Paolo and {Mahajan}, Aprajit and {Shorttle}, Oliver},
        title = "{From Stars to Diverse Mantles, Melts, Crusts, and Atmospheres of Rocky Exoplanets}",
      journal = {Reviews in Mineralogy and Geochemistry},
         year = 2024,
        month = jul,
       volume = {90},
       number = {1},
        pages = {259-300},
          doi = {10.2138/rmg.2024.90.08},
archivePrefix = {arXiv},
       eprint = {2404.15427},
 primaryClass = {astro-ph.EP},
       adsurl = {https://ui.adsabs.harvard.edu/abs/2024RvMG...90..259G}
}

@ARTICLE{Hakim2018,
       author = {{Hakim}, Kaustubh and {Rivoldini}, Attilio and {Van Hoolst}, Tim and {Cottenier}, Stefaan and {Jaeken}, Jan and {Chust}, Thomas and {Steinle-Neumann}, Gerd},
        title = "{A new ab initio equation of state of hcp-Fe and its implication on the interior structure and mass-radius relations of rocky super-Earths}",
      journal = {\icarus},
         year = 2018,
        month = oct,
       volume = {313},
        pages = {61-78},
          doi = {10.1016/j.icarus.2018.05.005},
archivePrefix = {arXiv},
       eprint = {1805.10530},
 primaryClass = {astro-ph.EP},
       adsurl = {https://ui.adsabs.harvard.edu/abs/2018Icar..313...61H}
}

@ARTICLE{Haldemann2020,
       author = {{Haldemann}, Jonas and {Alibert}, Yann and {Mordasini}, Christoph and {Benz}, Willy},
        title = "{AQUA: a collection of H$_{2}$O equations of state for planetary models}",
      journal = {\aap},
         year = 2020,
        month = nov,
       volume = {643},
          eid = {A105},
        pages = {A105},
          doi = {10.1051/0004-6361/202038367},
archivePrefix = {arXiv},
       eprint = {2009.10098},
 primaryClass = {astro-ph.EP},
       adsurl = {https://ui.adsabs.harvard.edu/abs/2020A&A...643A.105H}
}

@ARTICLE{Haldemann2024,
       author = {{Haldemann}, Jonas and {Dorn}, Caroline and {Venturini}, Julia and {Alibert}, Yann and {Benz}, Willy},
        title = "{BICEPS: An improved characterization model for low- and intermediate-mass exoplanets}",
      journal = {\aap},
         year = 2024,
        month = jan,
       volume = {681},
          eid = {A96},
        pages = {A96},
          doi = {10.1051/0004-6361/202346965},
       adsurl = {https://ui.adsabs.harvard.edu/abs/2024A&A...681A..96H}
}

@ARTICLE{Hallatt2026,
       author = {{Hallatt}, Tim and {Millholland}, Sarah},
        title = "{Coupled Planetary Interior and Tidal Evolution}",
      journal = {\apj},
         year = 2026,
        month = feb,
       volume = {997},
       number = {2},
          eid = {138},
        pages = {138},
          doi = {10.3847/1538-4357/ae129d},
archivePrefix = {arXiv},
       eprint = {2509.22923},
 primaryClass = {astro-ph.EP},
       adsurl = {https://ui.adsabs.harvard.edu/abs/2026ApJ...997..138H}
}

@ARTICLE{Hedges2021,
       author = {{Hedges}, Christina and {Hughes}, Alex and {Zhou}, George and {David}, Trevor J. and {Becker}, Juliette and {Giacalone}, Steven and {Vanderburg}, Andrew and {Rodriguez}, Joseph E. and {Bieryla}, Allyson and {Wirth}, Christopher and {Atherton}, Shaun and {Fetherolf}, Tara and {Collins}, Karen A. and {Price-Whelan}, Adrian M. and {Bedell}, Megan and {Quinn}, Samuel N. and {Gan}, Tianjun and {Ricker}, George R. and {Latham}, David W. and {Vanderspek}, Roland K. and {Seager}, Sara and {Winn}, Joshua N. and {Jenkins}, Jon M. and {Kielkopf}, John F. and {Schwarz}, Richard P. and {Dressing}, Courtney D. and {Gonzales}, Erica J. and {Crossfield}, Ian J.~M. and {Matthews}, Elisabeth C. and {Jensen}, Eric L.~N. and {Furlan}, Elise and {Gnilka}, Crystal L. and {Howell}, Steve B. and {Lester}, Kathryn V. and {Scott}, Nicholas J. and {Feliz}, Dax L. and {Lund}, Michael B. and {Siverd}, Robert J. and {Stevens}, Daniel J. and {Narita}, N. and {Fukui}, A. and {Murgas}, F. and {Palle}, Enric and {Sutton}, Phil J. and {Stassun}, Keivan G. and {Bouma}, Luke G. and {Vezie}, Michael and {Villase{\~n}or}, Jesus Noel and {Quintana}, Elisa V. and {Smith}, Jeffrey C.},
        title = "{TOI-2076 and TOI-1807: Two Young, Comoving Planetary Systems within 50 pc Identified by TESS that are Ideal Candidates for Further Follow Up}",
      journal = {\aj},
         year = 2021,
        month = aug,
       volume = {162},
       number = {2},
          eid = {54},
        pages = {54},
          doi = {10.3847/1538-3881/ac06cd},
archivePrefix = {arXiv},
       eprint = {2111.01311},
 primaryClass = {astro-ph.EP},
       adsurl = {https://ui.adsabs.harvard.edu/abs/2021AJ....162...54H}
}

@ARTICLE{Helled2011,
       author = {{Helled}, Ravit and {Anderson}, John D. and {Podolak}, Morris and {Schubert}, Gerald},
        title = "{Interior Models of Uranus and Neptune}",
      journal = {\apj},
         year = 2011,
        month = jan,
       volume = {726},
       number = {1},
          eid = {15},
        pages = {15},
          doi = {10.1088/0004-637X/726/1/15},
archivePrefix = {arXiv},
       eprint = {1010.5546},
 primaryClass = {astro-ph.EP},
       adsurl = {https://ui.adsabs.harvard.edu/abs/2011ApJ...726...15H}
}

@ARTICLE{Hellier2012,
       author = {{Hellier}, Coel and {Anderson}, D.~R. and {Collier Cameron}, A. and {Doyle}, A.~P. and {Fumel}, A. and {Gillon}, M. and {Jehin}, E. and {Lendl}, M. and {Maxted}, P.~F.~L. and {Pepe}, F. and {Pollacco}, D. and {Queloz}, D. and {S{\'e}gransan}, D. and {Smalley}, B. and {Smith}, A.~M.~S. and {Southworth}, J. and {Triaud}, A.~H.~M.~J. and {Udry}, S. and {West}, R.~G.},
        title = "{Seven transiting hot Jupiters from WASP-South, Euler and TRAPPIST: WASP-47b, WASP-55b, WASP-61b, WASP-62b, WASP-63b, WASP-66b and WASP-67b}",
      journal = {\mnras},
         year = 2012,
        month = oct,
       volume = {426},
       number = {1},
        pages = {739-750},
          doi = {10.1111/j.1365-2966.2012.21780.x},
archivePrefix = {arXiv},
       eprint = {1204.5095},
 primaryClass = {astro-ph.EP},
       adsurl = {https://ui.adsabs.harvard.edu/abs/2012MNRAS.426..739H}
}

@article{Hillert1978,
    title = {A model for alloying in ferromagnetic metals},
    volume = {2},
    issn = {0364-5916},
    url = {https://www.sciencedirect.com/science/article/pii/0364591678900111},
    doi = {10.1016/0364-5916(78)90011-1},
    number = {3},
    urldate = {2026-04-01},
    journal = {Calphad},
    author = {Hillert, Mats and Jarl, Magnus},
    month = jan,
    year = {1978},
    pages = {227--238},
}

@ARTICLE{Hirose2021,
       author = {{Hirose}, Kei and {Wood}, Bernard and {Vo{\v{c}}adlo}, Lidunka},
        title = "{Light elements in the Earth's core}",
      journal = {Nature Reviews Earth and Environment},
         year = 2021,
        month = aug,
       volume = {2},
       number = {9},
        pages = {645-658},
          doi = {10.1038/s43017-021-00203-6},
       adsurl = {https://ui.adsabs.harvard.edu/abs/2021NRvEE...2..645H}
}

@ARTICLE{Holzapfel1996,
       author = {{Holzapfel}, W.~B.},
        title = "{Physics of solids under strong compression}",
      journal = {Reports on Progress in Physics},
         year = 1996,
        month = jan,
       volume = {59},
       number = {1},
        pages = {29-90},
          doi = {10.1088/0034-4885/59/1/002},
       adsurl = {https://ui.adsabs.harvard.edu/abs/1996RPPh...59...29H}
}

@ARTICLE{Holzapfel1998,
       author = {{Holzapfel}, W.~B.},
        title = "{Equations of state for solids under strong compression}",
      journal = {High Pressure Research},
         year = 1998,
        month = oct,
       volume = {16},
       number = {2},
        pages = {81-126},
          doi = {10.1080/08957959808200283},
       adsurl = {https://ui.adsabs.harvard.edu/abs/1998HPR....16...81H}
}

@ARTICLE{Howard2025,
       author = {{Howard}, Andrew W. and {Sinukoff}, Evan and {Blunt}, Sarah and {Petigura}, Erik A. and {Crossfield}, Ian J.~M. and {Isaacson}, Howard and {Kosiarek}, Molly and {Rubenzahl}, Ryan A. and {Brewer}, John M. and {Fulton}, Benjamin J. and {Dressing}, Courtney D. and {Hirsch}, Lea A. and {Knutson}, Heather and {Livingston}, John H. and {Mills}, Sean M. and {Roy}, Arpita and {Weiss}, Lauren M. and {Benneke}, Bjorn and {Ciardi}, David R. and {Christiansen}, Jessie L. and {Cochran}, William D. and {Crepp}, Justin R. and {Gonzales}, Erica and {Hansen}, Brad M.~S. and {Hardegree-Ullman}, Kevin and {Howell}, Steve B. and {L{\'e}pine}, S{\'e}bastien and {Martinez}, Arturo O. and {Rogers}, Leslie A. and {Schlieder}, Joshua E. and {Werner}, Michael and {Polanski}, Alex S. and {Angelo}, Isabel and {Beard}, Corey and {Behmard}, Aida and {Bouma}, Luke G. and {Brinkman}, Casey L. and {Chontos}, Ashley and {Dai}, Fei and {Dalba}, Paul A. and {Giacalone}, Steven and {Grunblatt}, Samuel K. and {Hill}, Michelle L. and {Kane}, Stephen R. and {Lubin}, Jack and {Mayo}, Andrew W. and {Mocnik}, Teo and {Murphy}, Joseph M. Akana and {Rice}, Malena and {Rosenthal}, Lee J. and {Tyler}, Dakotah and {Van Zandt}, Judah and {Yee}, Samuel W.},
        title = "{Planet Masses, Radii, and Orbits from NASA's K2 Mission}",
      journal = {\apjs},
         year = 2025,
        month = jun,
       volume = {278},
       number = {2},
          eid = {52},
        pages = {52},
          doi = {10.3847/1538-4365/adc5e4},
archivePrefix = {arXiv},
       eprint = {2502.04436},
 primaryClass = {astro-ph.EP},
       adsurl = {https://ui.adsabs.harvard.edu/abs/2025ApJS..278...52H}
}

@ARTICLE{Hu2024,
       author = {{Hu}, Renyu and {Bello-Arufe}, Aaron and {Zhang}, Michael and {Paragas}, Kimberly and {Zilinskas}, Mantas and {van Buchem}, Christiaan and {Bess}, Michael and {Patel}, Jayshil and {Ito}, Yuichi and {Damiano}, Mario and {Scheucher}, Markus and {Oza}, Apurva V. and {Knutson}, Heather A. and {Miguel}, Yamila and {Dragomir}, Diana and {Brandeker}, Alexis and {Demory}, Brice-Olivier},
        title = "{A secondary atmosphere on the rocky exoplanet 55 Cancri e}",
      journal = {\nat},
         year = 2024,
        month = jun,
       volume = {630},
       number = {8017},
        pages = {609-612},
          doi = {10.1038/s41586-024-07432-x},
archivePrefix = {arXiv},
       eprint = {2405.04744},
 primaryClass = {astro-ph.EP},
       adsurl = {https://ui.adsabs.harvard.edu/abs/2024Natur.630..609H}
}

@ARTICLE{Huang2022,
       author = {{Huang}, Chenliang and {Rice}, David R. and {Steffen}, Jason H.},
        title = "{MAGRATHEA: an open-source spherical symmetric planet interior structure code}",
      journal = {\mnras},
         year = 2022,
        month = jul,
       volume = {513},
       number = {4},
        pages = {5256-5269},
          doi = {10.1093/mnras/stac1133},
archivePrefix = {arXiv},
       eprint = {2201.03094},
 primaryClass = {astro-ph.EP},
       adsurl = {https://ui.adsabs.harvard.edu/abs/2022MNRAS.513.5256H}
}

@ARTICLE{Ichikawa2020,
       author = {{Ichikawa}, Hiroki and {Tsuchiya}, Taku},
        title = "{Ab Initio Thermoelasticity of Liquid Iron-Nickel-Light Element Alloys}",
      journal = {Minerals},
         year = 2020,
        month = jan,
       volume = {10},
       number = {1},
          eid = {59},
        pages = {59},
          doi = {10.3390/min10010059},
       adsurl = {https://ui.adsabs.harvard.edu/abs/2020Mine...10...59I}
}

@ARTICLE{Iota2007,
       author = {{Iota}, Valentin and {Klepeis}, Jae-Hyun Park and {Yoo}, Choong-Shik and {Lang}, Jonathan and {Haskel}, Daniel and {Srajer}, George},
        title = "{Electronic structure and magnetism in compressed 3d transition metals}",
      journal = {Applied Physics Letters},
         year = 2007,
        month = jan,
       volume = {90},
       number = {4},
          eid = {042505},
        pages = {042505},
          doi = {10.1063/1.2434184},
       adsurl = {https://ui.adsabs.harvard.edu/abs/2007ApPhL..90d2505I}
}

@ARTICLE{Irifune1993,
       author = {{Irifune}, T. and {Ringwood}, A.~E.},
        title = "{Phase transformations in subducted oceanic crust and buoyancy relationships at depths of 600-800 km in the mantle}",
      journal = {Earth and Planetary Science Letters},
         year = 1993,
        month = may,
       volume = {117},
       number = {1-2},
        pages = {101-110},
          doi = {10.1016/0012-821X(93)90120-X},
       adsurl = {https://ui.adsabs.harvard.edu/abs/1993E&PSL.117..101I}
}

@ARTICLE{Jacobs2010,
       author = {{Jacobs}, Michel H.~G. and {Schmid-Fetzer}, Rainer},
        title = "{Thermodynamic properties and equation of state of fcc aluminum and bcc iron, derived from a lattice vibrational method}",
      journal = {Physics and Chemistry of Minerals},
         year = 2010,
        month = dec,
       volume = {37},
       number = {10},
        pages = {721-739},
          doi = {10.1007/s00269-010-0371-6},
       adsurl = {https://ui.adsabs.harvard.edu/abs/2010PCM....37..721J}
}

@ARTICLE{Jeanloz1981,
       author = {{Jeanloz}, Raymond},
        title = "{Finite-strain equation of state for high-pressure phases}",
      journal = {\grl},
         year = 1981,
        month = dec,
       volume = {8},
       number = {12},
        pages = {1219-1222},
          doi = {10.1029/GL008i012p01219},
       adsurl = {https://ui.adsabs.harvard.edu/abs/1981GeoRL...8.1219J}
}

@ARTICLE{Jeanloz1988,
       author = {{Jeanloz}, Raymond},
        title = "{Universal equation of state}",
      journal = {\prb},
         year = 1988,
        month = jul,
       volume = {38},
       number = {1},
        pages = {805-807},
          doi = {10.1103/PhysRevB.38.805},
       adsurl = {https://ui.adsabs.harvard.edu/abs/1988PhRvB..38..805J}
}

@ARTICLE{Journaux2020,
       author = {{Journaux}, B. and {Brown}, J.~M. and {Pakhomova}, A. and {Collings}, I.~E. and {Petitgirard}, S. and {Espinoza}, P. and {Boffa Ballaran}, T. and {Vance}, S.~D. and {Ott}, J. and {Cova}, F. and et al.},
        title = "{Holistic Approach for Studying Planetary Hydrospheres: Gibbs Representation of Ices Thermodynamics, Elasticity, and the Water Phase Diagram to 2,300 MPa}",
      journal = {Journal of Geophysical Research (Planets)},
         year = 2020,
        month = jan,
       volume = {125},
       number = {1},
          eid = {e06176},
        pages = {e06176},
          doi = {10.1029/2019JE006176},
       adsurl = {https://ui.adsabs.harvard.edu/abs/2020JGRE..12506176J}
}

@ARTICLE{Keane1954,
       author = {{Keane}, A.},
        title = "{An Investigation of Finite Strain in an Isotropic Material Subjected to Hydrostatic Pressure and its Seismological Applications}",
      journal = {Australian Journal of Physics},
         year = 1954,
        month = jun,
       volume = {7},
        pages = {322},
          doi = {10.1071/PH540322},
       adsurl = {https://ui.adsabs.harvard.edu/abs/1954AuJPh...7..322K}
}

@ARTICLE{Kite2020,
       author = {{Kite}, Edwin S. and {Barnett}, Megan N.},
        title = "{Exoplanet secondary atmosphere loss and revival}",
      journal = {Proceedings of the National Academy of Science},
         year = 2020,
        month = jul,
       volume = {117},
        pages = {18264-18271},
          doi = {10.1073/pnas.2006177117},
archivePrefix = {arXiv},
       eprint = {2006.02589},
 primaryClass = {astro-ph.EP},
       adsurl = {https://ui.adsabs.harvard.edu/abs/2020PNAS..11718264K}
}

@ARTICLE{Kite2021,
       author = {{Kite}, Edwin S. and {Schaefer}, Laura},
        title = "{Water on Hot Rocky Exoplanets}",
      journal = {\apjl},
         year = 2021,
        month = mar,
       volume = {909},
       number = {2},
          eid = {L22},
        pages = {L22},
          doi = {10.3847/2041-8213/abe7dc},
archivePrefix = {arXiv},
       eprint = {2103.07753},
 primaryClass = {astro-ph.EP},
       adsurl = {https://ui.adsabs.harvard.edu/abs/2021ApJ...909L..22K}
}

@ARTICLE{Knopoff1969,
       author = {{Knopoff}, L. and {Shapiro}, J.~N.},
        title = "{Comments on the interrelationships between Gr{\"u}neisen's parameter and shock and isothermal equations of state}",
      journal = {\jgr},
         year = 1969,
        month = mar,
       volume = {74},
       number = {6},
        pages = {1439-1450},
          doi = {10.1029/JB074i006p01439},
       adsurl = {https://ui.adsabs.harvard.edu/abs/1969JGR....74.1439K}
}

@ARTICLE{Knopoff1970,
       author = {{Knopoff}, L. and {Shapiro}, J.~N.},
        title = "{Pseudo-Gr{\"u}neisen Parameter for Liquids}",
      journal = {\prb},
         year = 1970,
        month = may,
       volume = {1},
       number = {10},
        pages = {3893-3895},
          doi = {10.1103/PhysRevB.1.3893},
       adsurl = {https://ui.adsabs.harvard.edu/abs/1970PhRvB...1.3893K}
}

@ARTICLE{Komabayashi2009,
       author = {{Komabayashi}, Tetsuya and {Fei}, Yingwei and {Meng}, Yue and {Prakapenka}, Vitali},
        title = "{In-situ X-ray diffraction measurements of the {\ensuremath{\gamma}}-{\ensuremath{\in}} transition boundary of iron in an internally-heated diamond anvil cell}",
      journal = {Earth and Planetary Science Letters},
         year = 2009,
        month = may,
       volume = {282},
       number = {1-4},
        pages = {252-257},
          doi = {10.1016/j.epsl.2009.03.025},
       adsurl = {https://ui.adsabs.harvard.edu/abs/2009E&PSL.282..252K}
}

@ARTICLE{KrissansenTotton2024,
       author = {{Krissansen-Totton}, Joshua and {Wogan}, Nicholas and {Thompson}, Maggie and {Fortney}, Jonathan J.},
        title = "{The erosion of large primary atmospheres typically leaves behind substantial secondary atmospheres on temperate rocky planets}",
      journal = {Nature Communications},
         year = 2024,
        month = dec,
       volume = {15},
       number = {1},
          eid = {8374},
        pages = {8374},
          doi = {10.1038/s41467-024-52642-6},
archivePrefix = {arXiv},
       eprint = {2409.18940},
 primaryClass = {astro-ph.EP},
       adsurl = {https://ui.adsabs.harvard.edu/abs/2024NatCo..15.8374K}
}

@ARTICLE{Kunc2003,
       author = {{Kunc}, K. and {Loa}, I. and {Syassen}, K.},
        title = "{Equation of state and phonon frequency calculations of diamond at high pressures}",
      journal = {\prb},
         year = 2003,
        month = sep,
       volume = {68},
       number = {9},
          eid = {094107},
        pages = {094107},
          doi = {10.1103/PhysRevB.68.094107},
archivePrefix = {arXiv},
       eprint = {cond-mat/0304692},
 primaryClass = {cond-mat},
       adsurl = {https://ui.adsabs.harvard.edu/abs/2003PhRvB..68i4107K}
}

@ARTICLE{Li2020,
       author = {{Li}, Zhi and {Caracas}, Razvan and {Soubiran}, Fran{\c{c}}ois},
        title = "{Partial core vaporization during Giant Impacts inferred from the entropy and the critical point of iron}",
      journal = {Earth and Planetary Science Letters},
         year = 2020,
        month = oct,
       volume = {547},
          eid = {116463},
        pages = {116463},
          doi = {10.1016/j.epsl.2020.116463},
       adsurl = {https://ui.adsabs.harvard.edu/abs/2020E&PSL.54716463L}
}

@ARTICLE{Lichtenberg2021,
       author = {{Lichtenberg}, Tim and {Bower}, Dan J. and {Hammond}, Mark and {Boukrouche}, Ryan and {Sanan}, Patrick and {Tsai}, Shang-Min and {Pierrehumbert}, Raymond T.},
        title = "{Vertically Resolved Magma Ocean-Protoatmosphere Evolution: H$_{2}$, H$_{2}$O, CO$_{2}$, CH$_{4}$, CO, O$_{2}$, and N$_{2}$ as Primary Absorbers}",
      journal = {Journal of Geophysical Research (Planets)},
         year = 2021,
        month = feb,
       volume = {126},
       number = {2},
          eid = {e06711},
        pages = {e06711},
          doi = {10.1029/2020JE006711},
archivePrefix = {arXiv},
       eprint = {2101.10991},
 primaryClass = {astro-ph.EP},
       adsurl = {https://ui.adsabs.harvard.edu/abs/2021JGRE..12606711L}
}

@ARTICLE{Lichtenberg2025,
       author = {{Lichtenberg}, Tim and {Miguel}, Yamila},
        title = "{Super-Earths and Earth-like Exoplanets}",
      journal = {Treatise on Geochemistry},
         year = 2025,
        month = jan,
       volume = {7},
        pages = {51-112},
          doi = {10.1016/B978-0-323-99762-1.00122-4},
archivePrefix = {arXiv},
       eprint = {2405.04057},
 primaryClass = {astro-ph.EP},
       adsurl = {https://ui.adsabs.harvard.edu/abs/2025TrGeo...7...51L}
}

@ARTICLE{Lichtenberg2025b,
       author = {{Lichtenberg}, Tim and {Shorttle}, Oliver and {Teske}, Johanna and {Kempton}, Eliza M.-R.},
        title = "{Constraining exoplanet interiors using observations of their atmospheres}",
      journal = {Science},
         year = 2025,
        month = oct,
       volume = {390},
       number = {6769},
          eid = {eads3660},
        pages = {eads3660},
          doi = {10.1126/science.ads3360},
archivePrefix = {arXiv},
       eprint = {2510.08844},
 primaryClass = {astro-ph.EP},
       adsurl = {https://ui.adsabs.harvard.edu/abs/2025Sci...390S3660L}
}

@ARTICLE{Lichtenberg2025c,
       author = {{Lichtenberg}, Tim and {Schaefer}, Laura and {Krissansen-Totton}, Joshua and {Miguel}, Yamila and {Sergeev}, Denis E. and {Baumeister}, Philipp and {Cmiel}, Jessica and {Janssen}, Leoni J. and {Nguyen}, T. Giang and {Miyazaki}, Yoshinori and {Nicholls}, Harrison and {Papesh}, Alexandra and {Pelissard}, Hugo and {Peng}, Bo and {Perez}, Junellie and {Postolec}, Emma and {Sastre}, Mariana and {Salvador}, Arnaud and {Spreeuw}, Hanno and {Zorzi}, Andrea and {Fauchez}, Thomas J. and {Hamano}, Keiko and {Leconte}, J{\'e}r{\'e}my and {Maurice}, Maxime and {Noack}, Lena and {Soucasse}, Laurent},
        title = "{Coupled atmospHere Interior modeL Intercomparison (CHILI)---Protocol Version 1.0: A CUISINES Intercomparison Project of Magma Ocean Models}",
      journal = {\psj},
         year = 2026,
        month = may,
       volume = {7},
       number = {5},
          eid = {108},
        pages = {108},
          doi = {10.3847/PSJ/ae593b},
}

@ARTICLE{Lodders2003,
       author = {{Lodders}, Katharina},
        title = "{Solar System Abundances and Condensation Temperatures of the Elements}",
      journal = {\apj},
         year = 2003,
        month = jul,
       volume = {591},
       number = {2},
        pages = {1220-1247},
          doi = {10.1086/375492},
       adsurl = {https://ui.adsabs.harvard.edu/abs/2003ApJ...591.1220L}
}

@ARTICLE{Loftus2025,
       author = {{Loftus}, Kaitlyn and {Luo}, Yangcheng and {Fan}, Bowen and {Kite}, Edwin S.},
        title = "{Extreme weather variability on hot rocky exoplanet 55 Cancri e explained by magma temperature-cloud feedback}",
      journal = {Proceedings of the National Academy of Science},
         year = 2025,
        month = apr,
       volume = {122},
       number = {17},
          eid = {e2423473122},
        pages = {e2423473122},
          doi = {10.1073/pnas.2423473122},
archivePrefix = {arXiv},
       eprint = {2409.16270},
 primaryClass = {astro-ph.EP},
       adsurl = {https://ui.adsabs.harvard.edu/abs/2025PNAS..12223473L}
}

@ARTICLE{Luo2024,
       author = {{Luo}, Haiyang and {Dorn}, Caroline and {Deng}, Jie},
        title = "{The interior as the dominant water reservoir in super-Earths and sub-Neptunes}",
      journal = {Nature Astronomy},
         year = 2024,
        month = nov,
       volume = {8},
        pages = {1399-1407},
          doi = {10.1038/s41550-024-02347-z},
archivePrefix = {arXiv},
       eprint = {2401.16394},
 primaryClass = {astro-ph.EP},
       adsurl = {https://ui.adsabs.harvard.edu/abs/2024NatAs...8.1399L}
}

@ARTICLE{Luo2025,
       author = {{Luo}, Haiyang and {Deng}, Jie},
        title = "{Thermophysical States of MgSiO$_{3}$ Liquid up to Terapascal Pressures: Implications for Magma Oceans in Super-Earths and Sub-Neptunes}",
      journal = {Journal of Geophysical Research (Planets)},
         year = 2025,
        month = apr,
       volume = {130},
       number = {4},
          eid = {e2024JE008678},
        pages = {e2024JE008678},
          doi = {10.1029/2024JE008678},
       adsurl = {https://ui.adsabs.harvard.edu/abs/2025JGRE..13008678L}
}

@ARTICLE{Man2025,
       author = {{Man}, Lianjie and {Li}, Xiang and {Boffa Ballaran}, Tiziana and {Zhou}, Wenju and {Chantel}, Julien and {N{\'e}ri}, Adrien and {Kupenko}, Ilya and {Aprilis}, Georgios and {Kurnosov}, Alexander and {Namur}, Olivier and et al.},
        title = "{The structure and stability of Fe$_{4+x}$S$_{3}$ and its potential to form a Martian inner core}",
      journal = {Nature Communications},
         year = 2025,
        month = feb,
       volume = {16},
       number = {1},
          eid = {1710},
        pages = {1710},
          doi = {10.1038/s41467-025-56220-2},
       adsurl = {https://ui.adsabs.harvard.edu/abs/2025NatCo..16.1710M}
}

@ARTICLE{Mazevet2019,
       author = {{Mazevet}, S. and {Licari}, A. and {Chabrier}, G. and {Potekhin}, A.~Y.},
        title = "{Ab initio based equation of state of dense water for planetary and exoplanetary modeling}",
      journal = {\aap},
         year = 2019,
        month = jan,
       volume = {621},
          eid = {A128},
        pages = {A128},
          doi = {10.1051/0004-6361/201833963},
archivePrefix = {arXiv},
       eprint = {1810.05658},
 primaryClass = {astro-ph.EP},
       adsurl = {https://ui.adsabs.harvard.edu/abs/2019A&A...621A.128M}
}

@ARTICLE{McDonough1995,
       author = {{McDonough}, W.~F. and {Sun}, S.-s.},
        title = "{The composition of the Earth}",
      journal = {Chemical Geology},
         year = 1995,
        month = jan,
       volume = {120},
       number = {3},
        pages = {223-253},
          doi = {10.1016/0009-2541(94)00140-4},
       adsurl = {https://ui.adsabs.harvard.edu/abs/1995ChGeo.120..223M}
}

@ARTICLE{McKenzie1988,
       author = {{McKenzie}, D. and {Bickle}, M.~J.},
        title = "{The volume and composition of melt generated by extension of the lithosphere}",
      journal = {Journal of Petrology},
         year = 1988,
        month = jan,
       volume = {29},
        pages = {625-679},
          doi = {10.1093/petrology/29.3.625},
       adsurl = {https://ui.adsabs.harvard.edu/abs/1988JPet...29..625M}
}

@ARTICLE{Meier2023,
       author = {{Meier}, Tobias G. and {Bower}, Dan J. and {Lichtenberg}, Tim and {Hammond}, Mark and {Tackley}, Paul J.},
        title = "{Interior dynamics of super-Earth 55 Cancri e}",
      journal = {\aap},
         year = 2023,
        month = oct,
       volume = {678},
          eid = {A29},
        pages = {A29},
          doi = {10.1051/0004-6361/202346950},
archivePrefix = {arXiv},
       eprint = {2308.00592},
 primaryClass = {astro-ph.EP},
       adsurl = {https://ui.adsabs.harvard.edu/abs/2023A&A...678A..29M}
}

@ARTICLE{Meier2026,
       author = {{Meier}, Tobias G. and {Guimond}, Claire Marie and {Pierrehumbert}, Raymond T. and {Birkby}, Jayne and {Chatterjee}, Richard D. and {Fisher}, Chloe E. and {Golabek}, Gregor J. and {Hammond}, Mark and {Komacek}, Thaddeus D. and {Lichtenberg}, Tim and {McGinty}, Alex and {Vald{\'e}s}, Erik Meier and {Nicholls}, Harrison and {Parker}, Luke T. and {Spaargaren}, Rob J. and {Tackley}, Paul J.},
        title = "{Mantle Convection and Nightside Volcanism on Lava World K2-141 b}",
      journal = {\mnras},
         year = 2026,
        month = feb,
       volume = {547},
       number = {3},
          eid = {stag390},
        pages = {stag390},
          doi = {10.1093/mnras/stag390},
}

@ARTICLE{Mie1903,
       author = {{Mie}, Gustav},
        title = "{Zur kinetischen Theorie der einatomigen K{\"o}rper}",
      journal = {Annalen der Physik},
         year = 1903,
        month = jan,
       volume = {316},
       number = {8},
        pages = {657-697},
          doi = {10.1002/andp.19033160802},
       adsurl = {https://ui.adsabs.harvard.edu/abs/1903AnP...316..657M}
}

@ARTICLE{Militzer2013,
       author = {{Militzer}, B. and {Hubbard}, W.~B.},
        title = "{Ab Initio Equation of State for Hydrogen-Helium Mixtures with Recalibration of the Giant-planet Mass-Radius Relation}",
      journal = {\apj},
         year = 2013,
        month = sep,
       volume = {774},
       number = {2},
          eid = {148},
        pages = {148},
          doi = {10.1088/0004-637X/774/2/148},
archivePrefix = {arXiv},
       eprint = {1302.4691},
 primaryClass = {astro-ph.EP},
       adsurl = {https://ui.adsabs.harvard.edu/abs/2013ApJ...774..148M}
}

@ARTICLE{Militzer2023,
       author = {{Militzer}, Burkhard and {Hubbard}, William B.},
        title = "{Relation of Gravity, Winds, and the Moment of Inertia of Jupiter and Saturn}",
      journal = {\psj},
         year = 2023,
        month = may,
       volume = {4},
       number = {5},
          eid = {95},
        pages = {95},
          doi = {10.3847/PSJ/acd2cd},
archivePrefix = {arXiv},
       eprint = {2308.04986},
 primaryClass = {astro-ph.EP},
       adsurl = {https://ui.adsabs.harvard.edu/abs/2023PSJ.....4...95M}
}

@ARTICLE{Miozzi2020,
       author = {{Miozzi}, Francesca and {Matas}, Jan and {Guignot}, Nicolas and {Badro}, James and {Siebert}, Julien and {Fiquet}, Guillaume},
        title = "{A New Reference for the Thermal Equation of State of Iron}",
      journal = {Minerals},
         year = 2020,
        month = jan,
       volume = {10},
       number = {2},
          eid = {100},
        pages = {100},
          doi = {10.3390/min10020100},
       adsurl = {https://ui.adsabs.harvard.edu/abs/2020Mine...10..100M}
}

@ARTICLE{Monaghan2025,
       author = {{Monaghan}, Christopher and {Roy}, Pierre-Alexis and {Benneke}, Bj{\"o}rn and {Crossfield}, Ian J.~M. and {Coulombe}, Louis-Philippe and {Piaulet-Ghorayeb}, Caroline and {Kreidberg}, Laura and {Dressing}, Courtney D. and {Kane}, Stephen R. and {Dragomir}, Diana and et al.},
        title = "{Low 4.5 {\ensuremath{\mu}}m Dayside Emission Disfavors a Dark Bare-rock Scenario for the Hot Super-Earth TOI-431 b}",
      journal = {\aj},
         year = 2025,
        month = may,
       volume = {169},
       number = {5},
          eid = {239},
        pages = {239},
          doi = {10.3847/1538-3881/adbe75},
archivePrefix = {arXiv},
       eprint = {2503.09698},
 primaryClass = {astro-ph.EP},
       adsurl = {https://ui.adsabs.harvard.edu/abs/2025AJ....169..239M}
}

@ARTICLE{Murakami2004,
       author = {{Murakami}, Motohiko and {Hirose}, Kei and {Kawamura}, Katsuyuki and {Sata}, Nagayoshi and {Ohishi}, Yasuo},
        title = "{Post-Perovskite Phase Transition in MgSiO$_{3}$}",
      journal = {Science},
         year = 2004,
        month = may,
       volume = {304},
       number = {5672},
        pages = {855-858},
          doi = {10.1126/science.1095932},
       adsurl = {https://ui.adsabs.harvard.edu/abs/2004Sci...304..855M}
}

@ARTICLE{Murakami2024,
       author = {{Murakami}, Motohiko and {Khan}, Amir and {Sossi}, Paolo A. and {Ballmer}, Maxim D. and {Saha}, Pinku},
        title = "{The Composition of Earth's Lower Mantle}",
      journal = {Annual Review of Earth and Planetary Sciences},
         year = 2024,
        month = aug,
       volume = {52},
       number = {1},
        pages = {605-638},
          doi = {10.1146/annurev-earth-031621-075657},
       adsurl = {https://ui.adsabs.harvard.edu/abs/2024AREPS..52..605M}
}

@ARTICLE{Murnaghan1944,
       author = {{Murnaghan}, F.~D.},
        title = "{The Compressibility of Media under Extreme Pressures}",
      journal = {Proceedings of the National Academy of Science},
         year = 1944,
        month = sep,
       volume = {30},
       number = {9},
        pages = {244-247},
          doi = {10.1073/pnas.30.9.244},
       adsurl = {https://ui.adsabs.harvard.edu/abs/1944PNAS...30..244M}
}

@book{Murnaghan1951,
  title={Finite Deformation of an Elastic Solid},
  author={Murnaghan, F.D.},
  lccn={51014896},
  series={Applied mathematics series},
  url={https://books.google.nl/books?id=4rq7AAAAIAAJ},
  year={1951},
  publisher={Wiley}
}

@ARTICLE{Nakajima2015,
       author = {{Nakajima}, Miki and {Stevenson}, David J.},
        title = "{Melting and mixing states of the Earth's mantle after the Moon-forming impact}",
      journal = {Earth and Planetary Science Letters},
         year = 2015,
        month = oct,
       volume = {427},
        pages = {286-295},
          doi = {10.1016/j.epsl.2015.06.023},
archivePrefix = {arXiv},
       eprint = {1506.04853},
 primaryClass = {astro-ph.EP},
       adsurl = {https://ui.adsabs.harvard.edu/abs/2015E&PSL.427..286N}
}

@ARTICLE{Nicholls2024,
       author = {{Nicholls}, Harrison and {Lichtenberg}, Tim and {Bower}, Dan J. and {Pierrehumbert}, Raymond},
        title = "{Magma Ocean Evolution at Arbitrary Redox State}",
      journal = {Journal of Geophysical Research (Planets)},
         year = 2024,
        month = dec,
       volume = {129},
       number = {12},
        pages = {2024JE008576},
          doi = {10.1029/2024JE008576},
archivePrefix = {arXiv},
       eprint = {2411.19137},
 primaryClass = {astro-ph.EP},
       adsurl = {https://ui.adsabs.harvard.edu/abs/2024JGRE..12908576N}
}

@ARTICLE{Nicholls2026,
       author = {{Nicholls}, Harrison and {Lichtenberg}, Tim and {Chatterjee}, Richard D. and {Guimond}, Claire Marie and {Postolec}, Emma and {Pierrehumbert}, Raymond T.},
        title = "{Volatile-rich evolution of molten super-Earth L 98-59 d}",
      journal = {Nature Astronomy},
         year = 2026,
        month = mar,
       volume = {10},
       number = {6},
        pages = {809},
          doi = {10.1038/s41550-026-02815-8},
}

@ARTICLE{Nimmo2015,
       author = {{Nimmo}, F. and {Kleine}, T.},
        title = "{Early Differentiation and Core Formation: Processes and Timescales}",
      journal = {Geophysical Monograph Series},
       editor = {{Badro}, J. and {Walter}, M.},
    booktitle = {The Early Earth: Accretion and Differentiation},
         year = 2015,
        month = sep,
       volume = {212},
        pages = {83-102},
          doi = {10.1002/9781118860359.ch5},
       adsurl = {https://ui.adsabs.harvard.edu/abs/2015GMS...212...83N}
}

@ARTICLE{Niu2015,
       author = {{Niu}, Haiyang and {Oganov}, Artem R. and {Chen}, Xing-Qiu and {Li}, Dianzhong},
        title = "{Prediction of novel stable compounds in the Mg-Si-O system under exoplanet pressures}",
      journal = {Scientific Reports},
         year = 2015,
        month = dec,
       volume = {5},
          eid = {18347},
        pages = {18347},
          doi = {10.1038/srep18347},
archivePrefix = {arXiv},
       eprint = {1510.03061},
 primaryClass = {physics.geo-ph},
       adsurl = {https://ui.adsabs.harvard.edu/abs/2015NatSR...518347N}
}

@ARTICLE{Ono2005,
       author = {{Ono}, Shigeaki and {Oganov}, Artem R.},
        title = "{In situ observations of phase transition between perovskite and CaIrO $_{3}$-type phase in MgSiO $_{3}$ and pyrolitic mantle composition}",
      journal = {Earth and Planetary Science Letters},
         year = 2005,
        month = aug,
       volume = {236},
       number = {3-4},
        pages = {914-932},
          doi = {10.1016/j.epsl.2005.06.001},
       adsurl = {https://ui.adsabs.harvard.edu/abs/2005E&PSL.236..914O}
}

@ARTICLE{Padovan2018,
       author = {{Padovan}, S. and {Spohn}, T. and {Baumeister}, P. and {Tosi}, N. and {Breuer}, D. and {Csizmadia}, Sz. and {Hellard}, H. and {Sohl}, F.},
        title = "{Matrix-propagator approach to compute fluid Love numbers and applicability to extrasolar planets}",
      journal = {\aap},
         year = 2018,
        month = dec,
       volume = {620},
          eid = {A178},
        pages = {A178},
          doi = {10.1051/0004-6361/201834181},
archivePrefix = {arXiv},
       eprint = {1810.10064},
 primaryClass = {astro-ph.EP},
       adsurl = {https://ui.adsabs.harvard.edu/abs/2018A&A...620A.178P}
}

@ARTICLE{Parc2024,
       author = {{Parc}, L{\'e}na and {Bouchy}, Fran{\c{c}}ois and {Venturini}, Julia and {Dorn}, Caroline and {Helled}, Ravit},
        title = "{From super-Earths to sub-Neptunes: Observational constraints and connections to theoretical models}",
      journal = {\aap},
         year = 2024,
        month = aug,
       volume = {688},
          eid = {A59},
        pages = {A59},
          doi = {10.1051/0004-6361/202449911},
archivePrefix = {arXiv},
       eprint = {2406.04311},
 primaryClass = {astro-ph.EP},
       adsurl = {https://ui.adsabs.harvard.edu/abs/2024A&A...688A..59P}
}

@ARTICLE{ParkCoy2026,
       author = {{Park Coy}, Brandon and {Xue}, Qiao and {Weiner Mansfield}, Megan and {Eastman}, Jason D. and {Piette}, Anjali A.~A. and {Fairnington}, Tyler and {Smith}, Cole and {Zhang}, Michael and {Kempton}, Eliza M.~R. and {Bean}, Jacob L. and {Ji}, Xuan and {Gao}, Peter and {Ih}, Jegug and {Koll}, Daniel D.~B. and {Luque}, Rafael and {Orell-Miquel}, Jaume and {Kite}, Edwin S.},
        title = "{Evidence for an Atmosphere on the Ultra-short-period Super-Earth HD 3167 b}",
      journal = {\apjl},
         year = 2026,
        month = jul,
       volume = {1005},
       number = {2},
          eid = {L77},
        pages = {L77},
          doi = {10.3847/2041-8213/ae7f23},
}

@ARTICLE{Piette2023,
       author = {{Piette}, Anjali A.~A. and {Gao}, Peter and {Brugman}, Kara and {Shahar}, Anat and {Lichtenberg}, Tim and {Miozzi}, Francesca and {Driscoll}, Peter},
        title = "{Rocky Planet or Water World? Observability of Low-density Lava World Atmospheres}",
      journal = {\apj},
         year = 2023,
        month = sep,
       volume = {954},
       number = {1},
          eid = {29},
        pages = {29},
          doi = {10.3847/1538-4357/acdef2},
archivePrefix = {arXiv},
       eprint = {2306.10100},
 primaryClass = {astro-ph.EP},
       adsurl = {https://ui.adsabs.harvard.edu/abs/2023ApJ...954...29P}
}

@ARTICLE{Poirier1991,
       author = {{Poirier}, J.-P.},
        title = "{Introduction to the physics of the Earth's interior.}",
      journal = {Cambridge Topics in Mineral Physics and Chemistry},
         year = 1991,
        month = jan,
       volume = {3},
       adsurl = {https://ui.adsabs.harvard.edu/abs/1991CTMPC...3.....P}
}

@ARTICLE{Polanski2024,
       author = {{Polanski}, Alex S. and {Lubin}, Jack and {Beard}, Corey and {Akana Murphy}, Joseph M. and {Rubenzahl}, Ryan and {Hill}, Michelle L. and {Crossfield}, Ian J.~M. and {Chontos}, Ashley and {Robertson}, Paul and {Isaacson}, Howard and {Kane}, Stephen R. and {Ciardi}, David R. and {Batalha}, Natalie M. and {Dressing}, Courtney and {Fulton}, Benjamin and {Howard}, Andrew W. and {Huber}, Daniel and {Petigura}, Erik A. and {Weiss}, Lauren M. and {Angelo}, Isabel and {Behmard}, Aida and {Blunt}, Sarah and {Brinkman}, Casey L. and {Dai}, Fei and {Dalba}, Paul A. and {Fetherolf}, Tara and {Giacalone}, Steven and {Hirsch}, Lea A. and {Holcomb}, Rae and {Kosiarek}, Molly R. and {Mayo}, Andrew W. and {MacDougall}, Mason G. and {Mo{\v{c}}nik}, Teo and {Pidhorodetska}, Daria and {Rice}, Malena and {Rosenthal}, Lee J. and {Scarsdale}, Nicholas and {Turtelboom}, Emma V. and {Tyler}, Dakotah and {Van Zandt}, Judah and {Yee}, Samuel W. and {Coria}, David R. and {Dulz}, Shannon D. and {Hartman}, Joel D. and {Householder}, Aaron and {Lange}, Sarah and {Langford}, Andrew and {Louden}, Emma M. and {Siegel}, Jared C. and {Gilbert}, Emily A. and {Gonzales}, Erica J. and {Schlieder}, Joshua E. and {Boyle}, Andrew W. and {Christiansen}, Jessie L. and {Clark}, Catherine A. and {Fernandes}, Rachel B. and {Lund}, Michael B. and {Savel}, Arjun B. and {Gill}, Holden and {Beichman}, Charles and {Matson}, Rachel and {Matthews}, Elisabeth C. and {Furlan}, E. and {Howell}, Steve B. and {Scott}, Nicholas J. and {Everett}, Mark E. and {Livingston}, John H. and {Ershova}, Irina O. and {Cheryasov}, Dmitry V. and {Safonov}, Boris and {Lillo-Box}, Jorge and {Barrado}, David and {Morales-Calder{\'o}n}, Mar{\'\i}a},
        title = "{The TESS-Keck Survey. XX. 15 New TESS Planets and a Uniform RV Analysis of All Survey Targets}",
      journal = {\apjs},
         year = 2024,
        month = jun,
       volume = {272},
       number = {2},
          eid = {32},
        pages = {32},
          doi = {10.3847/1538-4365/ad4484},
archivePrefix = {arXiv},
       eprint = {2405.14786},
 primaryClass = {astro-ph.EP},
       adsurl = {https://ui.adsabs.harvard.edu/abs/2024ApJS..272...32P}
}

@INPROCEEDINGS{Raymond2014,
       author = {{Raymond}, S.~N. and {Kokubo}, E. and {Morbidelli}, A. and {Morishima}, R. and {Walsh}, K.~J.},
        title = "{Terrestrial Planet Formation at Home and Abroad}",
    booktitle = {Protostars and Planets VI},
         year = 2014,
       editor = {{Beuther}, Henrik and {Klessen}, Ralf S. and {Dullemond}, Cornelis P. and {Henning}, Thomas},
        month = jan,
        pages = {595-618},
          doi = {10.2458/azu_uapress_9780816531240-ch026},
archivePrefix = {arXiv},
       eprint = {1312.1689},
 primaryClass = {astro-ph.EP},
       adsurl = {https://ui.adsabs.harvard.edu/abs/2014prpl.conf..595R}
}

@ARTICLE{Rauer2025,
       author = {{Rauer}, Heike and {Aerts}, Conny and {Cabrera}, Juan and {Deleuil}, Magali and {Erikson}, Anders and {Gizon}, Laurent and {Goupil}, Mariejo and {Heras}, Ana and {Walloschek}, Thomas and {Lorenzo-Alvarez}, Jose and et al.},
        title = "{The PLATO mission}",
      journal = {Experimental Astronomy},
         year = 2025,
        month = jun,
       volume = {59},
       number = {3},
          eid = {26},
        pages = {26},
          doi = {10.1007/s10686-025-09985-9},
archivePrefix = {arXiv},
       eprint = {2406.05447},
 primaryClass = {astro-ph.IM},
       adsurl = {https://ui.adsabs.harvard.edu/abs/2025ExA....59...26R}
}

@ARTICLE{Ringwood1991,
       author = {{Ringwood}, A.~E.},
        title = "{Phase transformations and their bearing on the constitution and dynamics of the mantle}",
      journal = {\gca},
         year = 1991,
        month = aug,
       volume = {55},
       number = {8},
        pages = {2083-2110},
          doi = {10.1016/0016-7037(91)90090-R},
       adsurl = {https://ui.adsabs.harvard.edu/abs/1991GeCoA..55.2083R}
}

@ARTICLE{Rogers2010,
       author = {{Rogers}, L.~A. and {Seager}, S.},
        title = "{A Framework for Quantifying the Degeneracies of Exoplanet Interior Compositions}",
      journal = {\apj},
         year = 2010,
        month = apr,
       volume = {712},
       number = {2},
        pages = {974-991},
          doi = {10.1088/0004-637X/712/2/974},
archivePrefix = {arXiv},
       eprint = {0912.3288},
 primaryClass = {astro-ph.EP},
       adsurl = {https://ui.adsabs.harvard.edu/abs/2010ApJ...712..974R}
}

@ARTICLE{Rose1984,
       author = {{Rose}, James H. and {Smith}, John R. and {Guinea}, Francisco and {Ferrante}, John},
        title = "{Universal features of the equation of state of metals}",
      journal = {\prb},
         year = 1984,
        month = mar,
       volume = {29},
       number = {6},
        pages = {2963-2969},
          doi = {10.1103/PhysRevB.29.2963},
       adsurl = {https://ui.adsabs.harvard.edu/abs/1984PhRvB..29.2963R}
}

@ARTICLE{Rosenfeld1998,
       author = {{Rosenfeld}, Yaakov and {Tarazona}, Pedro},
        title = "{Density functional theory and the asymptotic high density expansion of the free energy of classical solids and fluids}",
      journal = {Molecular Physics},
         year = 1998,
        month = oct,
       volume = {95},
       number = {2},
        pages = {141-150},
          doi = {10.1080/00268979809483145},
       adsurl = {https://ui.adsabs.harvard.edu/abs/1998MolPh..95..141R}
}

@ARTICLE{Rubie2016,
       author = {{Rubie}, D.~C. and {Jacobson}, S.~A.},
        title = "{Mechanisms and Geochemical Models of Core Formation}",
      journal = {Geophysical Monograph Series},
       editor = {{Terasaki}, H. and {Fischer}, R.~A.},
    booktitle = {Deep Earth: Physics and Chemistry of the Lower Mantle and Core},
         year = 2016,
        month = mar,
       volume = {217},
        pages = {181-190},
          doi = {10.1002/9781118992487.ch14},
archivePrefix = {arXiv},
       eprint = {1504.05417},
 primaryClass = {astro-ph.EP},
       adsurl = {https://ui.adsabs.harvard.edu/abs/2016GMS...217..191R}
}

@ARTICLE{Sakai2016,
       author = {{Sakai}, Takeshi and {Dekura}, Haruhiko and {Hirao}, Naohisa},
        title = "{Experimental and theoretical thermal equations of state of MgSiO$_{3}$ post-perovskite at multi-megabar pressures}",
      journal = {Scientific Reports},
         year = 2016,
        month = mar,
       volume = {6},
          eid = {22652},
        pages = {22652},
          doi = {10.1038/srep22652},
       adsurl = {https://ui.adsabs.harvard.edu/abs/2016NatSR...622652S}
}

@ARTICLE{Saurety2025,
       author = {{Saurety}, Adrien and {Caracas}, Razvan and {Raymond}, Sean N.},
        title = "{Impact-induced Vaporization during Accretion of Planetary Bodies}",
      journal = {\apjl},
         year = 2025,
        month = mar,
       volume = {981},
       number = {1},
          eid = {L13},
        pages = {L13},
          doi = {10.3847/2041-8213/adb30e},
archivePrefix = {arXiv},
       eprint = {2502.04787},
 primaryClass = {astro-ph.EP},
       adsurl = {https://ui.adsabs.harvard.edu/abs/2025ApJ...981L..13S}
}

@ARTICLE{Schaefer2012,
       author = {{Schaefer}, Laura and {Lodders}, Katharina and {Fegley}, Bruce},
        title = "{Vaporization of the Earth: Application to Exoplanet Atmospheres}",
      journal = {\apj},
         year = 2012,
        month = aug,
       volume = {755},
       number = {1},
          eid = {41},
        pages = {41},
          doi = {10.1088/0004-637X/755/1/41},
archivePrefix = {arXiv},
       eprint = {1108.4660},
 primaryClass = {astro-ph.EP},
       adsurl = {https://ui.adsabs.harvard.edu/abs/2012ApJ...755...41S}
}

@ARTICLE{Schaefer2018,
       author = {{Schaefer}, Laura and {Elkins-Tanton}, Linda T.},
        title = "{Magma oceans as a critical stage in the tectonic development of rocky planets}",
      journal = {Philosophical Transactions of the Royal Society of London Series A},
         year = 2018,
        month = nov,
       volume = {376},
       number = {2132},
          eid = {20180109},
        pages = {20180109},
          doi = {10.1098/rsta.2018.0109},
archivePrefix = {arXiv},
       eprint = {1809.01629},
 primaryClass = {astro-ph.EP},
       adsurl = {https://ui.adsabs.harvard.edu/abs/2018RSPTA.37680109S}
}

@ARTICLE{Seager2007,
       author = {{Seager}, S. and {Kuchner}, M. and {Hier-Majumder}, C.~A. and {Militzer}, B.},
        title = "{Mass-Radius Relationships for Solid Exoplanets}",
      journal = {\apj},
         year = 2007,
        month = nov,
       volume = {669},
       number = {2},
        pages = {1279-1297},
          doi = {10.1086/521346},
archivePrefix = {arXiv},
       eprint = {0707.2895},
 primaryClass = {astro-ph},
       adsurl = {https://ui.adsabs.harvard.edu/abs/2007ApJ...669.1279S}
}

@ARTICLE{Seidler2025,
       author = {{Seidler}, Fabian L. and {Sossi}, Paolo A. and {Bower}, Dan J. and {Demory}, Brice-Olivier},
        title = "{Volatile-bearing mineral atmospheres of hot rocky exoplanets as probes of interior state and composition}",
      journal = {\aap},
         year = 2026,
        month = jun,
       volume = {710},
          eid = {A359},
        pages = {A359},
          doi = {10.1051/0004-6361/202557276},
}

@ARTICLE{Senft2008,
       author = {{Senft}, L.~E. and {Stewart}, S.~T.},
        title = "{Impact crater formation in icy layered terrains on Mars}",
      journal = {Meteoritics \& Planetary Science},
         year = 2008,
        month = dec,
       volume = {43},
       number = {12},
        pages = {1993-2013},
          doi = {10.1111/j.1945-5100.2008.tb00657.x},
       adsurl = {https://ui.adsabs.harvard.edu/abs/2008M&PS...43.1993S}
}

@ARTICLE{Shahar2026,
       author = {{Shahar}, Anat and {Young}, Edward D. and {Hirose}, Kei and {Yokoo}, Shunpei},
        title = "{The Compositions of Planetary Cores}",
      journal = {Annual Review of Earth and Planetary Sciences},
         year = 2026,
        month = feb,
          doi = {10.1146/annurev-earth-040722-094945},
         note = {In press; volume and pagination pending. Crossref publication date 2026-02-23.}
}

@ARTICLE{Stebbins1984,
       author = {{Stebbins}, J.~F. and {Carmichael}, I.~S.~E. and {Moret}, L.~K.},
        title = "{Heat capacities and entropies of silicate liquids and glasses}",
      journal = {Contributions to Mineralogy and Petrology},
         year = 1984,
        month = may,
       volume = {86},
       number = {2},
        pages = {131-148},
          doi = {10.1007/BF00381840},
       adsurl = {https://ui.adsabs.harvard.edu/abs/1984CoMP...86..131S}
}

@ARTICLE{Sokolova2022,
       author = {{Sokolova}, Tatiana S. and {Dorogokupets}, Peter I. and {Filippova}, Alena I.},
        title = "{Equations of state of clino- and orthoenstatite and phase relations in the MgSiO$_{3}$ system at pressures up to 12 GPa and high temperatures}",
      journal = {Physics and Chemistry of Minerals},
         year = 2022,
        month = sep,
       volume = {49},
       number = {9},
          eid = {37},
        pages = {37},
          doi = {10.1007/s00269-022-01212-7},
       adsurl = {https://ui.adsabs.harvard.edu/abs/2022PCM....49...37S}
}

@ARTICLE{Stacey1981,
       author = {{Stacey}, F.~D. and {Brennan}, B.~J. and {Irvine}, R.~D.},
        title = "{Finite strain theories and comparisons with seismological data}",
      journal = {Geophysical Surveys},
         year = 1981,
        month = apr,
       volume = {4},
       number = {3},
        pages = {189-232},
          doi = {10.1007/BF01449185},
       adsurl = {https://ui.adsabs.harvard.edu/abs/1981GeoSu...4..189S}
}

@ARTICLE{Stacey2004,
       author = {{Stacey}, F.~D. and {Davis}, P.~M.},
        title = "{High pressure equations of state with applications to the lower mantle and core}",
      journal = {Physics of the Earth and Planetary Interiors},
         year = 2004,
        month = may,
       volume = {142},
       number = {3-4},
        pages = {137-184},
          doi = {10.1016/j.pepi.2004.02.003},
       adsurl = {https://ui.adsabs.harvard.edu/abs/2004PEPI..142..137S}
}

@ARTICLE{Stebbins1988,
       author = {{Stebbins}, Jonathan F.},
        title = "{Effects of temperature and composition on silicate glass structure and dynamics: SI-29 NMR results}",
      journal = {Journal of Non Crystalline Solids},
         year = 1988,
        month = dec,
       volume = {106},
       number = {1-3},
        pages = {359-369},
          doi = {10.1016/0022-3093(88)90289-X},
       adsurl = {https://ui.adsabs.harvard.edu/abs/1988JNCS..106..359S}
}

@ARTICLE{Steinmeyer2026,
       author = {{Steinmeyer}, Marie-Luise and {Dorn}, Caroline and {Werlen}, Aaron and {Grimm}, Simon L.},
        title = "{Coupled Thermal─Chemical Evolution Models of Sub-Neptunes Reveal Atmospheric Signatures of Their Formation Location}",
      journal = {\apj},
         year = 2026,
        month = apr,
       volume = {1001},
       number = {1},
          eid = {36},
        pages = {36},
          doi = {10.3847/1538-4357/ae4c47},
archivePrefix = {arXiv},
       eprint = {2601.21377},
 primaryClass = {astro-ph.EP},
       adsurl = {https://ui.adsabs.harvard.edu/abs/2026ApJ..1001...36S}
}

@ARTICLE{Stixrude2005,
       author = {{Stixrude}, Lars and {Lithgow-Bertelloni}, Carolina},
        title = "{Thermodynamics of mantle minerals - I. Physical properties}",
      journal = {Geophysical Journal International},
         year = 2005,
        month = aug,
       volume = {162},
       number = {2},
        pages = {610-632},
          doi = {10.1111/j.1365-246X.2005.02642.x},
       adsurl = {https://ui.adsabs.harvard.edu/abs/2005GeoJI.162..610S}
}

@ARTICLE{Stixrude2014,
       author = {{Stixrude}, L.},
        title = "{Melting in super-earths}",
      journal = {Philosophical Transactions of the Royal Society of London Series A},
         year = 2014,
        month = mar,
       volume = {372},
       number = {2014},
        pages = {20130076-20130076},
          doi = {10.1098/rsta.2013.0076},
       adsurl = {https://ui.adsabs.harvard.edu/abs/2014RSPTA.37230076S}
}

@ARTICLE{Stixrude2022,
       author = {{Stixrude}, Lars and {Lithgow-Bertelloni}, Carolina},
        title = "{Thermal expansivity, heat capacity and bulk modulus of the mantle}",
      journal = {Geophysical Journal International},
         year = 2022,
        month = feb,
       volume = {228},
       number = {2},
        pages = {1119-1149},
          doi = {10.1093/gji/ggab394},
       adsurl = {https://ui.adsabs.harvard.edu/abs/2022GeoJI.228.1119S}
}

@ARTICLE{Swesty1996,
       author = {{Swesty}, F. Douglas},
        title = "{Thermodynamically Consistent Interpolation for Equation of State Tables}",
      journal = {Journal of Computational Physics},
         year = 1996,
        month = aug,
       volume = {127},
       number = {1},
        pages = {118-127},
          doi = {10.1006/jcph.1996.0162},
       adsurl = {https://ui.adsabs.harvard.edu/abs/1996JCoPh.127..118S}
}

@ARTICLE{Swift2012,
       author = {{Swift}, D.~C. and {Eggert}, J.~H. and {Hicks}, D.~G. and {Hamel}, S. and {Caspersen}, K. and {Schwegler}, E. and {Collins}, G.~W. and {Nettelmann}, N. and {Ackland}, G.~J.},
        title = "{Mass-Radius Relationships for Exoplanets}",
      journal = {\apj},
         year = 2012,
        month = jan,
       volume = {744},
       number = {1},
          eid = {59},
        pages = {59},
          doi = {10.1088/0004-637X/744/1/5910.1086/141924},
archivePrefix = {arXiv},
       eprint = {1001.4851},
 primaryClass = {astro-ph.EP},
       adsurl = {https://ui.adsabs.harvard.edu/abs/2012ApJ...744...59S}
}

@ARTICLE{Tange2012,
       author = {{Tange}, Yoshinori and {Kuwayama}, Yasuhiro and {Irifune}, Tetsuo and {Funakoshi}, Ken-Ichi and {Ohishi}, Yasuo},
        title = "{P-V-T equation of state of MgSiO$_{3}$ perovskite based on the MgO pressure scale: A comprehensive reference for mineralogy of the lower mantle}",
      journal = {Journal of Geophysical Research (Solid Earth)},
         year = 2012,
        month = jun,
       volume = {117},
       number = {B6},
          eid = {B06201},
        pages = {B06201},
          doi = {10.1029/2011JB008988},
       adsurl = {https://ui.adsabs.harvard.edu/abs/2012JGRB..117.6201T}
}

@ARTICLE{Tang2025,
       author = {{Tang}, Yao and {Fortney}, Jonathan J. and {Nimmo}, Francis and {Thorngren}, Daniel and {Ohno}, Kazumasa and {Murray-Clay}, Ruth},
        title = "{Reassessing Sub-Neptune Structure, Radii, and Thermal Evolution}",
      journal = {\aj},
         year = 2025,
        month = aug,
       volume = {989},
       number = {1},
          eid = {28},
        pages = {28},
          doi = {10.3847/1538-3881/adc7f0},
}

@ARTICLE{Tateno2010,
       author = {{Tateno}, Shigehiko and {Hirose}, Kei and {Ohishi}, Yasuo and {Tatsumi}, Yoshiyuki},
        title = "{The Structure of Iron in Earth{\textquoteright}s Inner Core}",
      journal = {Science},
         year = 2010,
        month = oct,
       volume = {330},
       number = {6002},
        pages = {359},
          doi = {10.1126/science.1194662},
       adsurl = {https://ui.adsabs.harvard.edu/abs/2010Sci...330..359T}
}

@ARTICLE{Teske2025,
       author = {{Teske}, Johanna K. and {Wallack}, Nicole L. and {Piette}, Anjali A.~A. and {Dang}, Lisa and {Lichtenberg}, Tim and {Plotnykov}, Mykhaylo and {Pierrehumbert}, Raymond and {Postolec}, Emma and {Boucher}, Samuel and {McGinty}, Alex and {Peng}, Bo and {Valencia}, Diana and {Hammond}, Mark},
        title = "{A Thick Volatile Atmosphere on the Ultrahot Super-Earth TOI-561 b}",
      journal = {\apjl},
         year = 2025,
        month = dec,
       volume = {995},
       number = {2},
          eid = {L39},
        pages = {L39},
          doi = {10.3847/2041-8213/ae0a4c},
archivePrefix = {arXiv},
       eprint = {2509.17231},
 primaryClass = {astro-ph.EP},
       adsurl = {https://ui.adsabs.harvard.edu/abs/2025ApJ...995L..39T}
}

@ARTICLE{Timmes1999,
       author = {{Timmes}, F.~X. and {Arnett}, Dave},
        title = "{The Accuracy, Consistency, and Speed of Five Equations of State for Stellar Hydrodynamics}",
      journal = {\apjs},
         year = 1999,
        month = nov,
       volume = {125},
       number = {1},
        pages = {277-294},
          doi = {10.1086/313271},
       adsurl = {https://ui.adsabs.harvard.edu/abs/1999ApJS..125..277T}
}

@ARTICLE{Tinetti2018,
       author = {{Tinetti}, Giovanna and {Drossart}, Pierre and {Eccleston}, Paul and {Hartogh}, Paul and {Heske}, Astrid and {Leconte}, J{\'e}r{\'e}my and {Micela}, Giusi and {Ollivier}, Marc and {Pilbratt}, G{\"o}ran and {Puig}, Ludovic and et al.},
        title = "{A chemical survey of exoplanets with ARIEL}",
      journal = {Experimental Astronomy},
         year = 2018,
        month = nov,
       volume = {46},
       number = {1},
        pages = {135-209},
          doi = {10.1007/s10686-018-9598-x},
       adsurl = {https://ui.adsabs.harvard.edu/abs/2018ExA....46..135T}
}

@ARTICLE{Tosi2010,
       author = {{Tosi}, Nicola and {Yuen}, David A. and {{\v{C}}adek}, Ond{\v{r}}ej},
        title = "{Dynamical consequences in the lower mantle with the post-perovskite phase change and strongly depth-dependent thermodynamic and transport properties}",
      journal = {Earth and Planetary Science Letters},
         year = 2010,
        month = sep,
       volume = {298},
       number = {1-2},
        pages = {229-243},
          doi = {10.1016/j.epsl.2010.08.001},
       adsurl = {https://ui.adsabs.harvard.edu/abs/2010E&PSL.298..229T}
}

@ARTICLE{Tosi2013,
       author = {{Tosi}, Nicola and {Yuen}, David A. and {de Koker}, Nico and {Wentzcovitch}, Renata M.},
        title = "{Mantle dynamics with pressure- and temperature-dependent thermal expansivity and conductivity}",
      journal = {Physics of the Earth and Planetary Interiors},
         year = 2013,
        month = apr,
       volume = {217},
        pages = {48-58},
          doi = {10.1016/j.pepi.2013.02.004},
       adsurl = {https://ui.adsabs.harvard.edu/abs/2013PEPI..217...48T}
}

@ARTICLE{Tosi2017,
       author = {{Tosi}, N. and {Godolt}, M. and {Stracke}, B. and {Ruedas}, T. and {Grenfell}, J.~L. and {H{\"o}ning}, D. and {Nikolaou}, A. and {Plesa}, A.-C. and {Breuer}, D. and {Spohn}, T.},
        title = "{The habitability of a stagnant-lid Earth}",
      journal = {\aap},
         year = 2017,
        month = sep,
       volume = {605},
          eid = {A71},
        pages = {A71},
          doi = {10.1051/0004-6361/201730728},
archivePrefix = {arXiv},
       eprint = {1707.06051},
 primaryClass = {astro-ph.EP},
       adsurl = {https://ui.adsabs.harvard.edu/abs/2017A&A...605A..71T}
}

@ARTICLE{Tsuchiya2011,
       author = {{Tsuchiya}, Taku and {Tsuchiya}, Jun},
        title = "{Prediction of a hexagonal SiO$_{2}$ phase affecting stabilities of MgSiO$_{3}$ and CaSiO$_{3}$ at multimegabar pressures}",
      journal = {Proceedings of the National Academy of Science},
         year = 2011,
        month = jan,
       volume = {108},
       number = {4},
        pages = {1252-1255},
          doi = {10.1073/pnas.1013594108},
       adsurl = {https://ui.adsabs.harvard.edu/abs/2011PNAS..108.1252T}
}

@ARTICLE{Turbet2020,
       author = {{Turbet}, Martin and {Bolmont}, Emeline and {Ehrenreich}, David and {Gratier}, Pierre and {Leconte}, J{\'e}r{\'e}my and {Selsis}, Franck and {Hara}, Nathan and {Lovis}, Christophe},
        title = "{Revised mass-radius relationships for water-rich rocky planets more irradiated than the runaway greenhouse limit}",
      journal = {\aap},
         year = 2020,
        month = jun,
       volume = {638},
          eid = {A41},
        pages = {A41},
          doi = {10.1051/0004-6361/201937151},
}

@ARTICLE{Unterborn2023,
       author = {{Unterborn}, C.~T. and {Desch}, S.~J. and {Haldemann}, J. and {Lorenzo}, A. and {Schulze}, J.~G. and {Hinkel}, N.~R. and {Panero}, W.~R.},
        title = "{The Nominal Ranges of Rocky Planet Masses, Radii, Surface Gravities, and Bulk Densities}",
      journal = {\apj},
         year = 2023,
        month = feb,
       volume = {944},
       number = {1},
          eid = {42},
        pages = {42},
          doi = {10.3847/1538-4357/acaa3b},
archivePrefix = {arXiv},
       eprint = {2212.03934},
 primaryClass = {astro-ph.EP},
       adsurl = {https://ui.adsabs.harvard.edu/abs/2023ApJ...944...42U}
}

@ARTICLE{Valencia2006,
       author = {{Valencia}, Diana and {O'Connell}, Richard J. and {Sasselov}, Dimitar},
        title = "{Internal structure of massive terrestrial planets}",
      journal = {\icarus},
         year = 2006,
        month = apr,
       volume = {181},
       number = {2},
        pages = {545-554},
          doi = {10.1016/j.icarus.2005.11.021},
archivePrefix = {arXiv},
       eprint = {astro-ph/0511150},
 primaryClass = {astro-ph},
       adsurl = {https://ui.adsabs.harvard.edu/abs/2006Icar..181..545V}
}

@ARTICLE{Vidotto2011,
       author = {{Vidotto}, A.~A. and {Jardine}, M. and {Helling}, Ch.},
        title = "{Prospects for detection of exoplanet magnetic fields through bow-shock observations during transits}",
      journal = {\mnras},
         year = 2011,
        month = feb,
       volume = {411},
       number = {1},
        pages = {L46-L50},
          doi = {10.1111/j.1745-3933.2010.00991.x},
archivePrefix = {arXiv},
       eprint = {1011.3455},
 primaryClass = {astro-ph.EP},
       adsurl = {https://ui.adsabs.harvard.edu/abs/2011MNRAS.411L..46V}
}

@ARTICLE{Vinet1987,
       author = {{Vinet}, Pascal and {Ferrante}, John and {Smith}, John R. and {Rose}, James H.},
        title = "{Temperature effects on the universal equation of state of solids}",
      journal = {\prb},
         year = 1987,
        month = feb,
       volume = {35},
       number = {4},
        pages = {1945-1953},
          doi = {10.1103/PhysRevB.35.1945},
       adsurl = {https://ui.adsabs.harvard.edu/abs/1987PhRvB..35.1945V}
}

@ARTICLE{Umemoto2017,
       author = {{Umemoto}, Koichiro and {Wentzcovitch}, Renata M. and {Wu}, Shunqing and {Ji}, Min and {Wang}, Cai-Zhuang and {Ho}, Kai-Ming},
        title = "{Phase transitions in MgSiO$_{3}$ post-perovskite in super-Earth mantles}",
      journal = {Earth and Planetary Science Letters},
         year = 2017,
        month = nov,
       volume = {478},
        pages = {40-45},
          doi = {10.1016/j.epsl.2017.08.032},
archivePrefix = {arXiv},
       eprint = {1708.04767},
 primaryClass = {physics.geo-ph},
       adsurl = {https://ui.adsabs.harvard.edu/abs/2017E&PSL.478...40U}
}

@ARTICLE{Unterborn2019,
       author = {{Unterborn}, C.~T. and {Panero}, W.~R.},
        title = "{The Pressure and Temperature Limits of Likely Rocky Exoplanets}",
      journal = {Journal of Geophysical Research (Planets)},
         year = 2019,
        month = jul,
       volume = {124},
       number = {7},
        pages = {1704-1716},
          doi = {10.1029/2018JE005844},
archivePrefix = {arXiv},
       eprint = {1905.06530},
 primaryClass = {astro-ph.EP},
       adsurl = {https://ui.adsabs.harvard.edu/abs/2019JGRE..124.1704U}
}

@ARTICLE{Wagner2002,
       author = {{Wagner}, Wolfgang and {Pru{\ss}}, Andreas},
        title = "{The IAPWS Formulation 1995 for the Thermodynamic Properties of Ordinary Water Substance for General and Scientific Use}",
      journal = {Journal of Physical and Chemical Reference Data},
         year = 2002,
        month = jun,
       volume = {31},
       number = {2},
        pages = {387-535},
          doi = {10.1063/1.1461829},
       adsurl = {https://ui.adsabs.harvard.edu/abs/2002JPCRD..31..387W}
}

@ARTICLE{Wilkinson2024,
       author = {{Wilkinson}, C. and {Charnay}, B. and {Mazevet}, S. and {Lagrange}, A.-M. and {Chomez}, A. and {Squicciarini}, V. and {Panek}, E. and {Mazoyer}, J.},
        title = "{Breaking degeneracies in exoplanetary parameters through self-consistent atmosphere─interior modelling}",
      journal = {\aap},
         year = 2024,
        month = dec,
       volume = {692},
          eid = {A113},
        pages = {A113},
          doi = {10.1051/0004-6361/202348945},
archivePrefix = {arXiv},
       eprint = {2410.04470},
 primaryClass = {astro-ph.EP},
       adsurl = {https://ui.adsabs.harvard.edu/abs/2024A&A...692A.113W}
}

@ARTICLE{Wolf2015,
       author = {{Wolf}, Aaron S. and {Jackson}, Jennifer M. and {Dera}, Przemeslaw and {Prakapenka}, Vitali B.},
        title = "{The thermal equation of state of (Mg, Fe)SiO$_{3}$ bridgmanite (perovskite) and implications for lower mantle structures}",
      journal = {Journal of Geophysical Research (Solid Earth)},
         year = 2015,
        month = nov,
       volume = {120},
       number = {11},
        pages = {7460-7489},
          doi = {10.1002/2015JB012108},
       adsurl = {https://ui.adsabs.harvard.edu/abs/2015JGRB..120.7460W}
}

@ARTICLE{Wolf2018,
       author = {{Wolf}, Aaron S. and {Bower}, Dan J.},
        title = "{An equation of state for high pressure-temperature liquids (RTpress) with application to MgSiO$_{3}$ melt}",
      journal = {Physics of the Earth and Planetary Interiors},
         year = 2018,
        month = may,
       volume = {278},
        pages = {59-74},
          doi = {10.1016/j.pepi.2018.02.00410.31223/osf.io/4c2s5},
       adsurl = {https://ui.adsabs.harvard.edu/abs/2018PEPI..278...59W}
}

@ARTICLE{Zeng2013,
       author = {{Zeng}, Li and {Sasselov}, Dimitar},
        title = "{A Detailed Model Grid for Solid Planets from 0.1 through 100 Earth Masses}",
      journal = {\pasp},
         year = 2013,
        month = mar,
       volume = {125},
       number = {925},
        pages = {227},
          doi = {10.1086/669163},
archivePrefix = {arXiv},
       eprint = {1301.0818},
 primaryClass = {astro-ph.EP},
       adsurl = {https://ui.adsabs.harvard.edu/abs/2013PASP..125..227Z}
}

@ARTICLE{Zeng2016,
       author = {{Zeng}, Li and {Sasselov}, Dimitar D. and {Jacobsen}, Stein B.},
        title = "{Mass-Radius Relation for Rocky Planets Based on PREM}",
      journal = {\apj},
         year = 2016,
        month = mar,
       volume = {819},
       number = {2},
          eid = {127},
        pages = {127},
          doi = {10.3847/0004-637X/819/2/127},
archivePrefix = {arXiv},
       eprint = {1512.08827},
 primaryClass = {astro-ph.EP},
       adsurl = {https://ui.adsabs.harvard.edu/abs/2016ApJ...819..127Z}
}

@ARTICLE{Zeng2019,
       author = {{Zeng}, Li and {Jacobsen}, Stein B. and {Sasselov}, Dimitar D. and {Petaev}, Michail I. and {Vanderburg}, Andrew and {Lopez-Morales}, Mercedes and {Perez-Mercader}, Juan and {Mattsson}, Thomas R. and {Li}, Gongjie and {Heising}, Matthew Z. and et al.},
        title = "{Growth model interpretation of planet size distribution}",
      journal = {Proceedings of the National Academy of Science},
         year = 2019,
        month = may,
       volume = {116},
       number = {20},
        pages = {9723-9728},
          doi = {10.1073/pnas.1812905116},
archivePrefix = {arXiv},
       eprint = {1906.04253},
 primaryClass = {astro-ph.EP},
       adsurl = {https://ui.adsabs.harvard.edu/abs/2019PNAS..116.9723Z}
}

@ARTICLE{Zhang2015,
       author = {{Zhang}, Wen-Jin and {Liu}, Zhi-Yong and {Liu}, Zhong-Li and {Cai}, Ling-Cang},
        title = "{Melting curves and entropy of melting of iron under Earth's core conditions}",
      journal = {Physics of the Earth and Planetary Interiors},
         year = 2015,
        month = jul,
       volume = {244},
        pages = {69-77},
          doi = {10.1016/j.pepi.2014.10.011},
       adsurl = {https://ui.adsabs.harvard.edu/abs/2015PEPI..244...69Z}
}

@ARTICLE{Zhang2016,
       author = {{Zhang}, Dongzhou and {Jackson}, Jennifer M. and {Zhao}, Jiyong and {Sturhahn}, Wolfgang and {Alp}, E. Ercan and {Hu}, Michael Y. and {Toellner}, Thomas S. and {Murphy}, Caitlin A. and {Prakapenka}, Vitali B.},
        title = "{Temperature of Earth's core constrained from melting of Fe and Fe$_{0.9}$Ni$_{0.1}$ at high pressures}",
      journal = {Earth and Planetary Science Letters},
         year = 2016,
        month = aug,
       volume = {447},
        pages = {72-83},
          doi = {10.1016/j.epsl.2016.04.026},
       adsurl = {https://ui.adsabs.harvard.edu/abs/2016E&PSL.447...72Z}
}

@ARTICLE{Zhang2024,
       author = {{Zhang}, Michael and {Hu}, Renyu and {Inglis}, Julie and {Dai}, Fei and {Bean}, Jacob L. and {Knutson}, Heather A. and {Lam}, Kristine and {Goffo}, Elisa and {Gandolfi}, Davide},
        title = "{GJ 367b Is a Dark, Hot, Airless Sub-Earth}",
      journal = {\apjl},
         year = 2024,
        month = feb,
       volume = {961},
       number = {2},
          eid = {L44},
        pages = {L44},
          doi = {10.3847/2041-8213/ad1a07},
archivePrefix = {arXiv},
       eprint = {2401.01400},
 primaryClass = {astro-ph.EP},
       adsurl = {https://ui.adsabs.harvard.edu/abs/2024ApJ...961L..44Z}
}

@BOOK{Zharkov1971,
  title     = "Equations of state for solids at high pressures and temperatures",
  author    = "Zharkov, V. N.",
  publisher = "Springer",
  doi       =  {10.1007/978-1-4757-1517-0},
  month     =  nov,
  year      =  1971,
  address   = "New York, NY",
  language  = "en"
}

@ARTICLE{Zilinskas2022,
       author = {{Zilinskas}, M. and {van Buchem}, C.~P.~A. and {Miguel}, Y. and {Louca}, A. and {Lupu}, R. and {Zieba}, S. and {van Westrenen}, W.},
        title = "{Observability of evaporating lava worlds}",
      journal = {\aap},
         year = 2022,
        month = may,
       volume = {661},
          eid = {A126},
        pages = {A126},
          doi = {10.1051/0004-6361/202142984},
archivePrefix = {arXiv},
       eprint = {2202.04759},
 primaryClass = {astro-ph.EP},
       adsurl = {https://ui.adsabs.harvard.edu/abs/2022A&A...661A.126Z}
}

\begin{appendix}

\section{Thermodynamic derivations}
\label{app:thermo}

\subsection{Cold pressure derivatives}
\label{app:cold}

Throughout this subsection, $X = (V/V_0)^{1/3}$ denotes the linear compression ratio and $Y = V_0/V$ the volumetric compression ratio. $X$ is used in the Vinet, Holzapfel, and Kunc forms below, while $Y$ appears in the Keane form.

\paragraph{Third-order Birch--Murnaghan equation (Eq.~\eqref{eq:bm3}).}
\label{app:bm3}
Using the Eulerian finite strain $f_\mathrm{E} = \frac{1}{2}[(V_0/V)^{2/3} - 1]$ and the chain rule $\dd f_\mathrm{E}/\dd V = -(1+2f_\mathrm{E})/(3V)$, the isothermal bulk modulus is
\begin{equation}
\label{eq:kt_bm3}
K_T^\mathrm{cold} = K_0 (1 + 2f_\mathrm{E})^{5/2}\left[1 + \left(3K_0' - 5\right)f_\mathrm{E} + \frac{27}{2}(K_0' - 4)\,f_\mathrm{E}^2\right].
\end{equation}
The cold internal energy is
\begin{equation}
\label{eq:ucold_bm3}
U_\mathrm{cold} = \frac{9}{2}\,K_0 V_0\,f_\mathrm{E}^2\left[1 + (K_0' - 4)\,f_\mathrm{E}\right].
\end{equation}

\paragraph{Vinet equation (Eq.~\eqref{eq:vinet}).}
\label{app:vinet}
With $\eta_\mathrm{V} = \frac{3}{2}(K_0' - 1)$, the isothermal bulk modulus is
\begin{equation}
\label{eq:kt_vinet}
K_T^\mathrm{cold} = K_0 \frac{1 + (1 + \eta_\mathrm{V} X)(1-X)}{X^2}\exp\!\left[\eta_\mathrm{V}(1-X)\right].
\end{equation}
The cold internal energy admits the closed form
\begin{equation}
\label{eq:ucold_vinet}
U_\mathrm{cold} = \frac{9 K_0 V_0}{\eta_\mathrm{V}^2}\left\{1 - \left[1 - \eta_\mathrm{V}(1-X)\right]\exp\!\left[\eta_\mathrm{V}(1-X)\right]\right\}.
\end{equation}

\paragraph{Holzapfel equation (Eq.~\eqref{eq:holzapfel}).}
\label{app:holzapfel}
The bulk modulus in this formalism reads as
\begin{equation}
\label{eq:kt_holzapfel}
K_T^\mathrm{cold} = \frac{P_\mathrm{cold} - P_0}{3}\left[5 + \frac{X}{1 - X} - \frac{c_2 X(1 - 2X)}{1 + c_2 X(1-X)} + c_0 X\right],
\end{equation}
where $P_0$ is a reference pressure offset. The cold internal energy is evaluated by numerical quadrature, as no closed-form solution is available.

\paragraph{Keane equation (Eq.~\eqref{eq:keane}).}
\label{app:keane}
The isothermal bulk modulus is
\begin{equation}
\label{eq:kt_keane}
K_T^\mathrm{cold} = K_0\left[\frac{K_0'}{K_\infty'}\!\left(Y^{K_\infty'} - 1\right) + 1\right].
\end{equation}
The cold internal energy admits the closed form
\begin{multline}
\label{eq:ucold_keane}
U_\mathrm{cold} = K_0 V_0\left\{\frac{K_0'}{K_\infty'^2}\!\left[\frac{Y^{K_\infty'-1} - 1}{K_\infty' - 1} + \frac{1}{Y} - 1\right]\right. \\
\left.{}+ \left(\frac{K_0'}{K_\infty'} - 1\right)\!\left[\frac{\ln Y + 1}{Y} - 1\right]\right\}.
\end{multline}

\paragraph{Kunc equation (Eq.~\eqref{eq:kunc}).}
\label{app:kunc}
With $\eta_\mathrm{K} = \frac{3}{2}K_0' - \frac{9}{2}$, the isothermal bulk modulus is
\begin{equation}
\label{eq:kt_kunc}
K_T^\mathrm{cold} = K_0 X^{-5}\exp\!\left[\eta_\mathrm{K}(1-X)\right]\left[X + (1-X)(\eta_\mathrm{K} X + 5)\right].
\end{equation}
The cold internal energy $U_\mathrm{cold} = -\int_{V_0}^V P_\mathrm{cold}\,\dd V'$ is evaluated by numerical quadrature.

\subsection{Gr\"{u}neisen parameter and characteristic temperature}
\label{app:gruneisen}

The Gr\"{u}neisen parameter $\gamma(V) = -\dd\ln\Theta/\dd\ln V$, either for the MGD or Einstein models, controls the coupling between lattice vibrations and volume: compression raises vibrational frequencies (and hence $\Theta$) at a rate governed by $\gamma$, while $\gamma$ itself decreases under compression as the material stiffens and anharmonic effects weaken \citep{Anderson1995}. Given a functional form for $\gamma(V)$, the characteristic temperature follows by integration of $\dd\ln\Theta = -\gamma\,\dd\ln V$.

Two standard parametrizations are used in \paleos. The power-law form assumes $q = \dd\ln\gamma/\dd\ln V$ is constant, giving a monotonic decrease of $\gamma$ with compression:
\begin{align}
\label{eq:gamma_power}
\gamma(V) &= \gamma_0\left(\frac{V}{V_0}\right)^{\!q}, \\
\label{eq:theta_power}
\Theta(V) &= \Theta_0 \exp\!\left[\frac{\gamma_0 - \gamma(V)}{q}\right],
\end{align}
where $\gamma_0$ and $\Theta_0$ are reference values. The Al'tshuler parametrization \citep{Altshuler1987} introduces a finite high-compression asymptote $\gamma_\infty$, preventing $\gamma$ from vanishing at extreme compression. The exponent $\beta$ controls how rapidly $\gamma$ transitions from its ambient value $\gamma_0$ to $\gamma_\infty$, and $q(V)$ is now volume-dependent:
\begin{align}
\label{eq:gamma_altshuler}
\gamma(V) &= \gamma_\infty + (\gamma_0 - \gamma_\infty)\left(\frac{V}{V_0}\right)^{\!\beta}, \\
\label{eq:q_altshuler}
q(V) &= \frac{\beta\,(\gamma_0 - \gamma_\infty)(V/V_0)^\beta}{\gamma(V)}, \\
\label{eq:theta_altshuler}
\Theta(V) &= \Theta_0 \left(\frac{V}{V_0}\right)^{\!-\gamma_\infty} \exp\!\left[\frac{\gamma_0 - \gamma_\infty}{\beta}\left(1 - \left(\frac{V}{V_0}\right)^{\!\beta}\right)\right].
\end{align}

\subsection{Thermal models}
\label{app:thermal}

\paragraph{Mie--Gr\"{u}neisen--Debye model.}
\label{app:mgd}
The thermal energy in the Debye model is defined in Eq.~\eqref{eq:debye_energy}, complemented by the dependency of the Debye temperature to volume ($\Theta_\mathrm{D}(V)$, Appendix~\ref{app:gruneisen}). The thermal pressure follows from the Mie--Gr\"{u}neisen relation (Eq.~\eqref{eq:pth}). The stemming isochoric heat capacity is
\begin{equation}
\label{eq:cv_debye}
C_V^\mathrm{th} = 9nR\left(\frac{T}{\Theta_\mathrm{D}}\right)^{\!3}\int_0^{\Theta_\mathrm{D}/T} \frac{x^4 \mathrm{e}^x}{(\mathrm{e}^x - 1)^2}\,\dd x,
\end{equation}
and the entropy is \citep{Gopal1966}
\begin{equation}
\label{eq:s_debye}
S_\mathrm{th} = nR\left[4\,D_3\!\left(\frac{\Theta_\mathrm{D}}{T}\right) - 3\ln\!\left(1 - \mathrm{e}^{-\Theta_\mathrm{D}/T}\right)\right].
\end{equation}
We choose here to keep the explicit thermal annotation to differentiate this contribution from electronic and magnetic ones, as well as from the reference-state offset $S_0$ (Sect.~\ref{sec:eos:refstate}). The thermal bulk modulus $K_T^\mathrm{th} = -V(\partial P_\mathrm{th}/\partial V)_T$ requires the identity $(\partial U_\mathrm{th}/\partial V)_T = \gamma C_V^\mathrm{th} T/V - P_\mathrm{th}$, which follows from the Maxwell relation $(\partial S/\partial V)_T = (\partial P/\partial T)_V$. The result is
\begin{equation}
\label{eq:kt_thermal}
K_T^\mathrm{th} = (1 + \gamma - q)\,P_\mathrm{th} - \frac{\gamma^2}{V}\!\left[T\,C_V^\mathrm{th}(T) - T_0\,C_V^\mathrm{th}(T_0)\right],
\end{equation}
where $q = \dd\ln\gamma/\dd\ln V$. The second term is commonly omitted in the literature \citep[e.g.,][]{Sakai2016, Dorogokupets2017} but is required for thermodynamic consistency. Differentiation of the thermal pressure with respect to temperature at constant volume gives the thermal expansion $\alpha_\mathrm{th} K_T^\mathrm{th} = \gamma C_V^\mathrm{th}/V$.

\paragraph{Einstein model.}
\label{app:einstein}
The Einstein model replaces the Debye phonon spectrum with a single characteristic frequency ($\Theta_\mathrm{E}(V)$, Appendix~\ref{app:gruneisen}). The thermal energy, including zero-point energy, is given by Eq.~\eqref{eq:einstein_energy}. The thermal pressure follows from the same Mie--Gr\"{u}neisen relation (Eq.~\eqref{eq:pth}). The isochoric heat capacity and entropy are
\begin{align}
\label{eq:cv_einstein}
C_V^\mathrm{th} &= 3nR\left(\frac{\Theta_\mathrm{E}}{T}\right)^{\!2} \frac{\mathrm{e}^{\Theta_\mathrm{E}/T}}{(\mathrm{e}^{\Theta_\mathrm{E}/T} - 1)^2}, \\
\label{eq:s_einstein}
S^\mathrm{th} &= 3nR\left[\frac{\Theta_\mathrm{E}/T}{\mathrm{e}^{\Theta_\mathrm{E}/T} - 1} - \ln\!\left(1 - \mathrm{e}^{-\Theta_\mathrm{E}/T}\right)\right].
\end{align}
The thermal bulk modulus has the same form as Eq.~\eqref{eq:kt_thermal} with the Einstein $C_V^\mathrm{th}$. The thermal expansion is also identical in its form. For applications requiring two characteristic frequencies \citep[e.g., pyroxene phases in][]{Sokolova2022}, the thermal energy generalizes to $U_\mathrm{th} = \sum_i m_i R\,\Theta_i/[\exp(\Theta_i/T) - 1]$, where $m_i$ are mode multiplicities satisfying $\sum_i m_i = 3n$.

\paragraph{\citet{Luo2024} liquid iron model.}
\label{app:luo24}
For liquid iron at super-Earth core conditions, \citet{Luo2024} adopt a thermal pressure that is linear in temperature with a polynomial volume dependence:
\begin{align}
\label{eq:pth_luo24}
P_\mathrm{th}(V,T) &= \frac{T - T_0}{1000\;\mathrm{K}}\left[a + b\,Y + c\,Y^2\right], \\
\label{eq:uth_luo24}
U_\mathrm{th}(V) &= \frac{T_0}{1000\;\mathrm{K}}\,\Phi(V), \\
\label{eq:sth_luo24}
S_\mathrm{th}(V) &= \frac{\Phi(V)}{1000\;\mathrm{K}}, \\
\label{eq:kt_luo24}
K_T^\mathrm{th} &= \frac{Y(T - T_0)}{1000\;\mathrm{K}}\left[b + 2c\,Y\right],
\end{align}
where $Y = V_0/V$ is the compression ratio, $T_0$ is the reference temperature, $a$, $b$, $c$ are coefficients fitted to ab initio MD simulations, and $\Phi(V) = a(V - V_0) + b\,V_0\ln(V/V_0) + c\,V_0(1 - V_0/V)$ is the volume integral of the polynomial. The thermal expansion coefficient is $\alpha_\mathrm{th} K_T^\mathrm{th} = (\partial P_\mathrm{th}/\partial T)_V = [a + b\,Y + c\,Y^2]/(1000\;\mathrm{K})$. This formulation is constructed to fit simulated pressure data but is thermodynamically inconsistent: the thermal energy and entropy depend only on volume, yielding a vanishing heat capacity ($C_V^\mathrm{th} = \partial U_\mathrm{th}/\partial T = 0$). We restore consistency by adding Dulong--Petit correction terms that enforce an explicit temperature dependence,
\begin{align}
\label{eq:udp}
U_\mathrm{DP}(T) &= 3nR\,(T - T_0), \\
\label{eq:sdp}
S_\mathrm{DP}(T) &= 3nR\,\ln\!\left(\frac{T}{T_0}\right),
\end{align}
so that the corrected quantities become $\tilde{U}_\mathrm{th} = U_\mathrm{th} + U_\mathrm{DP}$ and $\tilde{S}_\mathrm{th} = S_\mathrm{th} + S_\mathrm{DP}$, with $C_V^\mathrm{th} = (\partial \tilde{U}_\mathrm{th}/\partial T)_V = T\,(\partial \tilde{S}_\mathrm{th}/\partial T)_V = 3nR$. The pressure and bulk modulus are unaffected since $U_\mathrm{DP}$ depends only on temperature. Additionally, a softplus floor asymptoting to $10^{-7}\;\mathrm{K}^{-1}$ is applied to the thermal expansion coefficient $\alpha$ to prevent discontinuities from negative values at extreme compressions.

\paragraph{Electronic pressure.}
\label{app:electronic}
The thermal excitation of conduction electrons in iron gives rise to a pressure contribution that scales as $T^2$ at temperatures below the Fermi temperature \citep{Zharkov1971}. It is parametrized through an electronic Gr\"{u}neisen parameter mathematically behaving as its thermal counterpart in its power-law form (Eq.~\eqref{eq:gamma_power})
\begin{equation}
\label{eq:grunel}
e(V) = e_0\,\left(\frac{V}{V_0}\right)^g,
\end{equation}
$e_0$ being a reference value and $g$ the exponent, which yields the expressions of the pressure, heat capacity, entropy, and bulk modulus contributions:
\begin{align}
\label{eq:pel}
P_\mathrm{el} &= \frac{3nR\,e_0\,g\,(V/V_0)^g\,T^2}{2V}, \\
\label{eq:cvel}
C_V^\mathrm{el} &= 3nR\,e_0\,(V/V_0)^g\,T, \\
\label{eq:sel}
S_\mathrm{el} &= 3nR\,e_0\,(V/V_0)^g\,T, \\
\label{eq:kt_el}
K_T^\mathrm{el} &= P_\mathrm{el}(1 - g).
\end{align}
The thermal expansion for this model is $\alpha_\mathrm{el} K_T^\mathrm{el} = g\,C_V^\mathrm{el}/V$, and the contribution to internal energy flows from the pressure (Eq.~\eqref{eq:pel}) via a mirror relation of Eq.~\eqref{eq:pth}, giving $U_\mathrm{el} = P_\mathrm{el}V/g$.

\paragraph{Magnetic contribution.}
\label{app:magnetic}
The magnetic contribution to the Helmholtz free energy is parametrized as follows \citep{Dorogokupets2017}:
\begin{equation}
\label{eq:fmag}
F_\mathrm{mag}(T) = RT\ln(B_0 + 1)\left[f(\tau) - 1\right],
\end{equation}
where $\tau = T/T_\mathrm{c}$, $T_\mathrm{c}$ is the Curie temperature, $B_0$ is the saturation magnetic moment, and $f(\tau)$ is the scaling function defined piecewise. Because $F_\mathrm{mag}$ depends only on temperature, the magnetic pressure vanishes ($P_\mathrm{mag} = 0$, which also cancels $K_T^\mathrm{mag}$ and $\alpha_\mathrm{mag}$), but the magnetic contributions to heat capacity $C_V^\mathrm{mag} = -T\,\partial^2 F_\mathrm{mag}/\partial T^2$ and entropy $S_\mathrm{mag} = -\partial F_\mathrm{mag}/\partial T$ follow analytically and are retained.

\section{Phase boundary expressions}
\label{app:boundaries}

The phase diagrams of Figs.~\ref{fig:iron_phase} and~\ref{fig:mgsio3_phase} are built from a small set of analytical boundaries that \paleos inherits, in the case of iron, from the \textsc{biceps} compilation of \citet{Haldemann2024} and, in the case of MgSiO$_3$, from the primary mineral-physics literature directly. The expressions are shown here for reproducibility, with parameters inline and validity ranges given where relevant.

\subsection{Iron}
\label{app:boundaries:iron}

The solid--solid boundaries of \citet{Dorogokupets2017} are parametrized as polynomials in pressure for the $\gamma \leftrightarrow \varepsilon$ line and as piecewise linear segments for the low-pressure bcc transitions. With $P$ in GPa and the resulting temperature in kelvin,
\begin{align}
\label{eq:Tge}
T_{\gamma \leftrightarrow \varepsilon}(P) &= 679.517 + 17.733\,P + 0.213\,P^2 - 8.17\times 10^{-4}\,P^3, \\
\label{eq:Tag}
T_{\alpha \leftrightarrow \gamma}(P) &= 1120 - 300\,\frac{P}{7.3\;\mathrm{GPa}}, \\
\label{eq:Tdg}
T_{\delta \leftrightarrow \gamma}(P) &= 1580 + 418\,\frac{P}{5.2\;\mathrm{GPa}}, \\
\label{eq:Tae}
T_{\alpha \leftrightarrow \varepsilon}(P) &= 820 - 520\,\frac{P - 7.3\;\mathrm{GPa}}{8.5\;\mathrm{GPa}}.
\end{align}
Equation~\eqref{eq:Tge} is not the original \textsc{biceps} fit: we keep its quadratic and cubic curvature terms, calibrated to the high-pressure end of the boundary, but resolve the constant and linear coefficients so that the $\gamma \leftrightarrow \varepsilon$ boundary passes through both \citet{Dorogokupets2017} triple points, the $\alpha$--$\gamma$--$\varepsilon$ point at $(7.3\;\mathrm{GPa},\, 820\;\mathrm{K})$ and the $\varepsilon$--$\gamma$--liquid point at $(98.5\;\mathrm{GPa},\, 3712\;\mathrm{K})$. The original fit reproduces the latter but undershoots the former by ${\sim}\,100$\,K, which pinches the $\varepsilon$ field shut below ${\sim}\,8.5$\,GPa; the refit opens it at 7.3\,GPa. Equations~\eqref{eq:Tag} and~\eqref{eq:Tdg} are restricted to $P \leq 7.3$\,GPa and $P \leq 5.2$\,GPa respectively, and Eq.~\eqref{eq:Tae} to $7.3\;\mathrm{GPa} \leq P \leq 15.8$\,GPa.

The iron melting curve of \citet{Anzellini2013} is a two-branch Simon--Glatzel parametrization anchored at the reference point $(P_0,\, T_0) = (5.2\;\mathrm{GPa},\, 1991\;\mathrm{K})$ and the $\varepsilon$--$\gamma$--liquid triple point $(P_\mathrm{t},\, T_\mathrm{t}) = (98.5\;\mathrm{GPa},\, 3712\;\mathrm{K})$:
\begin{equation}
\label{eq:Tmelt_Fe}
T_\mathrm{melt}^\mathrm{Fe}(P) =
\begin{cases}
\displaystyle T_0\left(1 + \frac{P - P_0}{27.39\;\mathrm{GPa}}\right)^{1/2.38}, & P_0 \leq P < P_\mathrm{t}, \\[8pt]
\displaystyle T_\mathrm{t}\left(1 + \frac{P - P_\mathrm{t}}{161.2\;\mathrm{GPa}}\right)^{1/1.72}, & P \geq P_\mathrm{t}.
\end{cases}
\end{equation}

\subsection{Magnesium silicate}
\label{app:boundaries:mgsio3}

The three pyroxene polymorphs of \citet{Sokolova2022} meet at a triple point $(6.5\;\mathrm{GPa},\, 1100\;\mathrm{K})$. Two linear boundaries emanate from it, together with a parabola in $P(T)$ separating lpcen from en. With $T$ in kelvin and the resulting pressure in GPa,
\begin{align}
\label{eq:Plc_hc}
P_{\mathrm{lpcen} \leftrightarrow \mathrm{hpcen}}(T) &= 6.94 - 4.0\times 10^{-4}\,T, \\
\label{eq:Pen_hc}
P_{\mathrm{en} \leftrightarrow \mathrm{hpcen}}(T) &= 4.2 + 2.1\times 10^{-3}\,T, \\
\label{eq:Plc_en}
P_{\mathrm{lpcen} \leftrightarrow \mathrm{en}}(T) &= 15.6 - 4.78\times 10^{-2}\,T + 3.59\times 10^{-5}\,T^2.
\end{align}
Equation~\eqref{eq:Plc_en} is a parabola with a minimum near $T \approx 665$\,K; only the high-$T$ branch with $\dd P/\dd T > 0$ is physical, sweeping from $P \approx 0$ at $T \approx 750$\,K up to the triple point.

Above the triple-point pressure the \citet{Sokolova2022} calibration loses validity, and \paleos switches directly to brg at a constant cutoff $P = 12$\,GPa. This cutoff is a validity-range boundary rather than a calibrated Clapeyron line and is roughly consistent with the absorption of majorite and akimotoite into the brg stability field. The brg--ppv transition follows \citet{Ono2005},
\begin{equation}
\label{eq:Pbrg_ppv}
P_{\mathrm{brg} \leftrightarrow \mathrm{ppv}}(T) = 130\;\mathrm{GPa} + 7.0\times 10^{-3}\,(T - 2500)\;\mathrm{GPa\,K}^{-1},
\end{equation}
with $T$ in kelvin.

The MgSiO$_3$ melting curve is stitched from two parametrizations. The lower-bound fit of \citet{Fei2021} captures the correct asymptotic slope in the megabar regime relevant for deep mantles, but it is a log-linear power law in $P$ that collapses toward unphysically low temperatures as $P \to 0$ and therefore misrepresents the low-pressure phase structure, where the ambient-pressure melt temperature of en is ${\sim}\,1831$\,K. We therefore replaced it at low pressure with the Simon--Glatzel form of \citet{Belonoshko2005}, which anchors the correct ambient melt temperature and effectively lifts the low-pressure branch of the curve to higher temperatures than a pure \citet{Fei2021} extrapolation would yield. The two parametrizations are joined at the crossover pressure $P_\star \approx 2.55$\,GPa,
\begin{equation}
\label{eq:Tmelt_Mg}
T_\mathrm{melt}^{\mathrm{MgSiO_3}}(P) =
\begin{cases}
\displaystyle 1831\;\mathrm{K}\,\left(1 + \frac{P}{4.6\;\mathrm{GPa}}\right)^{0.33}, & P < P_\star, \\[8pt]
\displaystyle 6000\;\mathrm{K}\,\left(\frac{P}{140\;\mathrm{GPa}}\right)^{0.26}, & P \geq P_\star,
\end{cases}
\end{equation}
with $P_\star$ determined numerically from the intersection of the two branches to ensure continuity.

\end{appendix}

\end{document}